\documentclass{linear}

\usepackage[english]{babel}

\usepackage{pgfplots}

\usepackage{algorithm}
\usepackage{algpseudocode}
\crefname{algorithm}{Algorithm}{Algorithms}
\Crefname{algorithm}{Algorithm}{Algorithms}

\usetikzlibrary{3d}
\usetikzlibrary{external}
\usetikzlibrary{hobby}
\usetikzlibrary{patterns}
\usetikzlibrary{perspective}
\usetikzlibrary{spy}

\title{Unfitted Multi-Level \textit{hp} Refinement for Localized and Moving Solution Features}
\date{\today}

\author[1]{Jan Niklas Schmäke\thanks{Corresponding author \href{mailto:jan.schmaeke@hs-duesseldorf.de}{jan.schmaeke@hs-duesseldorf.de}}}
\author[1, 2]{Martin Ruess}

\affil[1]{%
  Düsseldorf University of Applied Sciences\\%
  Faculty of Mechanical and Process Engineering\\%
  Münsterstr. 156\\%
  40476 Düsseldorf\\%
  Germany}

\affil[2]{%
  Computer-Aided Methods in Civil Engineering\\%
  Brandenburg University of Technology Cottbus-Senftenberg\\%
  Konrad-Wachsmann-Allee 2\\%
  03046 Cottbus\\%
  Germany}

\newcommand{\fracd}[2]{\ensuremath\frac{\mathrm{d}#1}{\mathrm{d}#2}}

\tikzstyle{surface} = [bottom color = lightblue!50!white, top color = lightblue!50!white, middle color = lightblue!25!white, shading angle = -20]
\tikzstyle{alternativesurface} = [bottom color = lightgreen!50!white, top color = lightgreen!50!white, middle color = lightgreen!25!white, shading angle = -20]
\tikzstyle{inactivesurface} = [bottom color = lightgray!50!white, top color = lightgray!50!white, middle color = white, shading angle = -20]
\tikzstyle{highlightedsurface} = [bottom color = lightred!50!white, top color = lightred!50!white, middle color = lightred!25!white, shading angle = -20]

\colorlet{plotblue}{sky-blue2}
\colorlet{plotorange}{orange3}
\colorlet{plotgreen}{chameleon3}
\colorlet{plotgold}{butter3}
\colorlet{plotpurple}{plum1}
\colorlet{plotskyblue}{sky-blue1}
\colorlet{plotyellow}{butter2}
\colorlet{plotgray}{aluminium4}

\colorlet{listingkeyword}{plum3}
\colorlet{listingtype}{sky-blue3}
\colorlet{listingfunction}{sky-blue2}
\colorlet{listingcomment}{chameleon4}
\colorlet{listingstring}{orange4}
\colorlet{listingemphasis}{chocolate3}
\colorlet{listingrule}{aluminium2}
\colorlet{listingbackground}{aluminium0!35!white}

\pgfplotscreateplotcyclelist{plot markers}{%
	{plotblue, line width = 0.9pt, mark = *, mark size = 2.2pt,
		mark options = {solid, draw = plotblue!55!black, fill = plotblue, line width = 0.35pt}}\\%
	{plotorange, line width = 0.9pt, mark = square*, mark size = 2.2pt,
		mark options = {solid, draw = plotorange!55!black, fill = plotorange, line width = 0.35pt}}\\%
	{plotgreen, line width = 0.9pt, mark = triangle*, mark size = 2.3pt,
		mark options = {solid, draw = plotgreen!55!black, fill = plotgreen, line width = 0.35pt}}\\%
	{plotgold, line width = 0.9pt, mark = diamond*, mark size = 2.4pt,
		mark options = {solid, draw = plotgold!55!black, fill = plotgold, line width = 0.35pt}}\\%
	{plotpurple, line width = 0.9pt, mark = pentagon*, mark size = 2.3pt,
		mark options = {solid, draw = plotpurple!55!black, fill = plotpurple, line width = 0.35pt}}\\%
	{plotskyblue, line width = 0.9pt, mark = *, mark size = 2.2pt,
		mark options = {solid, draw = plotskyblue!55!black, fill = plotskyblue, line width = 0.35pt}}\\%
	{plotyellow, line width = 0.9pt, mark = square*, mark size = 2.2pt,
		mark options = {solid, draw = plotyellow!45!black, fill = plotyellow, line width = 0.35pt}}\\%
	{plotgray, line width = 0.9pt, mark = triangle*, mark size = 2.3pt,
		mark options = {solid, draw = plotgray!55!black, fill = plotgray, line width = 0.35pt}}\\%
}

\pgfplotscreateplotcyclelist{plot no markers}{%
	{plotblue, line width = 0.9pt, mark = none}\\%
	{plotorange, line width = 0.9pt, mark = none}\\%
	{plotgreen, line width = 0.9pt, mark = none}\\%
	{plotgold, line width = 0.9pt, mark = none}\\%
	{plotpurple, line width = 0.9pt, mark = none}\\%
	{plotskyblue, line width = 0.9pt, mark = none}\\%
	{plotyellow, line width = 0.9pt, mark = none}\\%
	{plotgray, line width = 0.9pt, mark = none}\\%
}

\pgfplotsset{
	plot markers/.style    = {cycle list name = plot markers},
	plot no markers/.style = {cycle list name = plot no markers},
	plot markers,
}

\pgfplotsset{
	colormap={drainbow}{%
		rgb = (0, 0, 0.360784)
		rgb = (0, 1, 1)
		rgb = (0, 0.501961, 0)
		rgb = (1, 1, 0)
		rgb = (1, 0.380392, 0)
		rgb = (0.419608, 0, 0)
	},
}

\newcommand{\colorbar}[6]{
	\begin{axis}[
		shift = {(#1, #2)},
    hide axis,
    scale only axis,
    height=0pt,
    width=0pt,
    colorbar,
    point meta min=#3,
    point meta max=#4,
    colorbar style={
      width = 0.15cm,
			height = #5,
			xlabel = {#6},
			font={\footnotesize},
			scaled ticks = false,
      ticklabel style = {/pgf/number format/fixed},
      xlabel style={yshift=4mm}
    }]
    \addplot [draw=none] coordinates {(0,0)};
	\end{axis}
}

\begin{document}

  \maketitle
  \hrule
  \section*{Abstract}
Localized features such as singularities, sharp gradients, discontinuities, and moving sources require adaptive finite element discretizations.
Conventional refinement strategies introduce significant computational overhead through mesh-topology modifications, constraint handling for non-matching interfaces, and repeated remeshing with state transfer.
This work presents an unfitted multi-level \textit{hp}-refinement strategy that enriches a fixed base discretization by independently positioned overlay meshes.
The global approximation space is constructed by superposition of the active spaces across all refinement levels, while homogeneous constraints on artificial overlay boundaries ensure global $\mathcal{C}^0$ continuity.
Coupling between non-matching meshes is assembled over admissible integration regions defined by intersections of element partitions, enabling reuse of standard element-level finite element routines within a lightweight superposition framework.
In contrast to fitted multi-level approaches, overlay boundaries are not required to align with underlying mesh interfaces.
This reduces inter-level coupling and allows refinement zones to be inserted, translated, and removed without modifying the base discretization.
Numerical studies for discontinuous and singular benchmark problems, as well as a moving source, demonstrate the performance of the method.
The unfitted approach retains exponential convergence for non-smooth problems and achieves improved error-to-cost ratios compared to fitted multi-level \textit{hp}-refinement.
For representative cases, comparable accuracy is obtained with substantially fewer degrees of freedom, while localized high-order refinement accurately tracks moving features.

\section*{Keywords}
\begin{flushleft}
  Unfitted Refinement\keywordsep Multi-Level \textit{hp}-Refinement\keywordsep Overlay Meshes\keywordsep Moving Heat Sources
\end{flushleft}

  \clearpage

  \tableofcontents
  \clearpage

  \section{Introduction}

The accurate resolution of localized features such as singularities, sharp gradients, discontinuities, and moving source terms remains a central challenge in finite element analysis.
Prominent examples arise in crack propagation \cite{Borden:2012aa, Miehe:2010aa}, contact mechanics \cite{Wriggers:1998aa, Popp:2010aa}, and process-driven simulations such as additive manufacturing \cite{Moreira:2022aa, Kollmannsberger:2022aa}.
Conventional mesh-adaptive approaches, including conforming \textit{h}-refinement, incur significant algorithmic overhead due to mesh distortion, repeated mesh generation, and the projection of state variables, thereby limiting their robustness and efficiency in transient settings \cite{Demkowicz:2006aa}.

Unfitted element methods such as the finite cell method \cite{Parvizian:2007aa, Schillinger:2015aa} mitigate these limitations by decoupling the approximation space from the mesh topology.
In this context, structured, axis-aligned meshes offer clear advantages in terms of implementation simplicity and efficient, dimension-independent intersection algorithms.
However, within a fixed discretization, the accurate representation of localized or moving features requires enrichment mechanisms that can be introduced, moved, and removed without repeated remeshing or global changes to the discretization \cite{Zander:2016aa, Fries:2010aa}.

The present work revisits and extends the concept of refinement-by-superposition as a systematic strategy for elementwise \textit{hp}-enrichment \cite{Belytschko:1990aa, Fish:1992aa, Schillinger:2015aa, Rank:1992aa}.
In contrast to classical refinement approaches that modify the underlying mesh topology, superposition-based refinement constructs hierarchical approximation spaces through the overlay of locally enriched discretizations.
This paradigm enables strict locality of refinement while preserving the integrity of the base mesh \cite{Schillinger:2012ab}.
Although superposition techniques have been applied to crack modelling and transient elastodynamics \cite{Lee:2004aa, Yue:2005aa}, their use for dynamically evolving features has received limited attention, particularly in the context of unfitted discretization.

\begin{figure}[!ht]
  \centering
  \resizebox{\textwidth}{!}{\tikzsetnextfilename{superposition_concept}
\begin{tikzpicture}[
		approximation/.style = {blue, thick},
		exactsolution/.style = {black!55, thin, densely dashed},
		declare function = {
			ub(\x) = 1 / (2 + sin(90 * \x - 90)) + 0.06 * \x;
			uexact(\x) = ub(\x) + 0.42 * sin(180 * (\x + 1));
		},
	]
	\begin{scope}[shift = {(0, 0)}]
		\draw[gray, densely dotted] (-2, 0) -- ++(0, {ub(-2)});
		\draw[gray, densely dotted] (2, 0) -- ++(0, {ub(2)});

		\draw[black, densely dashed, -latex] (-2.4, 0) -- node[below, pos = 1] {$x$} ++(4.8, 0);
		\draw[black, thick] (-2, 0) -- ++(4, 0);
		\draw[black, thick] (-2, -0.1) -- ++(0, 0.2);
		\draw[black, thick] (2, -0.1) -- ++(0, 0.2);

		\draw[exactsolution] (0, 0) plot[domain=-2:-1, samples = 25, smooth] ({\x}, {ub(\x)});
		\draw[exactsolution] (0, 0) plot[domain=-1:1, samples = 45, smooth] ({\x}, {uexact(\x)});
		\draw[exactsolution] (0, 0) plot[domain=1:2, samples = 25, smooth] ({\x}, {ub(\x)});

		\draw[approximation] (0, 0) plot[domain=-2:2, samples = 45, smooth] ({\x}, {ub(\x)});
		\draw[approximation, fill = blue] (-2, {ub(-2)}) circle(1.5pt);
		\draw[approximation, fill = blue] ( 2, {ub(2)}) circle(1.5pt);

		\node[below right] at (-2.6, -0.3) {\footnotesize base approximation --- $u_b$};
	\end{scope}

	\begin{scope}[shift = {(6, 0)}]
		\draw[gray, densely dotted] ({-1/3}, 0) -- ++(0, 0.36);
		\draw[gray, densely dotted] ({1/3}, 0) -- ++(0, -0.36);

		\draw[black, densely dashed, -latex] (-2.4, 0) -- node[below, pos = 1] {$x$} ++(4.8, 0);
		\draw[black, thick] (-1, 0) -- ++(2, 0);
		\draw[black, thick] (-1, -0.1) -- ++(0, 0.2);
		\draw[black, thick] ({-1/3}, -0.1) -- ++(0, 0.2);
		\draw[black, thick] ({1/3}, -0.1) -- ++(0, 0.2);
		\draw[black, thick] (1, -0.1) -- ++(0, 0.2);

		\draw[approximation] (-1, 0) -- ++({2 / 3}, 0.36) -- ++({2 / 3}, -0.72) -- ++({2/3}, 0.36);

		\draw[approximation, fill = white] (-1, 0) circle(1.5pt);
		\draw[approximation, fill = blue] ({-1/3}, 0.36) circle(1.5pt);
		\draw[approximation, fill = blue] ({1/3}, -0.36) circle(1.5pt);
		\draw[approximation, fill = white] (1, 0) circle(1.5pt);

		\node[below right] at (-2.6, -0.3) {\footnotesize overlay contribution --- $u_o$};
	\end{scope}

	\begin{scope}[shift = {(12, 0)}]
		\draw[gray, densely dotted] (-2, 0) -- ++(0, {ub(-2)});
		\draw[gray, densely dotted] (-1, 0) -- ++(0, {ub(-1)});
		\draw[gray, densely dotted] ({-1/3}, 0) -- ++(0, {ub(-1/3) + 0.36});
		\draw[gray, densely dotted] ({1/3}, 0) -- ++(0, {ub(1/3) - 0.36});
		\draw[gray, densely dotted] (1, 0) -- ++(0, {ub(1)});
		\draw[gray, densely dotted] (2, 0) -- ++(0, {ub(2)});

		\draw[black, densely dashed, -latex] (-2.4, 0) -- node[below, pos = 1] {$x$} ++(4.8, 0);
		\draw[black, thick] (-2, 0) -- ++(4, 0);
		\draw[black, thick] (-2, -0.1) -- ++(0, 0.2);
		\draw[black, thick] (-1, -0.1) -- ++(0, 0.2);
		\draw[black, thick] ({-1/3}, -0.1) -- ++(0, 0.2);
		\draw[black, thick] ({1/3}, -0.1) -- ++(0, 0.2);
		\draw[black, thick] (1, -0.1) -- ++(0, 0.2);
		\draw[black, thick] (2, -0.1) -- ++(0, 0.2);

		\draw[exactsolution] (0, 0) plot[domain=-2:-1, samples = 25, smooth] ({\x}, {ub(\x)});
		\draw[exactsolution] (0, 0) plot[domain=-1:1, samples = 45, smooth] ({\x}, {uexact(\x)});
		\draw[exactsolution] (0, 0) plot[domain=1:2, samples = 25, smooth] ({\x}, {ub(\x)});

		\draw[approximation] (0, 0) plot[domain=-2:-1, samples = 25, smooth] ({\x}, {ub(\x)});

		\draw[approximation] (0, 0) plot[domain=-1:-0.333, samples = 25, smooth] ({\x}, {ub(\x) + 0.36 / 2 * 3 * (\x + 1)});
		\draw[approximation] (0, 0) plot[domain=-0.333:0.333, samples = 25, smooth] ({\x}, {ub(\x) - 0.36 * 3 * \x});
		\draw[approximation] (0, 0) plot[domain=0.333:1, samples = 25, smooth] ({\x}, {ub(\x) + 0.36 / 2 * 3 * (\x - 1)});

		\draw[approximation] (0, 0) plot[domain=1:2, samples = 25, smooth] ({\x}, {ub(\x)});

		\draw[approximation, fill = blue] (-2, {ub(-2)}) circle(1.5pt);
		\draw[approximation, fill = white] (-1, {ub(-1)}) circle(1.5pt);
		\draw[approximation, fill = blue] ({-1/3}, {ub(-1/3) + 0.36}) circle(1.5pt);
		\draw[approximation, fill = blue] ({1/3}, {ub(1/3) - 0.36}) circle(1.5pt);
		\draw[approximation, fill = white] (1, {ub(1)}) circle(1.5pt);
		\draw[approximation, fill = blue] (2, {ub(2)}) circle(1.5pt);

		\node[below right] at (-2.6, -0.3) {\footnotesize enriched approximation --- $u = u_b + u_o$};
	\end{scope}

	\node[] at (3, 0) {\Large $\boldsymbol{+}$};
	\node[] at (9, 0) {\Large $\boldsymbol{=}$};

	\draw[blue, thick, fill = blue] (-2, -1) circle (1.5pt) node[black, right] {\footnotesize Active node};
	\draw[blue, thick, fill = white] (1, -1) circle (1.5pt) node[black, right] {\footnotesize Constrained node};
	\draw[exactsolution] (4.5, -1) -- ++(0.6, 0) node[black, right] {\footnotesize Exact solution};
\end{tikzpicture} }
  \caption{Concept of refinement-by-superposition in one dimension. A locally refined overlay discretization is superimposed on a coarser base discretization, and the final coupled approximation $u$ is represented as the sum of the base contribution $u_b$ and the overlay contribution $u_o$. Filled nodes denote active degrees of freedom, while open nodes are constrained to zero such that the overlay contribution vanishes at the boundary of the refinement region.}
  \label{fig:superposition_concept}
\end{figure}
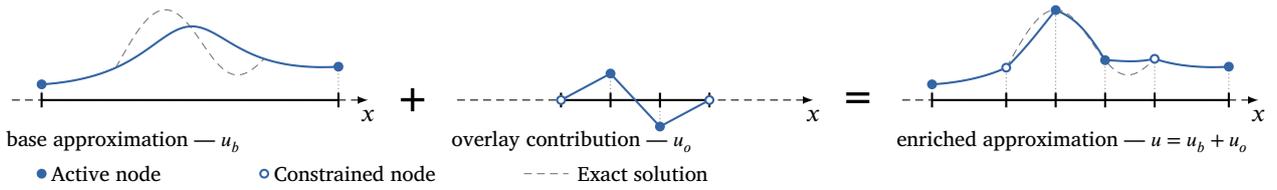

The basic construction of superposed refinement is illustrated in \Cref{fig:superposition_concept}.
A coarse base discretization is enriched by a locally supported overlay discretization, and the final approximation is obtained by summing both contributions.
This decomposition is assembled as one coupled variational problem; the separate curves in the figure indicate contributions to the same approximation rather than independently computed solutions.
Homogeneous constraints on the overlay boundary ensure that the local contribution vanishes at the boundary of the refinement region, such that global continuity of the approximation is preserved.

We propose a framework for unfitted multi-level \textit{hp}-refinement on axis-aligned finite element meshes, in which localized overlay meshes are positioned independently of the background discretization.
The method combines hierarchical superposition with non-matching overlay meshes to build an adaptable approximation space that can be aligned with discontinuities, singular gradients, and moving source terms.
The formulation supports multiple refinement levels, enforces the constraints required for global continuity, and admits a consistent variational treatment of overlapping approximation spaces.
Particular attention is given to the construction of integration regions for non-matching meshes, the role of deactivation rules in the unfitted setting, and conditioning effects caused by small overlaps.

The proposed approach is designed to operate without remeshing and without global modification of the underlying discretization.
Its behavior is demonstrated for benchmark problems with discontinuous or singular solution features and for a transient heat-transfer problem with a moving heat source motivated by additive manufacturing.
The numerical results show that the method can achieve high accuracy with substantially fewer degrees of freedom than fitted multi-level \textit{hp}-refinement or highly resolved reference meshes in the considered examples.

  \section{Conceptual aspects of refinement-by-superposition}
Since \textit{refinement-by-superposition} was pioneered for localized high-gradient problems \cite{Belytschko:1990aa}, two principal methodological lines have emerged: (i) the \textit{s}-version \cite{Fish:1992aa}, and (ii) the \textit{hp}-decomposition approach \cite{Rank:1992aa}.
The presented method shares important features of both approaches.
Therefore, we provide a concise overview of these approaches and their modern developments.

In contrast to conforming refinement-by-replacement schemes, superposition-based methods avoid the creation of hanging nodes and associated explicit constraint equations.
Compatibility between overlapping approximation spaces is instead enforced through overlay boundary conditions and, in fitted hierarchical variants, through deactivation of affected basis functions.

\subsection{The \textit{s}-version}
The \textit{s}-version \cite{Fish:1992aa} superimposes topologically independent meshes.
The general workflow applies a base mesh covering the analysis domain and superimposes a locally confined finer mesh in regions where enhanced resolution is required.
Global $\mathcal{C}^0$ compatibility is enforced by homogeneous Dirichlet boundary conditions on overlay boundaries that do not coincide with the boundaries of the global analysis domain, ensuring that the overlay contribution vanishes at the boundary of the refinement region.
This approach demonstrated that localized superposition reduces the global error for problems with non-smooth solutions and improves the effectiveness of global \textit{p}-refinement.
Subsequent developments extended the method to adaptive formulations for linear elastostatics \cite{Fish:1993aa}.

While the original formulation was restricted to a single overlay mesh, subsequent research extended the approach to multiple overlays \cite{Fish:1994aa}.
However, to simplify implementation, the applied method required the overlay boundaries to align with element boundaries of the underlying mesh.
This constraint precludes partial element coverage by overlays and introduces a topological coupling between refined levels.

Many recent \textit{s}-version approaches retain these restrictions, employing either a single unfitted overlay mesh \cite{He:2023aa, Shibanuma:2022aa, Tu:2024aa, Magome:2024aa} or multiple fitted overlays \cite{Yan:2023aa, Yan:2024aa}.

\subsection{The fitted multi-level \textit{hp} method}
The \textit{hp}-decomposition approach \cite{Rank:1992aa} adopts a similar superposition strategy, in which a high-order base mesh is locally covered by a finer, lower-order overlay mesh.
In this formulation, the overlay mesh must align with element boundaries and cannot partially cover individual base-mesh elements.
The original method was further limited to a single overlay mesh.

Subsequent work by Schillinger et al.\ \cite{Schillinger:2010aa, Schillinger:2011aa, Schillinger:2012aa} extended this method to multiple overlay meshes within hierarchical approximation spaces in the context of isogeometric and embedded-domain analysis.
The core principle adopted in these studies bisects each element selected for refinement once per spatial dimension, generating $2^d$ overlay elements per refined underlying element, where $d$ denotes the spatial dimension.
This results in refinement patterns closely resembling those of conventional refinement-by-replacement methods.
For high-order elements based on integrated Legendre shape functions, overlay meshes were initially restricted to linear basis functions.

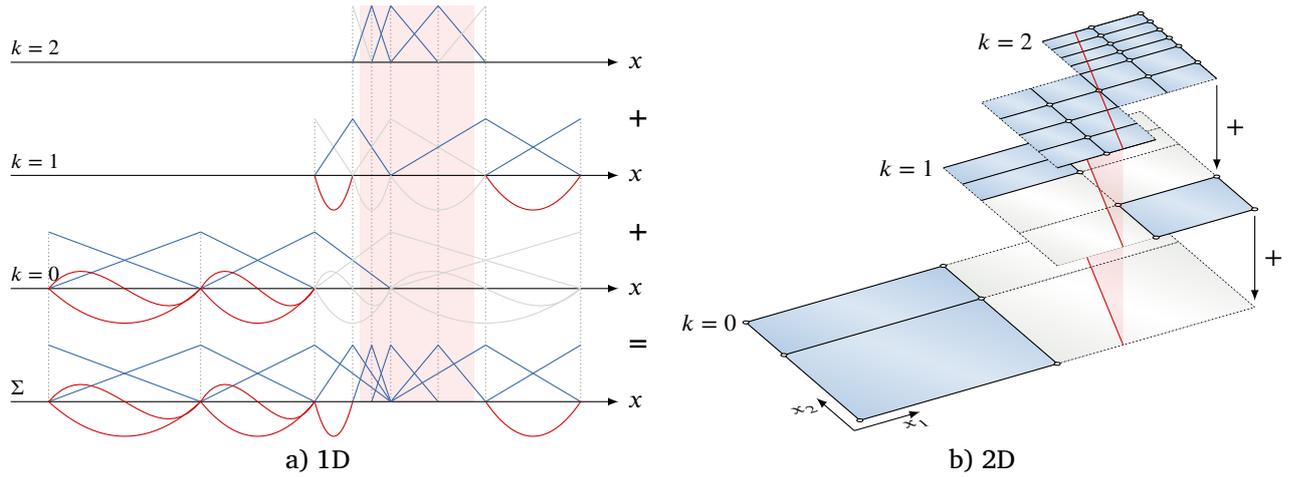
\begin{figure}[!ht]
  \centering
  \begin{tabular}{@{} c @{\hspace{0cm}} c @{}}
    \tikzsetnextfilename{multi_level_hp_shapes}

\newcommand{\linearshapel}[4]{
  \draw[#4] (#1, {#2 + 0.75cm}) -- ++(#3, -0.75cm);
}

\newcommand{\linearshaper}[4]{
  \draw[#4] (#1, #2) -- ++(#3, 0.75cm);
}

\newcommand{\quadraticshape}[4]{
  \draw[#4] plot[domain = 0:1] (%
    {#1 + #3 * \x},%
    {#2 + 0.75cm * (0.25 * sqrt(6) * ((2 * \x - 1)^2 - 1))}%
  );
}

\newcommand{\cubicshape}[4]{
  \draw[#4] plot[domain = 0:1] (%
    {#1 + #3 * \x},%
    {#2 + 0.75cm * (0.25 * sqrt(10) * ((2 * \x - 1)^2 - 1) * (2 * \x - 1))}%
  );
}

\begin{tikzpicture}

  \fill[lightred, opacity = 0.1] (4.1, -1.5) rectangle (5.6, 3.75);

  \draw[gray, densely dotted] (0, -1.5) -- ++(0, {0.75 + 1 * 1.5});
  \draw[gray, densely dotted] (2, -1.5) -- ++(0, {0.75 + 1 * 1.5});
  \draw[gray, densely dotted] (3.5, -1.5) -- ++(0, {0.75 + 2 * 1.5});
  \draw[gray, densely dotted] (4, -1.5) -- ++(0, {0.75 + 3 * 1.5});
  \draw[gray, densely dotted] (4.25, -1.5) -- ++(0, {0.75 + 3 * 1.5});
  \draw[gray, densely dotted] (4.5, -1.5) -- ++(0, {0.75 + 3 * 1.5});
  \draw[gray, densely dotted] (5.125, -1.5) -- ++(0, {0.75 + 3 * 1.5});
  \draw[gray, densely dotted] (5.75, -1.5) -- ++(0, {0.75 + 3 * 1.5});
  \draw[gray, densely dotted] (7, -1.5) -- ++(0, {0.75 + 2 * 1.5});

  \begin{scope}[xshift = 0, yshift = 0]
    \draw[black, -latex] (-0.5, 0) -- node[pos = 1, right] {$x$} ++(8cm, 0cm);
    \node[above right, inner sep = 0, outer sep = 0, yshift = 1mm] at (-0.5, 0) {\footnotesize$k = 0$};

    \linearshapel{0cm}{0cm}{2cm}{blue}
    \linearshaper{0cm}{0cm}{2cm}{blue}
    \quadraticshape{0cm}{0cm}{2cm}{red}
    \cubicshape{0cm}{0cm}{2cm}{red}

    \linearshapel{2cm}{0cm}{1.5cm}{blue}
    \linearshaper{2cm}{0cm}{1.5cm}{blue}
    \quadraticshape{2cm}{0cm}{1.5cm}{red}
    \cubicshape{2cm}{0cm}{1.5cm}{red}

    \linearshapel{3.5cm}{0cm}{1cm}{blue}
    \linearshaper{3.5cm}{0cm}{1cm}{lightgray}
    \quadraticshape{3.5cm}{0cm}{1cm}{lightgray}
    \cubicshape{3.5cm}{0cm}{1cm}{lightgray}

    \linearshapel{4.5cm}{0cm}{2.5cm}{lightgray}
    \linearshaper{4.5cm}{0cm}{2.5cm}{lightgray}
    \quadraticshape{4.5cm}{0cm}{2.5cm}{lightgray}
    \cubicshape{4.5cm}{0cm}{2.5cm}{lightgray}
  \end{scope}

  \begin{scope}[xshift = 0, yshift = 1.5cm]
    \draw[black, -latex] (-0.5, 0) -- node[pos = 1, right] {$x$} ++(8cm, 0cm);
    \node[above right, inner sep = 0, outer sep = 0, yshift = 1mm] at (-0.5, 0) {\footnotesize$k = 1$};

    \linearshapel{3.5cm}{0cm}{0.5cm}{lightgray}
    \linearshaper{3.5cm}{0cm}{0.5cm}{blue}
    \quadraticshape{3.5cm}{0cm}{0.5cm}{red}

    \linearshapel{4cm}{0cm}{0.5cm}{blue}
    \linearshaper{4cm}{0cm}{0.5cm}{lightgray}
    \quadraticshape{4cm}{0cm}{0.5cm}{lightgray}

    \linearshapel{4.5cm}{0cm}{1.25cm}{lightgray}
    \linearshaper{4.5cm}{0cm}{1.25cm}{blue}
    \quadraticshape{4.5cm}{0cm}{1.25cm}{lightgray}

    \linearshapel{5.75cm}{0cm}{1.25cm}{blue}
    \linearshaper{5.75cm}{0cm}{1.25cm}{blue}
    \quadraticshape{5.75cm}{0cm}{1.25cm}{red}
  \end{scope}

  \begin{scope}[xshift = 0, yshift = 3cm]
    \draw[black, -latex] (-0.5, 0) -- node[pos = 1, right] {$x$} ++(8cm, 0cm);
    \node[above right, inner sep = 0, outer sep = 0, yshift = 1mm] at (-0.5, 0) {\footnotesize$k = 2$};

    \linearshapel{4.0cm}{0cm}{0.25cm}{lightgray}
    \linearshaper{4.0cm}{0cm}{0.25cm}{blue}

    \linearshapel{4.25cm}{0cm}{0.25cm}{blue}
    \linearshaper{4.25cm}{0cm}{0.25cm}{blue}

    \linearshapel{4.5cm}{0cm}{0.625cm}{blue}
    \linearshaper{4.5cm}{0cm}{0.625cm}{blue}

    \linearshapel{5.125cm}{0cm}{0.625cm}{blue}
    \linearshaper{5.125cm}{0cm}{0.625cm}{lightgray}
  \end{scope}

  \begin{scope}[xshift = 0, yshift = -1.5cm]
    \draw[black, -latex] (-0.5, 0) -- node[pos = 1, right] {$x$} ++(8cm, 0cm);
    \node[above right, inner sep = 0, outer sep = 0, yshift = 1mm] at (-0.5, 0) {\footnotesize$\Sigma$};

    \linearshapel{0cm}{0cm}{2cm}{blue}
    \linearshaper{0cm}{0cm}{2cm}{blue}
    \quadraticshape{0cm}{0cm}{2cm}{red}
    \cubicshape{0cm}{0cm}{2cm}{red}

    \linearshapel{2cm}{0cm}{1.5cm}{blue}
    \linearshaper{2cm}{0cm}{1.5cm}{blue}
    \quadraticshape{2cm}{0cm}{1.5cm}{red}
    \cubicshape{2cm}{0cm}{1.5cm}{red}

    \linearshapel{3.5cm}{0cm}{1cm}{blue}

    \linearshaper{3.5cm}{0cm}{0.5cm}{blue}
    \quadraticshape{3.5cm}{0cm}{0.5cm}{red}

    \linearshapel{4cm}{0cm}{0.5cm}{blue}

    \linearshaper{4.5cm}{0cm}{1.25cm}{blue}

    \linearshapel{5.75cm}{0cm}{1.25cm}{blue}
    \linearshaper{5.75cm}{0cm}{1.25cm}{blue}
    \quadraticshape{5.75cm}{0cm}{1.25cm}{red}
    \linearshaper{4.0cm}{0cm}{0.25cm}{blue}

    \linearshapel{4.25cm}{0cm}{0.25cm}{blue}
    \linearshaper{4.25cm}{0cm}{0.25cm}{blue}

    \linearshapel{4.5cm}{0cm}{0.625cm}{blue}
    \linearshaper{4.5cm}{0cm}{0.625cm}{blue}

    \linearshapel{5.125cm}{0cm}{0.625cm}{blue}
  \end{scope}

  \node[right] at (7.5, 0.75) {$\boldsymbol{+}$};
  \node[right] at (7.5, 2.25) {$\boldsymbol{+}$};
  \node[right] at (7.5, -0.75) {$\boldsymbol{=}$};

\end{tikzpicture} &
    \tikzsetnextfilename{multi_level_hp_concept}
\begin{tikzpicture}[3d view = {-30}{30}]

  \begin{scope}[canvas is xy plane at z = 0]

    \draw[black, -latex] (-0.2, -0.2) -- node[pos = 0.8, below, transform shape] {$x_1$} ++(1, 0);
    \draw[black, -latex] (-0.2, -0.2) -- node[pos = 0.8, left, transform shape] {$x_2$} ++(0, 1);

    \fill[surface] (0, 0) rectangle ++(3, 2);
    \fill[surface] (0, 2) rectangle ++(3, 1);

    \fill[inactivesurface] (3, 0) rectangle ++(3, 2);
    \fill[inactivesurface] (3, 2) rectangle ++(3, 1);

    \draw[black] (0, 0) rectangle ++(3, 3);
    \draw[black] (0, 2) -- ++(3, 0);
    \draw[black, densely dotted] (3, 0) -- ++(3, 0) -- ++(0, 3) -- ++(-3, 0);
    \draw[black, densely dotted] (3, 2) -- ++(3, 0);

    \foreach \x in {0, 3}
    {
      \foreach \y in {0, 2, 3}
      {
        \draw[black, fill = white] (\x, \y) circle (0.4mm);
      }
    }

    \draw[red] (4, 0) -- ++(1, 3);
  \end{scope}

  \fill[lightred, fill opacity = 0.1] (4, 0, 0) -- ++(1, 3, 0) -- ++(0, 0, 1.5) -- ++(-1, -3, 0) -- cycle;
  \draw[black, latex-] (6, 0, 0.1) -- node[right] {$+$} ++(0, 0, 1.3);

  \begin{scope}[canvas is xy plane at z = 1.5]

    \fill[surface] (4.5, 0) rectangle ++(1.5, 1);
    \fill[surface] (3, 2) rectangle ++(1.5, 0.5);
    \fill[surface] (3, 2.5) rectangle ++(1.5, 0.5);

    \fill[inactivesurface] (3, 0) rectangle ++(1.5, 1);
    \fill[inactivesurface] (3, 1) rectangle ++(1.5, 1);
    \fill[inactivesurface] (4.5, 1) rectangle ++(1.5, 1);
    \fill[inactivesurface] (4.5, 2) rectangle ++(1.5, 0.5);
    \fill[inactivesurface] (4.5, 2.5) rectangle ++(1.5, 0.5);

    \draw[black] (4.5, 0) rectangle ++(1.5, 1);
    \draw[black] (3, 2) -- ++(1.5, 0) -- ++(0, 1) -- ++(-1.5, 0);
    \draw[black] (3, 2.5) -- ++(1.5, 0);

    \draw[black, densely dotted] (3, 0) -- ++(0, 3);
    \draw[black, densely dotted] (6, 1) -- ++(0, 2) -- ++(-1.5, 0);
    \draw[black, densely dotted] (3, 1) -- ++(1.5, 0) -- ++(0, 1) -- ++(1.5, 0);
    \draw[black, densely dotted] (4.5, 2.5) -- ++(1.5, 0);

    \foreach \x in {4.5}
    {
      \foreach \y in {0, 1, 2, 2.5, 3}
      {
        \draw[black, fill = white] (\x, \y) circle (0.4mm);
      }
    }
    \draw[black, fill = white] (6, 0) circle (0.4mm);
    \draw[black, fill = white] (6, 1) circle (0.4mm);

    \draw[red] (4, 0) -- ++(1, 3);
  \end{scope}

  \fill[lightred, fill opacity = 0.1] (4, 0, 1.5) -- ++(1, 3, 0) -- ++(0, 0, 1.5) -- ++(-1, -3, 0) -- cycle;

  \draw[black, latex-] (6, 1, 1.6) -- node[right] {$+$} ++(0, 0, 1.3);

  \begin{scope}[canvas is xy plane at z = 3]
    \foreach \x in {3, 3.75}
    {
      \foreach \y in {0, 0.5, 1, 1.5}
      {
        \fill[surface] (\x, \y) rectangle ++(0.75, 0.5);
      }
    }

    \foreach \x in {4.5, 5.25}
    {
      \foreach \y in {1, 1.5}
      {
        \fill[surface] (\x, \y) rectangle ++(0.75, 0.5);
      }
    }

    \foreach \x in {4.5, 5.25}
    {
      \foreach \y in {2, 2.25, 2.5, 2.75}
      {
        \fill[surface] (\x, \y) rectangle ++(0.75, 0.25);
      }
    }

    \draw[black] (3.75, 0) -- ++(0, 2);
    \draw[black] (4.5, 1) -- ++(0, 1);
    \draw[black] (5.25, 1) -- ++(0, 2);
    \foreach \y in {0.5, 1, 1.5}
    {
      \draw[black] (3, \y) -- ++(1.5, 0);
      \draw[black, fill = white] (3.75, \y) circle (0.4mm);
    }
    \foreach \y in {1.5, 2, 2.25, 2.5, 2.75}
    {
      \draw[black] (4.5, \y) -- ++(1.5, 0);
    }
    \draw[black, fill = white] (4.5, 1.5) circle (0.4mm);
    \draw[black, densely dotted] (3, 0) -- ++(1.5, 0) -- ++(0, 1) -- ++(1.5, 0) -- ++(0, 2) -- ++(-1.5, 0) -- ++(0, -1) -- ++(-1.5, 0) -- cycle;

    \draw[black] (3, 0) -- ++(1.5, 0);
    \draw[black] (6, 1) -- ++(0, 2) -- ++(-1.5, 0);

    \foreach \x in {5.25, 6}
    {
      \foreach \y in {1.5, 2, 2.25, 2.5, 2.75, 3}
      {
        \draw[black, fill = white] (\x, \y) circle (0.4mm);
      }
    }
    \draw[black, fill = white] (3.75, 0) circle (0.4mm);

    \draw[red] (4, 0) -- ++(1, 3);
  \end{scope}

  \node[left] at (0, 3, 0) {\small$k = 0$};
  \node[left] at (3, 3, 1.5) {\small$k = 1$};
  \node[left] at (4.5, 3, 3) {\small$k = 2$};

\end{tikzpicture} \\
    a) 1D & b) 2D \\
  \end{tabular}
  \caption{Schematic illustration of the fitted \cite{Zander:2015aa, Zander:2016aa, Zander:2017ab, Zander:2022aa, Kopp:2022aa} multi-level \textit{hp}-refinement in one dimension (a) and two dimensions (b). Global continuity and linear independence are enforced by deactivating selected shape functions, shown in gray in (a) and as gray surfaces, dotted lines, and missing vertices in (b).}
  \label{fig:fitted_mlhp_concept}
\end{figure}

Zander later generalized this method by enabling high-order basis functions across all levels of the hierarchical finite element discretization \cite{Zander:2015aa, Zander:2016aa, Zander:2017ab}.
A schematic illustration of the method is shown in \Cref{fig:fitted_mlhp_concept}.

This extension is commonly referred to as the multi-level \textit{hp} method.
\Cref{fig:fitted_mlhp_concept} (a) shows a one-dimensional example, starting from a base mesh ($k = 0$) consisting of four high-order elements of unequal size and polynomial degree $p = 3$.
The region shaded in red indicates a domain requiring increased resolution.
In the first overlay layer ($k = 1$), all elements partially or fully intersecting the red region are bisected into two elements of equal size.
The polynomial degree on each refinement layer is defined by the user.
In the example shown, the polynomial degree is reduced by one with each refinement step, starting from $p = 3$ on the base mesh.

All shape functions located on the boundary of a given mesh level that do not coincide with the boundary of the analysis domain are deactivated to enforce global $\mathcal{C}^0$ continuity of the approximation space.
Here, deactivation denotes removal of the corresponding functions from the active approximation space, which is equivalent to imposing zero coefficients; deactivated functions are indicated in gray.
Linear independence is ensured by additionally deactivating shape functions associated with vertices or edges whenever these topological entities are refined and their refined counterparts remain active.
This additional deactivation is applied only to shape functions that have not already been eliminated by continuity constraints.
The procedure is then repeated for subsequent refinement levels, e.g., $k = 2$.
The bottom layer of \Cref{fig:fitted_mlhp_concept} (a), labeled $\Sigma$, visualizes the sum of all active shape functions and represents the final approximation space used to approximate the solution.
\Cref{fig:fitted_mlhp_concept} (b) illustrates the corresponding two-dimensional refinement, again with a red region indicating the need for additional resolution.
Instead of individual shape functions, topological mesh entities, namely faces, edges, and vertices, are shown, with deactivated components indicated by missing vertices, dotted edges, and gray faces.

The rules for continuity and linear independence extend to arbitrary spatial dimensions and enable the construction of a dimension-independent hypercube-based finite element framework, as demonstrated by Kopp et al.\ \cite{Kopp:2022aa}.
In this context, a hypercube refers to the tensor-product cell in dimension $d$: an interval in one dimension, a quadrilateral in two dimensions, a hexahedron in three dimensions, and, in general, the $d$-cube.

A remaining limitation of this approach is the strong coupling of topological entities across different mesh levels.
This coupling arises from the fitted refinement pattern: vertices, edges, faces, and their refined counterparts must be tracked across levels to enforce both continuity and linear independence.
While the method can be adapted to anisotropic refinement in two dimensions \cite{Zander:2022aa}, its extension to three or higher dimensions becomes increasingly complex due to the growing number of refinement configurations that must be handled.

\section{The unfitted multi-level \textit{hp} method}
The \textit{unfitted} multi-level \textit{hp}-refinement concept introduced in this section constitutes the central contribution of this work.
Building on the framework of the fitted multi-level \textit{hp} method, it removes the requirement that overlay boundaries coincide with element boundaries of the underlying mesh.

The method adopts an element-independent refinement paradigm.
Refined meshes are superimposed independently of the base discretization, rather than being generated through successive element-wise bisection.
This decouples mesh topology across refinement levels and enables greater flexibility in positioning refinement regions.
As a result, localized features can be resolved more accurately and efficiently by aligning the refinement with the underlying solution structure.

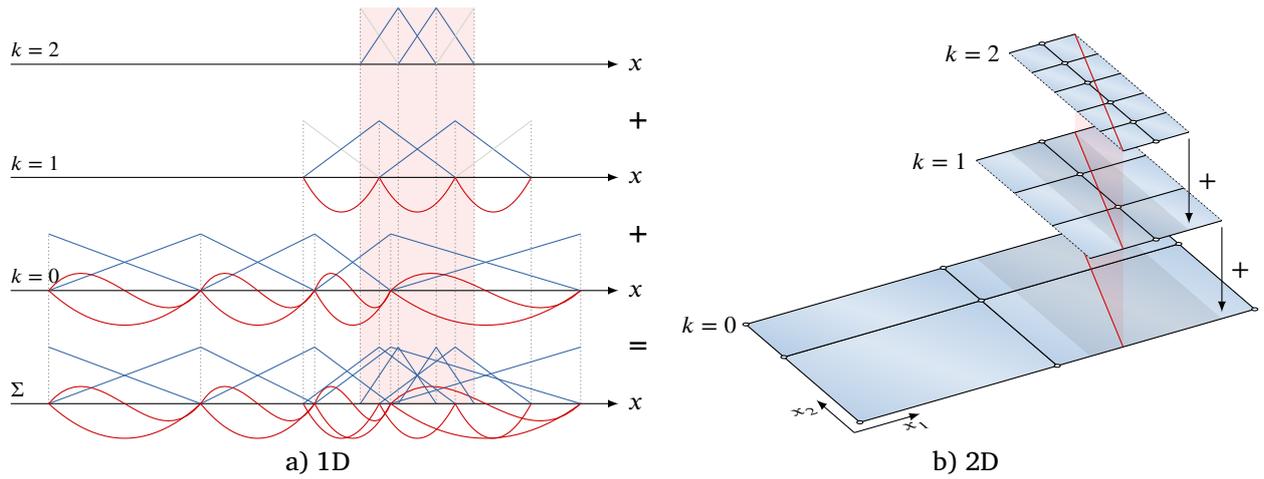
\begin{figure}[!ht]
  \centering
  \begin{tabular}{@{} c @{\hspace{0cm}} c @{}}
    \tikzsetnextfilename{unfitted_multi_level_hp_shapes}

\newcommand{\linearshapel}[4]{
  \draw[#4] (#1, {#2 + 0.75cm}) -- ++(#3, -0.75cm);
}

\newcommand{\linearshaper}[4]{
  \draw[#4] (#1, #2) -- ++(#3, 0.75cm);
}

\newcommand{\quadraticshape}[4]{
  \draw[#4] plot[domain = 0:1] (%
    {#1 + #3 * \x},%
    {#2 + 0.75cm * (0.25 * sqrt(6) * ((2 * \x - 1)^2 - 1))}%
  );
}

\newcommand{\cubicshape}[4]{
  \draw[#4] plot[domain = 0:1] (%
    {#1 + #3 * \x},%
    {#2 + 0.75cm * (0.25 * sqrt(10) * ((2 * \x - 1)^2 - 1) * (2 * \x - 1))}%
  );
}

\begin{tikzpicture}

  \fill[lightred, opacity = 0.1] (4.1, -1.5) rectangle (5.6, 3.75);

  \draw[gray, densely dotted] (0, -1.5) -- ++(0, {0.75 + 1 * 1.5});
  \draw[gray, densely dotted] (2, -1.5) -- ++(0, {0.75 + 1 * 1.5});
  \draw[gray, densely dotted] (3.5, -1.5) -- ++(0, {0.75 + 1 * 1.5});
  \draw[gray, densely dotted] (4.5, -1.5) -- ++(0, {0.75 + 1 * 1.5});
  \draw[gray, densely dotted] (7, -1.5) -- ++(0, {0.75 + 1 * 1.5});

  \draw[gray, densely dotted] (3.35, -1.5) -- ++(0, {0.75 + 2 * 1.5});
  \draw[gray, densely dotted] (4.35, -1.5) -- ++(0, {0.75 + 2 * 1.5});
  \draw[gray, densely dotted] (5.35, -1.5) -- ++(0, {0.75 + 2 * 1.5});
  \draw[gray, densely dotted] (6.35, -1.5) -- ++(0, {0.75 + 2 * 1.5});

  \draw[gray, densely dotted] (4.1, -1.5) -- ++(0, {0.75 + 3 * 1.5});
  \draw[gray, densely dotted] (4.6, -1.5) -- ++(0, {0.75 + 3 * 1.5});
  \draw[gray, densely dotted] (5.1, -1.5) -- ++(0, {0.75 + 3 * 1.5});
  \draw[gray, densely dotted] (5.6, -1.5) -- ++(0, {0.75 + 3 * 1.5});

  \begin{scope}[xshift = 0, yshift = 0]
    \draw[black, -latex] (-0.5, 0) -- node[pos = 1, right] {$x$} ++(8cm, 0cm);
    \node[above right, inner sep = 0, outer sep = 0, yshift = 1mm] at (-0.5, 0) {\footnotesize$k = 0$};

    \linearshapel{0cm}{0cm}{2cm}{blue}
    \linearshaper{0cm}{0cm}{2cm}{blue}
    \quadraticshape{0cm}{0cm}{2cm}{red}
    \cubicshape{0cm}{0cm}{2cm}{red}

    \linearshapel{2cm}{0cm}{1.5cm}{blue}
    \linearshaper{2cm}{0cm}{1.5cm}{blue}
    \quadraticshape{2cm}{0cm}{1.5cm}{red}
    \cubicshape{2cm}{0cm}{1.5cm}{red}

    \linearshapel{3.5cm}{0cm}{1cm}{blue}
    \linearshaper{3.5cm}{0cm}{1cm}{blue}
    \quadraticshape{3.5cm}{0cm}{1cm}{red}
    \cubicshape{3.5cm}{0cm}{1cm}{red}

    \linearshapel{4.5cm}{0cm}{2.5cm}{blue}
    \linearshaper{4.5cm}{0cm}{2.5cm}{blue}
    \quadraticshape{4.5cm}{0cm}{2.5cm}{red}
    \cubicshape{4.5cm}{0cm}{2.5cm}{red}
  \end{scope}

  \begin{scope}[xshift = 0, yshift = 1.5cm]
    \draw[black, -latex] (-0.5, 0) -- node[pos = 1, right] {$x$} ++(8cm, 0cm);
    \node[above right, inner sep = 0, outer sep = 0, yshift = 1mm] at (-0.5, 0) {\footnotesize$k = 1$};

    \linearshapel{3.35cm}{0cm}{1cm}{lightgray}
    \linearshaper{3.35cm}{0cm}{1cm}{blue}
    \quadraticshape{3.35cm}{0cm}{1cm}{red}

    \linearshapel{4.35cm}{0cm}{1cm}{blue}
    \linearshaper{4.35cm}{0cm}{1cm}{blue}
    \quadraticshape{4.35cm}{0cm}{1cm}{red}

    \linearshapel{5.35cm}{0cm}{1cm}{blue}
    \linearshaper{5.35cm}{0cm}{1cm}{lightgray}
    \quadraticshape{5.35cm}{0cm}{1cm}{red}

  \end{scope}

  \begin{scope}[xshift = 0, yshift = 3cm]
    \draw[black, -latex] (-0.5, 0) -- node[pos = 1, right] {$x$} ++(8cm, 0cm);
    \node[above right, inner sep = 0, outer sep = 0, yshift = 1mm] at (-0.5, 0) {\footnotesize$k = 2$};

    \linearshapel{4.1cm}{0cm}{0.5cm}{lightgray}
    \linearshaper{4.1cm}{0cm}{0.5cm}{blue}

    \linearshapel{4.6cm}{0cm}{0.5cm}{blue}
    \linearshaper{4.6cm}{0cm}{0.5cm}{blue}

    \linearshapel{5.1cm}{0cm}{0.5cm}{blue}
    \linearshaper{5.1cm}{0cm}{0.5cm}{lightgray}

  \end{scope}

  \begin{scope}[xshift = 0, yshift = -1.5cm]
    \draw[black, -latex] (-0.5, 0) -- node[pos = 1, right] {$x$} ++(8cm, 0cm);
    \node[above right, inner sep = 0, outer sep = 0, yshift = 1mm] at (-0.5, 0) {\footnotesize$\Sigma$};

    \linearshapel{0cm}{0cm}{2cm}{blue}
    \linearshaper{0cm}{0cm}{2cm}{blue}
    \quadraticshape{0cm}{0cm}{2cm}{red}
    \cubicshape{0cm}{0cm}{2cm}{red}

    \linearshapel{2cm}{0cm}{1.5cm}{blue}
    \linearshaper{2cm}{0cm}{1.5cm}{blue}
    \quadraticshape{2cm}{0cm}{1.5cm}{red}
    \cubicshape{2cm}{0cm}{1.5cm}{red}

    \linearshapel{3.5cm}{0cm}{1cm}{blue}
    \linearshaper{3.5cm}{0cm}{1cm}{blue}
    \quadraticshape{3.5cm}{0cm}{1cm}{red}
    \cubicshape{3.5cm}{0cm}{1cm}{red}

    \linearshapel{4.5cm}{0cm}{2.5cm}{blue}
    \linearshaper{4.5cm}{0cm}{2.5cm}{blue}
    \quadraticshape{4.5cm}{0cm}{2.5cm}{red}
    \cubicshape{4.5cm}{0cm}{2.5cm}{red}

    \linearshaper{3.35cm}{0cm}{1cm}{blue}
    \quadraticshape{3.35cm}{0cm}{1cm}{red}

    \linearshapel{4.35cm}{0cm}{1cm}{blue}
    \linearshaper{4.35cm}{0cm}{1cm}{blue}
    \quadraticshape{4.35cm}{0cm}{1cm}{red}

    \linearshapel{5.35cm}{0cm}{1cm}{blue}
    \quadraticshape{5.35cm}{0cm}{1cm}{red}

    \linearshaper{4.1cm}{0cm}{0.5cm}{blue}

    \linearshapel{4.6cm}{0cm}{0.5cm}{blue}
    \linearshaper{4.6cm}{0cm}{0.5cm}{blue}

    \linearshapel{5.1cm}{0cm}{0.5cm}{blue}

  \end{scope}

  \node[right] at (7.5, 0.75) {$\boldsymbol{+}$};
  \node[right] at (7.5, 2.25) {$\boldsymbol{+}$};
  \node[right] at (7.5, -0.75) {$\boldsymbol{=}$};

\end{tikzpicture} &
    \tikzsetnextfilename{unfitted_multi_level_concept}
\begin{tikzpicture}[3d view = {-30}{30}]

  \begin{scope}[canvas is xy plane at z = 0]

    \draw[black, -latex] (-0.2, -0.2) -- node[pos = 0.8, below, transform shape] {$x_1$} ++(1, 0);
    \draw[black, -latex] (-0.2, -0.2) -- node[pos = 0.8, left, transform shape] {$x_2$} ++(0, 1);

    \foreach \x in {0, 3}
    {
      \fill[draw = black, surface] (\x, 0) rectangle ++(3, 2);

      \fill[draw = black, surface] (\x, 2) rectangle ++(3, 1);
    }

    \fill[black, opacity = 0.075] (3.5, 0) rectangle ++(2, 3);

    \foreach \x in {0, 3, 6}
    {
      \foreach \y in {0, 2, 3}
      {
        \draw[black, fill = white] (\x, \y) circle (0.4mm);
      }
    }

    \draw[red] (4, 0) -- ++(1, 3);
  \end{scope}

  \draw[black, latex-] (5.5, 0, 0.1) -- node[right] {$+$} ++(0, 0, 1.3);
  \fill[lightred, fill opacity = 0.1] (4, 0, 0) -- ++(1, 3, 0) -- ++(0, 0, 1.5) -- ++(-1, -3, 0) -- cycle;

  \begin{scope}[canvas is xy plane at z = 1.5]
    \foreach \x in {3.5, 4.5}
    {
      \foreach \y in {0, 1, 2}
      {
        \fill[surface] (\x, \y) rectangle ++(1, 1);
      }
    }

    \fill[black, opacity = 0.075] (4, 0) rectangle ++(1, 3);

    \draw[black] (4.5, 0) -- ++(0, 3);
    \draw[black, densely dotted] (3.5, 0) -- ++(0, 3);
    \draw[black, densely dotted] (5.5, 0) -- ++(0, 3);
    \foreach \y in {0, 1, 2, 3}
    {
      \draw[black] (3.5, \y) -- ++(2, 0);
      \draw[black, fill = white] (4.5, \y) circle (0.4mm);
    }

    \draw[red] (4, 0) -- ++(1, 3);
  \end{scope}

  \draw[black, latex-] (5, 0, 1.6) -- node[right] {$+$} ++(0, 0, 1.3);
  \fill[lightred, fill opacity = 0.1] (4, 0, 1.5) -- ++(1, 3, 0) -- ++(0, 0, 1.5) -- ++(-1, -3, 0) -- cycle;

  \begin{scope}[canvas is xy plane at z = 3]
    \foreach \x in {4, 4.5}
    {
      \foreach \y in {0, 0.6, 1.2, 1.8, 2.4}
      {
        \fill[surface] (\x, \y) rectangle ++(0.5, 0.6);
      }
    }

    \draw[black] (4.5, 0) -- ++(0, 3);
    \draw[black, densely dotted] (4, 0) -- ++(0, 3);
    \draw[black, densely dotted] (5, 0) -- ++(0, 3);
    \foreach \y in {0, 0.6, 1.2, 1.8, 2.4, 3}
    {
      \draw[black] (4, \y) -- ++(1, 0);
      \draw[black, fill = white] (4.5, \y) circle (0.4mm);
    }

    \draw[red] (4, 0) -- ++(1, 3);
  \end{scope}

  \node[left] at (0, 3, 0) {\small$k = 0$};
  \node[left] at (3.5, 3, 1.5) {\small$k = 1$};
  \node[left] at (4, 3, 3) {\small$k = 2$};

\end{tikzpicture} \\
    a) 1D & b) 2D \\
  \end{tabular} 
  \caption{Schematics of the unfitted multi-level \textit{hp} refinement in one dimension (a) and two dimensions (b). Blue and solid black components are active, while gray and dotted ones are deactivated. The vertical red surface indicates a line within the domain requiring increased resolution.}
  \label{fig:unfitted_mlhp_concept}
\end{figure}

Let $\Omega^{(0)} = \Omega$ denote the domain of the base mesh and let $\Omega^{(k)} \subseteq \Omega$, $k = 1,\ldots,L$, denote the domains associated with the overlay meshes.
Each level is equipped with its own finite element mesh $\mathcal{T}_h^{(k)}$ and approximation space $\widetilde{V}_h^{(k)}$.
For overlay levels, global continuity is enforced by restricting the active overlay space to functions that vanish on artificial overlay boundaries,
\begin{equation}
  V_h^{(0)} = \widetilde{V}_h^{(0)}, 
  \qquad
  V_h^{(k)} =
  \left\{
    v \in \widetilde{V}_h^{(k)}
    \,\middle|\,
    v = 0 \text{ on } \Gamma_\mathrm{o}^{(k)}
  \right\},
  \qquad
  \Gamma_\mathrm{o}^{(k)} = \partial\Omega^{(k)} \setminus \partial\Omega .
\end{equation}
The functions on overlay levels are extended by zero outside their respective domains $\Omega^{(k)}$.
The resulting trial space is the sum of the active spaces across all levels,
\begin{equation}
  V_h = V_h^{(0)} + V_h^{(1)} + \cdots + V_h^{(L)},
  \qquad
  u_h(\boldsymbol{x}) = \sum_{k = 0}^{L} u_h^{(k)}(\boldsymbol{x}),
  \qquad
  u_h^{(k)} \in V_h^{(k)} .
\end{equation}
The plus sign denotes a superposition of spaces and does not imply a direct sum for arbitrary overlay configurations; consequently, the representation of $u_h$ is not necessarily unique.
For a problem with bilinear form $a(\cdot,\cdot)$ and linear form $\ell(\cdot)$, the resulting Galerkin problem reads: find $u_h \in V_h$ such that
\begin{equation}
  a(u_h, v_h) = \ell(v_h)
  \qquad \forall\, v_h \in V_h .
\end{equation}
Writing $u_h = \sum_{i = 0}^{L} u_h^{(i)}$ and $v_h = \sum_{j = 0}^{L} v_h^{(j)}$ makes the coupling between levels explicit,
\begin{equation}
  a(u_h, v_h)
  =
  \sum_{i = 0}^{L}
  \sum_{j = 0}^{L}
  a\left(u_h^{(i)}, v_h^{(j)}\right).
\end{equation}
Only pairs of basis functions with overlapping support contribute to these terms, which is why the implementation must construct integration regions shared by the involved mesh levels.
The unfitted character of the method lies in the fact that overlay boundaries are not required to coincide with element boundaries of lower levels, i.e.,
\begin{equation}
  \partial\Omega^{(k)}
  \not\subseteq
  \bigcup_{E \in \mathcal{T}_h^{(j)}} \partial E
  \quad \text{in general for } j < k .
\end{equation}
The concept is shown in \Cref{fig:unfitted_mlhp_concept}.

\Cref{fig:unfitted_mlhp_concept} (a) illustrates a one-dimensional example using the same base mesh and refinement region as in \Cref{fig:fitted_mlhp_concept}.
In contrast to the splitting strategy of the fitted approach, the first overlay mesh ($k = 1$) consists of three equally sized elements that fully cover the refinement region without aligning with base-mesh vertices.
A second refinement layer ($k = 2$) introduces three smaller elements, again covering the refinement region and providing additional resolution.

A key distinction across the fitted and unfitted approaches is the reduced topological coupling between mesh levels.
Since overlay elements may partially cover elements on other mesh levels, the deactivation rules used in fitted schemes to enforce linear independence do not transfer directly.
In the unfitted formulation, only constraints required for global $\mathcal{C}^0$ continuity are imposed as a general principle.
This does not guarantee linear independence for arbitrary overlay configurations; pathological or nearly dependent configurations may still require additional checks or stabilization, as discussed in the context of the \textit{s}-version FEM and related partition-of-unity or enriched finite element spaces \cite{Ooya:2009aa, Tian:2006aa, Babuska:2012aa}.
No such issues were observed in the numerical experiments presented here, and a systematic detection and treatment of dependent configurations is therefore beyond the scope of this work.
This distinction can also be observed in the two-dimensional example shown in \Cref{fig:unfitted_mlhp_concept} (b).

  \section{Implementation Aspects}
Building on the framework introduced in the previous section, this section outlines key implementation aspects of the proposed method.
A central advantage over conventional approaches is that it requires no modifications to existing finite element code components.
Instead, the method can be realized by introducing only a minimal set of additional data structures and routines.

\subsection{Construction of \texorpdfstring{$C^\infty$}{C-infinity} integration regions}
\begin{figure}[!ht]
  \centering
  \tikzsetnextfilename{integration_regions}

\newcommand{\quadpoints}[5]{
  \foreach \x in {-0.5773502692, 0.5773502692}
  {
    \foreach \y in {-0.5773502692, 0.5773502692}
    {
      \draw[#5, line width = 0.2mm] ({#1 + 0.5 * \x * #3}, {#2 + 0.5 * \y * #4}) circle (0.3mm);
    }
  }
}

\newcommand{\quadpointsc}[5]{
  \foreach \x in {-0.5773502692, 0.5773502692}
  {
    \foreach \y in {-0.5773502692, 0.5773502692}
    {
      \draw[#5, line width = 0.2mm] ({#1 + 0.5 * \x * #3}, {#2 + 0.5 * \y * #4}) -- ++({-0.4mm / sqrt(2)}, {-0.4mm / sqrt(2)}) -- ++({0.8mm / sqrt(2)}, {0.8mm / sqrt(2)});
      \draw[#5, line width = 0.2mm] ({#1 + 0.5 * \x * #3}, {#2 + 0.5 * \y * #4}) -- ++({0.4mm / sqrt(2)}, {-0.4mm / sqrt(2)}) -- ++({-0.8mm / sqrt(2)}, {0.8mm / sqrt(2)});
    }
  }
}

\begin{tikzpicture}[scale = 1.3]

  \draw[black, -latex] (-2mm, -2mm) -- node[below, pos = 0.8] {$x_1$} ++(10mm, 0mm); 
  \draw[black, -latex] (-2mm, -2mm) -- node[left, pos = 0.8] {$x_2$} ++(0mm, 10mm);
  \fill[black] (-2mm, -2mm) circle (0.5mm);

  \foreach \x in {0mm, 30mm}
  {
    \foreach \y in {0mm}
    {
      \fill[draw = black, surface] (\x, \y) rectangle ++(30mm, 20mm); 
    }
  }

  \foreach \x in {0mm, 30mm}
  {
    \foreach \y in {20mm}
    {
      \fill[draw = black, surface] (\x, \y) rectangle ++(30mm, 10mm); 
    }
  }

  \foreach \x in {0mm, 30mm, 60mm}
  {
    \foreach \y in {0mm, 20mm, 30mm}
    {
      \draw[black, fill = white] (\x, \y) circle (0.4mm);
    }
  }

  \fill[black, opacity = 0.1] (26mm, 4mm) rectangle ++(20mm, 20mm);

  \foreach \x in {25mm, 35mm}
  {
    \foreach \y in {5mm, 15mm}
    {
      \fill[draw = black, densely dotted, surface] (\x, \y) rectangle ++(10mm, 10mm); 
    }
  }

  \draw[black] (35mm, 5mm) -- ++(0mm, 20mm);
  \draw[black] (25mm, 15mm) -- ++(20mm, 0mm);

  \draw[black, fill = white] (35mm, 15mm) circle (0.4mm);

  \draw[black, densely dotted] (30mm, 5mm) -- (30mm, 25mm);
  \draw[black, densely dotted] (25mm, 20mm) -- (45mm, 20mm);

  \quadpoints{27.5mm}{10mm}{5mm}{10mm}{red}
  \quadpointsc{32.5mm}{10mm}{5mm}{10mm}{blue}

  \quadpointsc{27.5mm}{17.5mm}{5mm}{5mm}{blue}
  \quadpoints{27.5mm}{22.5mm}{5mm}{5mm}{red}
  \quadpoints{32.5mm}{17.5mm}{5mm}{5mm}{red}
  \quadpointsc{32.5mm}{22.5mm}{5mm}{5mm}{blue}

  \quadpoints{40mm}{10mm}{10mm}{10mm}{red}

  \quadpointsc{40mm}{17.5mm}{10mm}{5mm}{blue}
  \quadpoints{40mm}{22.5mm}{10mm}{5mm}{red} 

  \node[above right] at (0mm, 30mm) {Base mesh};

  \draw[densely dashed, latex-] (24mm, 15mm) to[out = 190, in = 80] node[below, pos = 1, inner sep =0] {Overlay mesh} (15mm, 5mm);

\end{tikzpicture}
  \caption{Visualization of the quadrature regions used for the evaluation of coupling terms between two finite element meshes. Each individual quadrature region is shown in a distinct color.}
  \label{fig:integration_regions}
\end{figure}
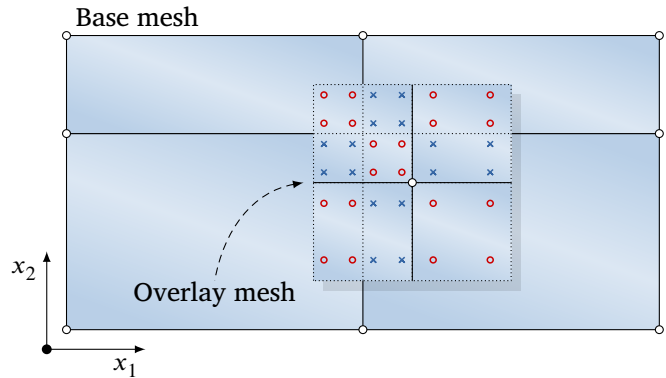

The first implementation aspect concerns the construction of integration regions for the evaluation of coupling terms between overlapping meshes.
While Gaussian quadrature is efficient for smooth integrands, its accuracy deteriorates in the presence of limited regularity.
Therefore, to ensure reliable integration, the intersection of two meshes must be partitioned into sub-regions that are free of shape-function discontinuities or gradient kinks in their interior.
An example of an admissible integration rule is shown in \Cref{fig:integration_regions}.

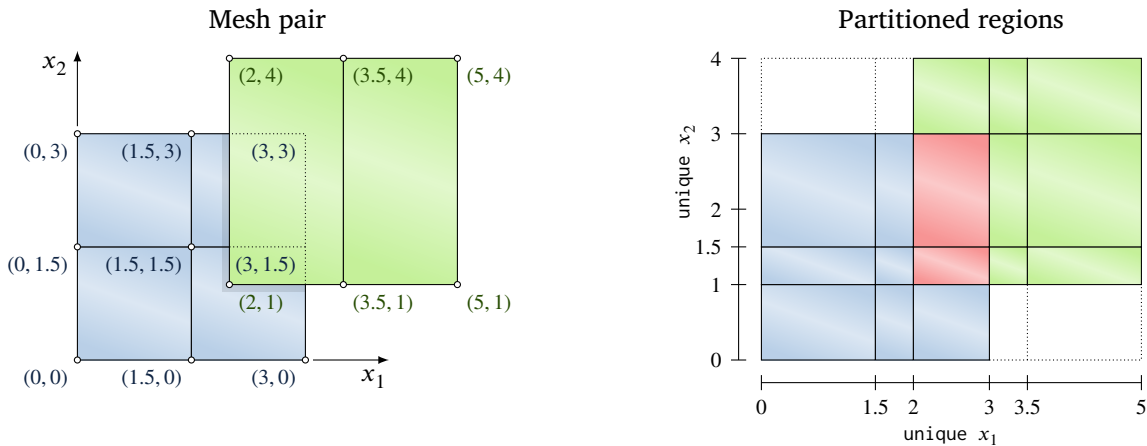
\begin{figure}[!ht]
  \centering
  \tikzsetnextfilename{partitioning_algorithm}
\begin{tikzpicture}

  \begin{scope}

    \draw[black, -latex] (3.1cm, 0cm) -- node[below, pos = 0.8] {$x_1$} ++(1cm, 0cm);
    \draw[black, -latex] (0cm, 3.1cm) -- node[left, pos = 0.8] {$x_2$} ++(0cm, 1cm);
  
    \foreach \x in {0cm, 1.5cm}
    {
      \foreach \y in {0cm, 1.5cm}
      {
        \fill[draw = black, surface] (\x, \y) rectangle ++(1.5cm, 1.5cm);
      }
    }

    \foreach \x in {0cm, 1.5cm, 3cm}
    {
      \foreach \y in {0cm, 1.5cm, 3cm}
      {
        \draw[black, fill = white] (\x, \y) circle (0.4mm);
      }
    }

    \fill[black, opacity = 0.1] (1.9cm, 0.9cm) rectangle (3cm, 3cm);

    \foreach \x in {2cm, 3.5cm}
    {
      \foreach \y in {1cm}
      {
        \fill[draw = black, alternativesurface] (\x, \y) rectangle ++(1.5cm, 3cm);
      }
    }

    \foreach \x in {2cm, 3.5cm, 5cm}
    {
      \foreach \y in {1cm, 4cm}
      {
        \draw[black, fill = white] (\x, \y) circle (0.4mm);
      }
    }

    \draw[black, densely dotted] (3cm, 1cm) -- ++(0cm, 2cm) -- ++(-1cm, 0cm);
    \draw[black, densely dotted] (2cm, 1.5cm) -- ++(1cm, 0cm);

    \foreach \x in {0, 1.5, 3}
    {
      \foreach \y in {0, 1.5, 3}
      {
        \node[below left, darkblue!50!black] at ({\x * 1cm}, {\y * 1cm}) {\footnotesize$(\x, \y)$}; 
      }
    }

    \foreach \x in {2, 3.5, 5}
    {
      \foreach \y in {1, 4}
      {
        \node[below right, darkgreen!50!black] at ({\x * 1cm}, {\y * 1cm}) {\footnotesize$(\x, \y)$};
      }
    }

    \node[above] at (2.5cm, 4.2cm) {Mesh pair};
  \end{scope}

  \begin{scope}[xshift = 9cm]

    \draw[black] (0cm, -0.3cm) -- ++(5cm, 0cm);
    \foreach \x in {0, 1.5, 2, 3, 3.5, 5}
    {
      \draw[black] (\x, -0.2cm) -- node[pos = 1, below] {\footnotesize$\x$} ++(0, -0.2cm);
    }
    \node[below] at (2.5, -0.75) {\footnotesize\texttt{unique $x_1$}};

    \draw[black] (-0.3cm, 0cm) -- ++(0cm, 4cm);
    \foreach \x in {0, 1, 1.5, 2, 3, 4}
    {
      \draw[black] (-0.2cm, \x) -- node[pos = 1, left] {\footnotesize$\x$} ++(-0.2cm, 0);
    }
    \node[above, rotate = 90] at (-0.75, 2.5) {\footnotesize\texttt{unique $x_2$}};

    \fill[draw = black, surface] (0cm, 0cm) rectangle ++(1.5cm, 1cm);

    \fill[draw = black, surface] (0cm, 1cm) rectangle ++(1.5cm, 0.5cm);

    \fill[draw = black, surface] (0cm, 1.5cm) rectangle ++(1.5cm, 1.5cm);

    \fill[draw = black, surface] (1.5cm, 0cm) rectangle ++(0.5cm, 1cm);

    \fill[draw = black, surface] (1.5cm, 1cm) rectangle ++(0.5cm, 0.5cm);

    \fill[draw = black, surface] (1.5cm, 1.5cm) rectangle ++(0.5cm, 1.5cm);

    \fill[draw = black, surface] (2cm, 0cm) rectangle ++(1cm, 1cm);

    \fill[draw = black, highlightedsurface] (2cm, 1cm) rectangle ++(1cm, 0.5cm);

    \fill[draw = black, highlightedsurface] (2cm, 1.5cm) rectangle ++(1cm, 1.5cm);

    \fill[draw = black, alternativesurface] (2cm, 3cm) rectangle ++(1cm, 1cm);

    \fill[draw = black, alternativesurface] (3cm, 1cm) rectangle ++(0.5cm, 0.5cm);

    \fill[draw = black, alternativesurface] (3cm, 1.5cm) rectangle ++(0.5cm, 1.5cm);

    \fill[draw = black, alternativesurface] (3cm, 3cm) rectangle ++(0.5cm, 1cm);

    \fill[draw = black, alternativesurface] (3.5cm, 1cm) rectangle ++(1.5cm, 0.5cm);

    \fill[draw = black, alternativesurface] (3.5cm, 1.5cm) rectangle ++(1.5cm, 1.5cm);

    \fill[draw = black, alternativesurface] (3.5cm, 3cm) rectangle ++(1.5cm, 1cm);

    \draw[black, densely dotted] (3cm, 0cm) -- ++(2cm, 0cm) -- ++(0cm, 1cm);
    \draw[black, densely dotted] (3.5cm, 0cm) -- ++(0cm, 1cm);

    \draw[black, densely dotted] (0cm, 3cm) -- ++(0cm, 1cm) -- ++(2cm, 0cm);
    \draw[black, densely dotted] (1.5cm, 3cm) -- ++(0cm, 1cm);

    \node[above] at (2.5cm, 4.2cm) {Partitioned regions};

  \end{scope}

\end{tikzpicture}
  \caption{Visualization of the dimension-independent algorithm used to determine the intersection regions of two axis-aligned finite element meshes. 
  The example shows two two-dimensional meshes (green and blue) on the left and the resulting $C^\infty$ integration regions on the right. 
  Regions highlighted in red contain elements from both meshes.}
  \label{fig:integration_regions_algorithm}
\end{figure}

Since this work focuses exclusively on axis-aligned meshes, the construction of integration regions can be realized by a simple dimension-independent algorithm, illustrated in \Cref{fig:integration_regions_algorithm}.

The example considers two partially overlapping meshes with non-matching element boundaries: a $2 \times 2$ mesh shown in green and a $2 \times 1$ mesh shown in blue.
This setting is representative of superimposed discretizations on a common base mesh (not shown in the sketch).
All vertices are labeled by their coordinates.

The partitioning algorithm is depicted on the right.
First, all unique vertex coordinates from both meshes are collected independently for each axis, yielding
$\boldsymbol{x}_{1,\mathrm{unique}} = [0, 1.5, 2, 3, 3.5, 5]$ for the first dimension and
$\boldsymbol{x}_{2,\mathrm{unique}} = [0, 1, 1.5, 2, 3, 5]$ for the second dimension.
The Cartesian product of the resulting intervals in each coordinate direction defines a set of axis-aligned boxes that form candidate integration regions.

For the computation of coupling terms between the two meshes, only the red boxes need to be considered, as these are the only regions that contain elements from both meshes.
For the assembly of superposed fields, the same construction can also be applied with modified selection criteria, for example by retaining all boxes covered by at least one active finite element field.
Beyond the global integration regions, the implementation must also identify the contributing parent elements and the corresponding local sub-regions required to evaluate their shape functions.
The resulting procedure is summarized in \Cref{algo:integration_regions}.
\begin{algorithm}[!ht]
  \caption{Construction of admissible integration regions for non-matching, axis-aligned finite element meshes.}
  \label{algo:integration_regions}
  \begin{algorithmic}[1]
    \Require meshes $\mathcal{M}_1,\ldots,\mathcal{M}_n$, criterion $\mathcal{C}$, tolerance $\varepsilon$
    \Ensure integration regions $\mathcal{T}_{\cap}$
    \State $\mathcal{T}_{\cap} \gets \emptyset$
    \For{$i = 1,\ldots,d$}
      \State $\mathcal{P}_i \gets$ element-boundary coordinates in direction $i$
      \State sort $\mathcal{P}_i$ and merge entries closer than $\varepsilon$
      \State $\mathcal{I}_i \gets$ non-degenerate intervals between adjacent entries of $\mathcal{P}_i$
    \EndFor
    \State $\mathcal{B} \gets \mathcal{I}_1 \times \cdots \times \mathcal{I}_d$
    \ForAll{$B \in \mathcal{B}$}
      \State $\mathcal{E}_B \gets \emptyset$, \quad $x_B \gets$ midpoint of $B$
      \For{$m = 1,\ldots,n$}
        \State add the element of $\mathcal{M}_m$ containing $x_B$ to $\mathcal{E}_B$, if present
      \EndFor
      \If{$\mathcal{E}_B$ satisfies $\mathcal{C}$}
        \State $\mathcal{S}_B \gets \emptyset$
        \ForAll{$E \in \mathcal{E}_B$}
          \State add $(E,\Phi_E^{-1}(B))$ to $\mathcal{S}_B$
        \EndFor
        \State add $(B,\mathcal{S}_B)$ to $\mathcal{T}_{\cap}$
      \EndIf
    \EndFor
    \State \Return $\mathcal{T}_{\cap}$
  \end{algorithmic}
\end{algorithm}

\subsection{Integration into existing finite element codes}
The second implementation aspect concerns the integration of the unfitted multi-level \textit{hp} formulation into existing finite element codes.
A practical strategy is to treat each mesh level as an independent discretization and introduce a thin superposition layer solely for integration and assembly.
The base mesh and overlay meshes retain their own elements, basis functions, and degrees of freedom, without any topological merging.
Instead, the previously defined intersection regions serve as a common integration grid on which the coupled variational forms are evaluated.

For each integration region, an auxiliary element is constructed that aggregates all active finite element contributions whose supports cover that region.
Its local approximation is given by the concatenation of the shape functions from the contributing elements across mesh levels.
Quadrature points on the auxiliary integration region are mapped to the local coordinates of each contributing element prior to evaluating shape functions and gradients.
With this construction, existing routines for element interpolation, gradient evaluation, local matrix construction, and weak-form integration can be reused through the auxiliary element interface; only the mapping between the integration region and the contributing element domains is introduced as an additional component.

Care is required in the application of constraints and the enumeration of degrees of freedom.
Homogeneous constraints must be imposed on artificial overlay boundaries to ensure vanishing overlay contributions on artificial overlay interfaces and to preserve global $\mathcal{C}^0$ continuity.
In contrast, physical boundary conditions are applied exclusively on the domain boundary and must be clearly distinguished from overlay-boundary constraints.
The global degree-of-freedom enumeration should therefore be performed only after all physical boundary conditions and overlay constraints have been assigned.
If constrained degrees of freedom are eliminated from the governing linear system of equations, their column contributions must be transferred consistently to the right-hand side, particularly in the presence of nonzero physical boundary data.

In transient settings with moving overlays, the intersection regions and corresponding element couplings need to be updated when overlays are inserted, moved, or removed.
For fixed overlay configurations, geometry-dependent operators such as stiffness or mass matrices are reused, while time-dependent load vectors are updated independently.
When the approximation space changes, state variables must be transferred between the old and new spaces by a variational projection, such as an $L^2$ or problem-dependent energy projection, rather than by pointwise interpolation.

Finally, implementations should account for the conditioning effects arising from very small overlaps between superimposed elements.
Such regions may lead to poorly scaled basis contributions and deteriorated solver convergence, especially for high polynomial degrees.
It is therefore advisable to monitor overlap sizes and solver behavior, and to handle vanishingly small intersections with care.
Possible remedies include scaling, suitable preconditioning, or the removal or aggregation of negligible intersection regions, but such modifications should be checked for consistency with the variational formulation.

  \section{Numerical Results}
In this section, the unfitted multi-level \textit{hp} approach is first assessed using benchmark problems with closed-form solutions, allowing reliable quantification of the numerical approximation error.
The method is further demonstrated on a transient heat-transfer problem of practical relevance.
For Galerkin discretizations, the relative energy-norm error is evaluated using the standard energy identity as
\begin{equation}
  \|e\|_r =
  \sqrt{
    \frac{
      \left| a\left( u_\mathrm{ex}, u_\mathrm{ex}\right) -
      a\left( u_\mathrm{num}, u_\mathrm{num} \right) \right|
    }{
      \left| a\left( u_\mathrm{ex}, u_\mathrm{ex} \right) \right|
    }
  } \cdot 100\,\%,
\end{equation}
where $a(\cdot, \cdot)$ denotes the bilinear form of the governing weak form, $u_\mathrm{ex}$ denotes the exact solution, and $u_\mathrm{num}$ denotes the numerical approximation.
This expression relies on Galerkin orthogonality; for non-Galerkin approximations, the error would need to be evaluated directly through $a(u_\mathrm{ex} - u_\mathrm{num}, u_\mathrm{ex} - u_\mathrm{num})$.

\subsection{Elastic bar with discontinuous strains}
\begin{figure}[!ht]
  \centering
  \begin{tabular}{c @{\hspace*{0.3cm}} c @{\hspace*{0.3cm}} c}
    \tikzsetnextfilename{one_dimensional_rod_model}
\begin{tikzpicture}

  \draw[background] (0, 0.3cm) -- ++(0.1cm, 0cm);
  \fill[draw = foreground, surface] (0cm, -0.1cm) rectangle ++(5cm, 0.2cm);

  \foreach \y in {-0.1cm, -0.05cm, 0cm, 0.05cm, 0.1cm}
  {
    \draw[foreground] (0cm, \y) -- ++(-0.1cm, -0.05cm);
  }

  \foreach \x in {0cm, 1cm, ..., 4cm}
  {
    \draw[foreground, -latex] ({\x + 0.02cm}, 0.2cm) -- ++(0.96cm, 0cm);
  }
  \node[above] at (2.5cm, 0.2cm) {$q(x)$};

  \fill[foreground] (3.3333cm, 0cm) circle (0.04cm);
  \draw[foreground, -latex] (3.3333cm, 0cm) -- node[below, yshift = -0.05cm] {$f$} ++(0.5cm, 0cm);

  \draw[foreground] (0cm, -0.2cm) -- ++(0cm, -0.5cm);
  \draw[foreground] (3.3333cm, -0.2cm) -- ++(0cm, -0.5cm);
  \draw[foreground] (5cm, -0.2cm) -- ++(0cm, -0.5cm);
  \draw[foreground, latex-latex] (0cm, -0.6cm) -- node[below] {$\sfrac{2}{3}\,L$} ++(3.3333cm, 0cm);
  \draw[foreground, latex-latex] (3.3333cm, -0.6cm) -- node[below] {$\sfrac{1}{3}\,L$} ++(1.6667cm, 0cm);

\end{tikzpicture} &
    \tikzsetnextfilename{one_dimensional_rod_mesh_mlhp}
\begin{tikzpicture}

  \draw[foreground] (0cm, 0cm) -- ++(5cm, 0cm);
  \draw[red, fill = background] (0cm, 0cm) circle (0.06cm);
  \draw[red, fill = background] (5cm, 0cm) circle (0.06cm);

  \draw[foreground, densely dotted] (3.333cm, 0cm) -- node[above, pos = 1] {$\vdots$} ++(0cm, 3.5cm);

  \draw[foreground] (0cm, -0.1cm) -- ++(0cm, -0.2cm);
  \draw[foreground] (3.3333cm, -0.1cm) -- ++(0cm, -0.2cm);
  \draw[foreground] (5cm, -0.1cm) -- ++(0cm, -0.2cm);
  \draw[foreground, latex-latex] (0cm, -0.2cm) -- node[below] {$\sfrac{2}{3}\,L$} ++(3.3333cm, 0cm);
  \draw[foreground, latex-latex] (3.3333cm, -0.2cm) -- node[below] {$\sfrac{1}{3}\,L$} (5cm, -0.2cm);

  \begin{scope}[xshift = 0cm, yshift = 1cm]
    \draw[foreground] (0cm, 0cm) -- ++(5cm, 0cm);
    \draw[black, fill = background] (0cm, 0cm) circle (0.06cm);
    \draw[foreground, fill = background] ( 2.5cm, 0cm) circle (0.06cm);
    \draw[red, fill = background] ( 5cm, 0cm) circle (0.06cm);
  \end{scope}

  \begin{scope}[xshift = 2.5cm, yshift = 2cm]
    \draw[foreground] (0cm, 0cm) -- ++(2.5cm, 0cm);
    \draw[red, fill = background] (0cm, 0cm) circle (0.06cm);
    \draw[foreground, fill = background] (1.25cm, 0cm) circle (0.06cm);
    \draw[black, fill = background] (2.5cm, 0cm) circle (0.06cm);
  \end{scope}

  \begin{scope}[xshift = 2.5cm, yshift = 3cm]
    \draw[foreground] (0cm, 0cm) -- ++(1.25cm, 0cm);
    \draw[red, fill = background] (0cm, 0cm) circle (0.06cm);
    \draw[foreground, fill = background] (0.625cm, 0cm) circle (0.06cm);
    \draw[red, fill = background] (1.25cm, 0cm) circle (0.06cm);
  \end{scope}

  \node[fill = background] at (3.3333cm, 0.5cm) {$+$};
  \node[fill = background] at (3.3333cm, 1.5cm) {$+$};
  \node[fill = background] at (3.3333cm, 2.5cm) {$+$};

  \node[above right] at (0cm, 0cm) {\footnotesize$k = 0$};
  \node[above right] at (0cm, 1cm) {\footnotesize$k = 1$};
  \node[above left] at (5cm, 2cm) {\footnotesize$k = 2$};
  \node[above] at (3.75cm, 3cm) {\footnotesize$k = 3$};

\end{tikzpicture} &
    \tikzsetnextfilename{one_dimensional_rod_mesh}
\begin{tikzpicture}

  \draw[foreground] (0cm, 0cm) -- ++(5cm, 0cm);
  \draw[foreground, fill = background] (0cm, 0cm) circle (0.06cm);
  \draw[foreground, fill = background] (5cm, 0cm) circle (0.06cm);

  \draw[foreground, densely dotted] (3.333cm, 0cm) -- node[above, pos = 1] {$\vdots$} ++(0cm, 3.5cm);

  \draw[foreground] (0cm, -0.1cm) -- ++(0cm, -0.2cm);
  \draw[foreground] (3.3333cm, -0.1cm) -- ++(0cm, -0.2cm);
  \draw[foreground] (5cm, -0.1cm) -- ++(0cm, -0.2cm);
  \draw[foreground, latex-latex] (0cm, -0.2cm) -- node[below] {$\sfrac{2}{3}\,L$} ++(3.3333cm, 0cm);
  \draw[foreground, latex-latex] (3.3333cm, -0.2cm) -- node[below] {$\sfrac{1}{3}\,L$} (5cm, -0.2cm);

  \begin{scope}[xshift = 3.333cm, yshift = 1cm]
    \draw[foreground] (-1.25cm, 0cm) -- ++(2.5cm, 0cm);
    \draw[red, fill = background] (-1.25cm, 0cm) circle (0.06cm);
    \draw[foreground, fill = background] ( 0.00cm, 0cm) circle (0.06cm);
    \draw[red, fill = background] ( 1.25cm, 0cm) circle (0.06cm);

    \draw[foreground] (-1.25cm, -0.1cm) -- ++(0cm, -0.2cm);
    \draw[foreground] ( 1.25cm, -0.1cm) -- ++(0cm, -0.2cm);
    \draw[foreground, latex-latex] (-1.25cm, -0.2cm) -- node[below, pos = 0.1] {$\alpha^1\,L$} ++(2.5cm, 0cm);
  \end{scope}

  \begin{scope}[xshift = 3.333cm, yshift = 2cm]
    \draw[foreground] (-0.625cm, 0cm) -- ++(1.25cm, 0cm);
    \draw[red, fill = background] (-0.625cm, 0cm) circle (0.06cm);
    \draw[foreground, fill = background] ( 0.00cm, 0cm) circle (0.06cm);
    \draw[red, fill = background] ( 0.625cm, 0cm) circle (0.06cm);

    \draw[foreground] (-0.625cm, -0.1cm) -- ++(0cm, -0.2cm);
    \draw[foreground] ( 0.625cm, -0.1cm) -- ++(0cm, -0.2cm);
    \draw[foreground, latex-latex] (-0.625cm, -0.2cm) -- node[below, pos = 0.1] {$\alpha^2\,L$} ++(1.25cm, 0cm);
  \end{scope}

  \begin{scope}[xshift = 3.333cm, yshift = 3cm]
    \draw[foreground] (-0.3125cm, 0cm) -- ++(0.625cm, 0cm);
    \draw[red, fill = background] (-0.3125cm, 0cm) circle (0.06cm);
    \draw[foreground, fill = background] ( 0.00cm, 0cm) circle (0.06cm);
    \draw[red, fill = background] ( 0.3125cm, 0cm) circle (0.06cm);

    \draw[foreground] (-0.3125cm, -0.1cm) -- ++(0cm, -0.2cm);
    \draw[foreground] ( 0.3125cm, -0.1cm) -- ++(0cm, -0.2cm);
    \draw[foreground, latex-latex] (-0.3125cm, -0.2cm) -- node[below left, pos = 0.3] {$\alpha^3\,L$} ++(0.625cm, 0cm);
  \end{scope}

  \node[fill = background] at (3.3333cm, 0.5cm) {$+$};
  \node[fill = background] at (3.3333cm, 1.5cm) {$+$};
  \node[fill = background] at (3.3333cm, 2.5cm) {$+$};

  \node[above right] at (0cm, 0cm) {\footnotesize$k = 0$};
  \node[above] at ({3.333cm + 1.25cm}, 1cm) {\footnotesize$k = 1$};
  \node[above] at ({3.333cm + 0.625cm}, 2cm) {\footnotesize$k = 2$};
  \node[above] at ({3.333cm + 0.3125cm}, 3cm) {\footnotesize$k = 3$};

\end{tikzpicture} \\
    a) Geometric setup. & b) Fitted multi-level \textit{hp}. & c) Unfitted multi-level \textit{hp}. \\
  \end{tabular}
  \caption{The one-dimensional bar problem. Shown in (a) is the geometric setup including boundary conditions; (b) displays the corresponding fitted multi-level \textit{hp} mesh; and (c) shows the unfitted version.}
  \label{fig:benchmark_elastic_bar}
\end{figure}
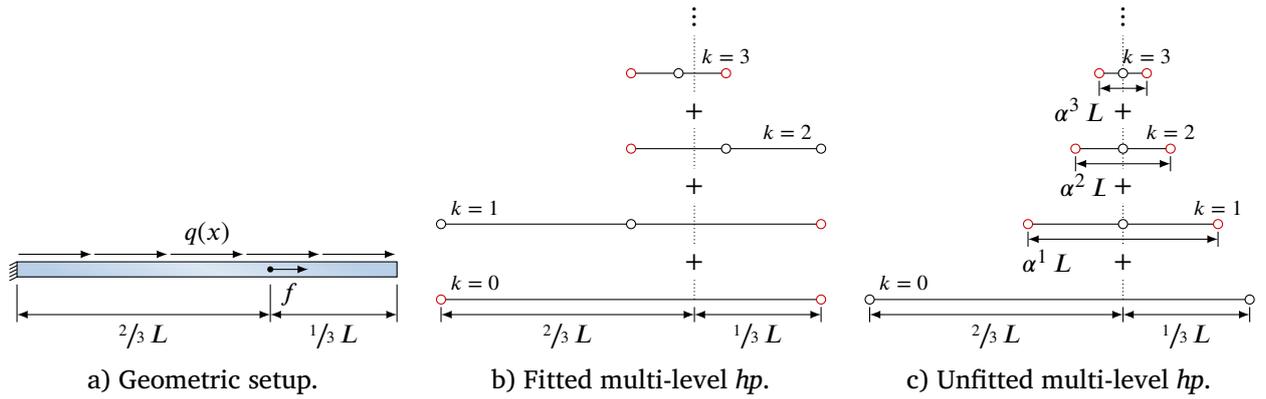

The benchmark problem considered in this subsection is adopted from \cite{Schmake:2023aa} and builds on the classic \textit{p}-FEM example presented in \cite{Szabo:2004aa}.
It considers a one-dimensional, linear elastic bar of length $L$, shown in \Cref{fig:benchmark_elastic_bar} (a).

The bar is clamped at the left end and traction-free at the right.
In its original form, the problem was subjected only to a sinusoidal volumetric force, resulting in a solution that cannot be represented exactly by polynomial basis functions.
The modified version considered here additionally includes a point load at $x = \sfrac{2}{3}\,L$, introducing a discontinuity in the strains.

The domain and boundary sets are
\begin{equation}
  \Omega := [0, L], \qquad
  \Gamma_D := \{0\}, \qquad
  \Gamma_N := \{L\}.
\end{equation}
The governing equations read
\begin{subequations}
  \begin{align}
    -\fracd{}{x} \left[ E\,A\,\fracd{u}{x} \right] &= q(x) &&\forall x \in \Omega \setminus \{\sfrac{2}{3} L\}, \\
    u &= 0 &&\forall x \in \Gamma_D, \\
    n\,\sigma &= 0 &&\forall x \in \Gamma_N, \\
    q(x) &= \sin(8\,x) &&\forall x \in \Omega, \\
    f &= \frac{1}{5} &&\text{at } x = \sfrac{2}{3} L.
  \end{align}
\end{subequations}
Equivalently, the point load contributes $f\,v(\sfrac{2}{3}L)$ to the weak form for a test function $v$.
The computations below use the non-dimensional values $L = 1$ and $E A = 1$.

For constant $E A$, the corresponding closed-form solution for the displacement is
\begin{equation}
  u(x) =
  \frac{1}{E A}
  \left[
    \frac{1}{8} \left( \frac{\sin(8\,x)}{8} - x\,\cos(8\,L) \right)
    +
    \begin{cases}
      f\,x, & x \leq \sfrac{2}{3}L, \\
      f\,\sfrac{2}{3}L, & x > \sfrac{2}{3}L,
    \end{cases}
  \right],
\end{equation}
and the strains are given by
\begin{equation}
  \fracd{u}{x} =
  \frac{1}{E A}
  \left[
    \frac{1}{8} \left( \cos(8\,x) - \cos(8\,L) \right)
    +
    \begin{cases}
      f, & x \leq \sfrac{2}{3}L, \\
      0, & x > \sfrac{2}{3}L,
    \end{cases}
  \right],
  \label{eq:elastic_bar_benchmark_exact_strains}
\end{equation}
which exhibits a discontinuity at $x = \sfrac{2}{3}\,L$.

Solution features such as discontinuities and kinks typically degrade the convergence behavior of high-order finite elements.
Local refinement therefore provides a natural mechanism to restore high approximation quality in their vicinity.

We discretized this model using both the fitted and unfitted multi-level \textit{hp} approaches shown in \Cref{fig:benchmark_elastic_bar} (b) and (c), respectively.
Both analyses started with a single-element mesh.
In the fitted approach, this element was bisected into two equally sized elements; in each subsequent refinement step, the element of the highest overlay level containing $x = \sfrac{2}{3}\,L$ is further subdivided.
In contrast, the unfitted approach superimposed a two-element mesh of size $\alpha^k\,L$, with $k$ denoting the refinement level and the mesh centroid located at $x = \sfrac{2}{3}\,L$.

Since the fitted approach covers entire elements at each refinement step, the size of each overlay mesh is fixed to $(\sfrac{1}{2})^{\,k - 1}$.
In contrast, the unfitted approach permits arbitrary mesh sizes.
A parameter $\alpha$ is introduced to control the size of the overlays, providing an additional degree of freedom for optimizing the approximation quality.

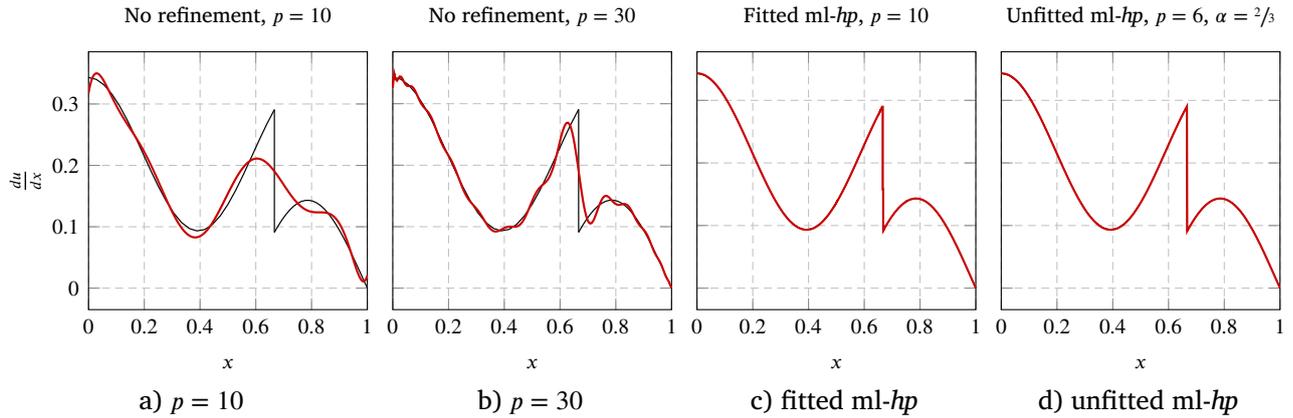
\begin{figure}[!ht]
	  \centering
	  \begin{tabular}{c @{\hspace*{-3mm}} c @{\hspace*{-3mm}} c @{\hspace*{-3mm}} c}
	    \tikzsetnextfilename{one_dimensiona_rod_gradient_p10}
\begin{tikzpicture}[spy using outlines={black, line width = 1, densely dashed, thick, circle, size=2cm, magnification=3, connect spies}]

	\begin{axis} [
			axis background/.style = {fill = black!0!white},
			font={\footnotesize},
			axis lines = box,
			title = {No refinement, $p = 10$},
			xlabel = $x$,
			ylabel = $\frac{du}{dx}$,
			ylabel style = {yshift = -4mm},
			width = 5.25cm,
			height = 5cm,
			max space between ticks = 40,
			grid = major,
			grid style = {densely dashed, line width = 0.1pt},
			minor x tick num = 0,
			minor y tick num = 0,
			legend pos = south west,
			legend style = {nodes = {scale = 0.75, transform shape}},
			xmin = 0,
			xmax = 1
		]

		\addplot[black, line width = 0.45pt, domain = 0:0.6666] {1/8 * (cos(8 * x / pi * 180) - cos(8 / pi * 180)) + 0.2};
		\addplot[black, line width = 0.45pt, domain = 0.6666:1] {1/8 * (cos(8 * x / pi * 180) - cos(8 / pi * 180))};
		\addplot[black, line width = 0.45pt] coordinates {
			(0.6666, {1/8 * (cos(8 * 0.6666 / pi * 180) - cos(8 / pi * 180)) + 0.2})
			(0.6666, {1/8 * (cos(8 * 0.6666 / pi * 180) - cos(8 / pi * 180))})
		};
		\addplot[red, line width = 0.8pt] table [x = x, y = y, col sep = comma] {04_Z_data_one_dimensional_rod_solution_p10.csv};

	\end{axis}

\end{tikzpicture}  &
	    \tikzsetnextfilename{one_dimensional_rod_gradient_p30}
\begin{tikzpicture}[spy using outlines={black, line width = 1, densely dashed, thick, circle, size=2cm, magnification=3, connect spies}]

	\begin{axis} [
			axis background/.style = {fill = black!0!white},
			font={\footnotesize},
			axis lines = box,
			title = {No refinement, $p = 30$},
			xlabel = $x$,
			yticklabel = \empty,
			ylabel style = {yshift = -4mm},
			width = 5.25cm,
			height = 5cm,
			max space between ticks = 40,
			grid = major,
			grid style = {densely dashed, line width = 0.1pt},
			minor x tick num = 0,
			minor y tick num = 0,
			legend pos = south west,
			legend style = {nodes = {scale = 0.75, transform shape}},
			xmin = 0,
			xmax = 1
		]

		\addplot[black, line width = 0.45pt, domain = 0:0.6666] {1/8 * (cos(8 * x / pi * 180) - cos(8 / pi * 180)) + 0.2};
		\addplot[black, line width = 0.45pt, domain = 0.6666:1] {1/8 * (cos(8 * x / pi * 180) - cos(8 / pi * 180))};
		\addplot[black, line width = 0.45pt] coordinates {
			(0.6666, {1/8 * (cos(8 * 0.6666 / pi * 180) - cos(8 / pi * 180)) + 0.2})
			(0.6666, {1/8 * (cos(8 * 0.6666 / pi * 180) - cos(8 / pi * 180))})
		};
		\addplot[red, line width = 0.8pt] table [x = x, y = y, col sep = comma] {04_Z_data_one_dimensional_rod_solution_p30.csv};

	\end{axis}

\end{tikzpicture}  &
	    \tikzsetnextfilename{one_dimensional_rod_gradient_mlhp10}
\begin{tikzpicture}[spy using outlines={black, line width = 1, densely dashed, thick, circle, size=2cm, magnification=3, connect spies}]

	\begin{axis} [
			axis background/.style = {fill = black!0!white},
			font={\footnotesize},
			axis lines = box,
			title = {Fitted ml-\textit{hp}, $p = 10$},
			xlabel = $x$,
			yticklabel = \empty,
			ylabel style = {yshift = -4mm},
			width = 5.25cm,
			height = 5cm,
			max space between ticks = 40,
			grid = major,
			grid style = {densely dashed, line width = 0.1pt},
			minor x tick num = 0,
			minor y tick num = 0,
			legend pos = south west,
			legend style = {nodes = {scale = 0.75, transform shape}},
			xmin = 0,
			xmax = 1
		]

		\addplot[black, line width = 0.45pt, domain = 0:0.6666] {1/8 * (cos(8 * x / pi * 180) - cos(8 / pi * 180)) + 0.2};
		\addplot[black, line width = 0.45pt, domain = 0.6666:1] {1/8 * (cos(8 * x / pi * 180) - cos(8 / pi * 180))};
		\addplot[black, line width = 0.45pt] coordinates {
			(0.6666, {1/8 * (cos(8 * 0.6666 / pi * 180) - cos(8 / pi * 180)) + 0.2})
			(0.6666, {1/8 * (cos(8 * 0.6666 / pi * 180) - cos(8 / pi * 180))})
		};
		\addplot[red, line width = 0.8pt] table [x = x, y = y, col sep = comma] {04_Z_data_one_dimensional_rod_solution_mlhp10.csv};

	\end{axis}

\end{tikzpicture}  &
	    \tikzsetnextfilename{one_dimensional_rod_gradient_ushp_6}
\begin{tikzpicture}[spy using outlines={black, line width = 1, densely dashed, thick, circle, size=2cm, magnification=3, connect spies}]

	\begin{axis} [
			axis background/.style = {fill = black!0!white},
			font={\footnotesize},
			axis lines = box,
			title = {Unfitted ml-\textit{hp}, $p = 6$, $\alpha = \sfrac{2}{3}$},
			xlabel = $x$,
			yticklabel = \empty,
			ylabel style = {yshift = -4mm},
			width = 5.25cm,
			height = 5cm,
			max space between ticks = 40,
			grid = major,
			grid style = {densely dashed, line width = 0.1pt},
			minor x tick num = 0,
			minor y tick num = 0,
			legend pos = south west,
			legend style = {nodes = {scale = 0.75, transform shape}},
			xmin = 0,
			xmax = 1
		]

		\addplot[black, line width = 0.45pt, domain = 0:0.6666] {1/8 * (cos(8 * x / pi * 180) - cos(8 / pi * 180)) + 0.2};
		\addplot[black, line width = 0.45pt, domain = 0.6666:1] {1/8 * (cos(8 * x / pi * 180) - cos(8 / pi * 180))};
		\addplot[black, line width = 0.45pt] coordinates {
			(0.6666, {1/8 * (cos(8 * 0.6666 / pi * 180) - cos(8 / pi * 180)) + 0.2})
			(0.6666, {1/8 * (cos(8 * 0.6666 / pi * 180) - cos(8 / pi * 180))})
		};
		\addplot[red, line width = 0.8pt] table [x = arc_length, y = u Gradient, col sep = comma] {04_Z_data_one_dimensional_rod_solution_ushp6.csv};

	\end{axis}

\end{tikzpicture}  \\
	    a) $p = 10$ & b) $p = 30$ & c) fitted ml-\textit{hp} & d) unfitted ml-\textit{hp} \\
	  \end{tabular}
	  \caption{Numerical strain solutions of the one-dimensional bar problem for four different meshes. 
	  Panels (a) and (b) correspond to single high-order elements with degrees $p = 10$ and $p = 30$, respectively. 
	  Panels (c) and (d) use local refinements: fitted multi-level \textit{hp} in (c) and unfitted multi-level \textit{hp} in (d).}
  \label{fig:benchmark_elastic_bar_gradients}
\end{figure}

\paragraph{Visual assessment}
Before assessing the two methods in terms of computational cost versus error, we first compared their performance visually.
For this purpose, the gradient solutions $\sfrac{\mathrm{d}u}{\mathrm{d}x}$ obtained from four different mesh configurations depicted in \Cref{fig:benchmark_elastic_bar_gradients} were examined.

In all four plots, the exact solution from \Cref{eq:elastic_bar_benchmark_exact_strains} is shown in black and the numerical approximation in red.

The two solutions on the left were obtained using a single high-order element with polynomial degrees $p = 10$ and $p = 30$, respectively.
These two cases illustrate the well-known difficulties of approximating non-smooth solution features, such as the discontinuity present here, using a smooth polynomial basis.
Attempts to represent the gradient jump not only yield a poor approximation near $x = \sfrac{2}{3}\,L$ but also induce pronounced oscillations across the entire element domain.
Increasing the polynomial degree does not mitigate this effect.

The two solutions on the right correspond to meshes refined with fitted multi-level \textit{hp} refinements in panel (c) and unfitted refinements in panel (d).
In both approaches, the polynomial degree varied across refinement levels, starting with $p_0 = p$ on the base mesh and decreasing the order by one on each refinement level, $p_i = p - i$.
This rule also determines the maximum number of refinements, since at $i = p - 1$ the basis becomes linear, and no further degree reduction is possible.
In the fitted solution shown, we used a base mesh with degree $p = 10$ and $k = 9$ overlay meshes, whereas the unfitted example with $p = 6$ used $k = 5$ refinement levels.

These two refined solutions are shown because they exhibit comparable visual quality, each providing an almost exact approximation of the gradient.
A quantitative comparison is discussed next.

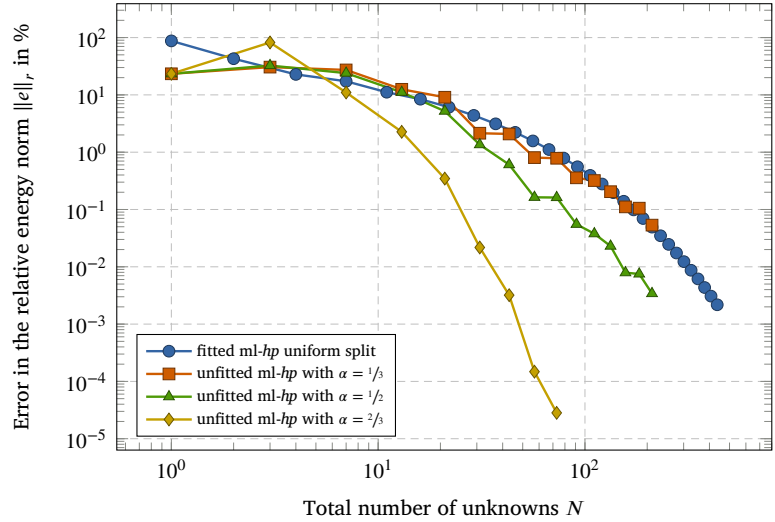
\begin{figure}[!ht]
  \centering
  \tikzsetnextfilename{one_dimensional_rod_convergence_plot}
\begin{tikzpicture}

	\begin{axis} [
		font={\footnotesize},
		axis lines = box,
		xlabel = Total number of unknowns $N$,
		ylabel = Error in the relative energy norm $\|e\|_r$ in \%,
		width = 0.6\textwidth,
		height = 7.5cm,
		max space between ticks = 40,
		grid = major,
		grid style = {densely dashed, line width = 0.1pt},
		minor x tick num = 9,
		minor y tick num = 9,
		legend pos = south west,
		legend style = {
			nodes = {scale = 0.75, transform shape},
			anchor = south west,
		},
		xmode = log,
		ymode = log,
		legend cell align={left},
	]

		\addplot table [x = N, y = E, col sep = comma] {04_Z_data_one_dimensional_rod_errors_mlhp.csv};
		\addplot table [x = N, y = E, col sep = comma] {04_Z_data_one_dimensional_rod_errors_ushp_r0.333.csv};
		\addplot table [x = N, y = E, col sep = comma] {04_Z_data_one_dimensional_rod_errors_ushp_r0.5.csv};
		\addplot table [x = N, y = E, col sep = comma, restrict expr to domain={\coordindex}{0:8}] {04_Z_data_one_dimensional_rod_errors_ushp_r0.666.csv};

		\legend{
      fitted ml-\textit{hp} uniform split,
      unfitted ml-\textit{hp} with $\alpha = \sfrac{1}{3}$,
      unfitted ml-\textit{hp} with $\alpha = \sfrac{1}{2}$,
      unfitted ml-\textit{hp} with $\alpha = \sfrac{2}{3}$
		};

	\end{axis}


\end{tikzpicture}
  \caption{Convergence of the relative energy-norm error $\|e\|_r$ for different refinement strategies applied to the one-dimensional bar problem.}
  \label{fig:benchmark_elastic_bar_errors}
\end{figure}

\paragraph{Quantitative assessment}
To evaluate the approximation quality of the methods, we conducted a convergence study.
Both the fitted and unfitted methods started with a single-element mesh of polynomial degree $p = 1$ without refinements.
With each refinement cycle, the polynomial order on every mesh level was increased by one, and an additional refinement layer was introduced.
The results of this study are presented in \Cref{fig:benchmark_elastic_bar_errors}.

The fitted multi-level \textit{hp} approach (red) exhibits exponential convergence, significantly exceeding the algebraic convergence rates typically observed for global \textit{h}- or \textit{p}-refinements in the presence of non-smooth solutions.

Since the unfitted approach permits explicit control over the size of the overlay meshes, three separate studies were performed using $\alpha = \{ \sfrac{1}{3},\ \sfrac{1}{2},\ \sfrac{2}{3} \}$, where $\alpha = \sfrac{2}{3}$ represents the largest permissible value for the given geometric configuration.
The results for $\alpha = \sfrac{1}{3}$ achieved exponential convergence comparable to that of the fitted approach.
Increasing $\alpha$ to $\sfrac{1}{2}$ and $\sfrac{2}{3}$ further improved the results, yielding higher convergence and, notably, lower errors per degree of freedom relative to the fitted method.

The final fitted solution resulted in an error of $2.17\cdot 10^{-3}\,\%$ with $n = 436$ unknowns, whereas the unfitted solution with $\alpha = \sfrac{2}{3}$ attained a significantly smaller error of $4.34\cdot 10^{-5}\,\%$ at less than half the number of unknowns ($n = 211$).

\subsection{Singular corner problem}
The second benchmark extends the investigation to a two-dimensional setting, illustrating conceptual and behavioral properties in higher dimensions.

\begin{figure}[!ht]
  \centering
  \begin{tabular}{c @{\hspace*{0cm}} c @{\hspace*{0cm}} c}
    \tikzsetnextfilename{singular_corner_model}
\begin{tikzpicture}[3d view = {-30}{30}]

  \begin{scope}[canvas is xy plane at z = 0]
    \fill[draw = black, surface] (0, 0) rectangle ++(3, 3);

    \draw[red, densely dashed] (3.1, 0.1) -- node[right, transform shape] {$\Gamma_N$} ++(0, 3) -- ++(-3, 0); 
    \draw[blue, densely dashed] (-0.1, 2.9) -- node[left, transform shape] {$\Gamma_D$} ++(0, -3) -- ++(3, 0); 

    \node[transform shape] at (1.5, 1.5) {$\Omega = [0, 1]^2$};

    \draw[black, -latex] (3, 0) -- node[transform shape, right] {$x_1$} ++(0, -1);
    \draw[black, -latex] (0, 3) -- node[transform shape, above] {$x_2$} ++(-1, 0);

    \node[purple, transform shape] (singularity) at (1.5, 2.25) {singularity};
    \draw[purple, -latex] (singularity) to[out = 60, in = -135] (3, 3);
  \end{scope}

\end{tikzpicture} &
    \tikzsetnextfilename{singular_corner_mesh_fitted_multi_level}
\begin{tikzpicture}[3d view = {-30}{30}]

  \begin{scope}[canvas is xy plane at z = 0]
    \fill[inactivesurface, draw = black, densely dotted] (0, 0) rectangle (3, 3);
  \end{scope}

  \begin{scope}[canvas is xy plane at z = 1]
    \fill[surface, draw = black] (0, 0) rectangle ++(1.5, 1.5);
    \fill[surface, draw = black] (0, 1.5) rectangle ++(1.5, 1.5);
    \fill[surface, draw = black] (1.5, 0) rectangle ++(1.5, 1.5);
    \fill[inactivesurface, draw = black, densely dotted] (1.5, 1.5) rectangle ++(1.5, 1.5);

    \foreach \x in {0, 1.5, 3}
      \foreach \y in {0, 1.5}
        \draw[black, fill = white] (\x, \y) circle (0.4mm);

    \foreach \x in {0, 1.5}
      \draw[black, fill = white] (\x, 3) circle (0.4mm);
  \end{scope}
  
  \begin{scope}[canvas is xy plane at z = 2]
    \begin{scope}[xshift = 1.5cm, yshift = 1.5cm]
      \fill[surface, draw = black, densely dotted] (0, 0) rectangle ++(0.75, 0.75);
      \fill[surface, draw = black, densely dotted] (0, 0.75) rectangle ++(0.75, 0.75);
      \fill[surface, draw = black, densely dotted] (0.75, 0) rectangle ++(0.75, 0.75);
      \fill[inactivesurface, draw = black, densely dotted] (0.75, 0.75) rectangle ++(0.75, 0.75);

      \draw[black] (0, 0.75) -- ++(1.5, 0) -- ++(0, -0.75);
      \draw[black] (0.75, 0) -- ++(0, 1.5) -- ++(-0.75, 0);

      \draw[black, fill = white] (0.75, 0.75) circle (0.4mm);
      \draw[black, fill = white] (0.75, 1.5) circle (0.4mm);
      \draw[black, fill = white] (1.5, 0.75) circle (0.4mm);
    \end{scope}
  \end{scope}

  \begin{scope}[canvas is xy plane at z = 3]
    \begin{scope}[xshift = 2.25cm, yshift = 2.25cm]
      \fill[surface, draw = black, densely dotted] (0, 0) rectangle ++(0.375, 0.375);
      \fill[surface, draw = black, densely dotted] (0, 0.375) rectangle ++(0.375, 0.375);
      \fill[surface, draw = black, densely dotted] (0.375, 0) rectangle ++(0.375, 0.375);
      \fill[surface, draw = black, densely dotted] (0.375, 0.375) rectangle ++(0.375, 0.375);

      \draw[black] (0, 0.375) -- ++(0.75, 0);
      \draw[black] (0.375, 0) -- ++(0, 0.75);
      \draw[black] (0, 0.75) -- ++(0.75, 0) -- ++(0, -0.75);

      \foreach \x in {0.375, 0.75}
        \foreach \y in {0.375, 0.75}
          \draw[black, fill = white] (\x, \y) circle (0.4mm);
    \end{scope}
  \end{scope}

  \draw[black, latex-] (3, 0, 0.1) -- node[right] {$+$} ++(0, 0, 0.8);
  \draw[black, latex-] (3, 1.5, 1.1) -- node[right] {$+$} ++(0, 0, 0.8);
  \draw[black, latex-] (3, 2.25, 2.1) -- node[right] {$+$} ++(0, 0, 0.8);

  \node[left] at (0, 3, 0) {\footnotesize$k = 0$};
  \node[left] at (0, 3, 1) {\footnotesize$k = 1$};
  \node[left] at (1.5, 3, 2) {\footnotesize$k = 2$};
  \node[left] at (2.25, 3, 3) {\footnotesize$k = 3$};

\end{tikzpicture} &
    \tikzsetnextfilename{singular_corner_mesh_unfitted_multi_level}
\begin{tikzpicture}[3d view = {-30}{30}]

  \begin{scope}[canvas is xy plane at z = 0]
    \fill[surface, draw = black] (0, 0) rectangle (3, 3);

    \fill[black, opacity = 0.075] (1.5, 1.5) rectangle ++(1.5, 1.5);

    \foreach \x in {0, 3}
      \foreach \y in {0, 3}
        \draw[black, fill = white] (\x, \y) circle (0.4mm);
  \end{scope}

  \begin{scope}[canvas is xy plane at z = 1, xshift = 1.5cm, yshift = 1.5cm]
    \fill[surface, draw = black, densely dashed] (0, 0) rectangle (1.5, 1.5);
    \draw[black] (1.5, 0) -- ++(0, 1.5) -- ++(-1.5, 0);
    \draw[black, fill = white] (1.5, 1.5) circle (0.4mm);
    \fill[black, opacity = 0.075] (0.75, 0.75) rectangle ++(0.75, 0.75);

    \draw[black] (-0.1, 0) -- ++(-0.3, 0);
    \draw[black] (-0.1, 1.5) -- ++(-0.3, 0);
    \draw[black, latex-latex] (-0.3, 0) -- node[transform shape, left] {$\alpha^1$} ++(0, 1.5);
  \end{scope}

  \begin{scope}[canvas is xy plane at z = 2, xshift = 2.25cm, yshift = 2.25cm]
    \fill[surface, draw = black, densely dashed] (0, 0) rectangle (0.75, 0.75);
    \draw[black] (0.75, 0) -- ++(0, 0.75) -- ++(-0.75, 0);
    \draw[black, fill = white] (0.75, 0.75) circle (0.4mm);
    \fill[black, opacity = 0.075] (0.375, 0.375) rectangle ++(0.375, 0.375);

    \draw[black] (-0.1, 0) -- ++(-0.3, 0);
    \draw[black] (-0.1, 0.75) -- ++(-0.3, 0);
    \draw[black, latex-latex] (-0.3, 0) -- node[transform shape, left] {$\alpha^2$} ++(0, 0.75);
  \end{scope}

  \begin{scope}[canvas is xy plane at z = 3, xshift = 2.625cm, yshift = 2.625cm]
    \fill[surface, draw = black, densely dashed] (0, 0) rectangle (0.375, 0.375);
    \draw[black] (0.375, 0) -- ++(0, 0.375) -- ++(-0.375, 0);
    \draw[black, fill = white] (0.375, 0.375) circle (0.4mm);

    \draw[black] (-0.1, 0) -- ++(-0.3, 0);
    \draw[black] (-0.1, 0.375) -- ++(-0.3, 0);
    \draw[black, latex-latex] (-0.3, 0) -- node[transform shape, left] {$\alpha^3$} ++(0, 0.375);
  \end{scope}

  \draw[black, latex-] (3, 1.5, 0.1) -- node[right] {$+$} ++(0, 0, 0.8);
  \draw[black, latex-] (3, 2.25, 1.1) -- node[right] {$+$} ++(0, 0, 0.8);
  \draw[black, latex-] (3, 2.625, 2.1) -- node[right] {$+$} ++(0, 0, 0.8);

  \node[left] at (0, 3, 0) {\footnotesize$k = 0$};
  \node[left] at (1.5, 3, 1.2) {\footnotesize$k = 1$};
  \node[left] at (2.25, 3, 2.2) {\footnotesize$k = 2$};
  \node[left] at (2.625, 3, 3.2) {\footnotesize$k = 3$};

\end{tikzpicture} \\
    a) Geometric setup. & b) Fitted multi-level \textit{hp}. & c) Unfitted multi-level \textit{hp}. \\
  \end{tabular}
  \caption{Geometric setup of the singular corner problem \cite{Kopp:2022aa} in (a). 
  (b) and (c) show the fitted and unfitted multi-level \textit{hp} meshing approaches, respectively, each with three refinement levels toward the singularity. 
  In the unfitted approach the overlay size is controlled by the parameter $\alpha$.}
  \label{fig:benchmark_singular_square}
\end{figure}
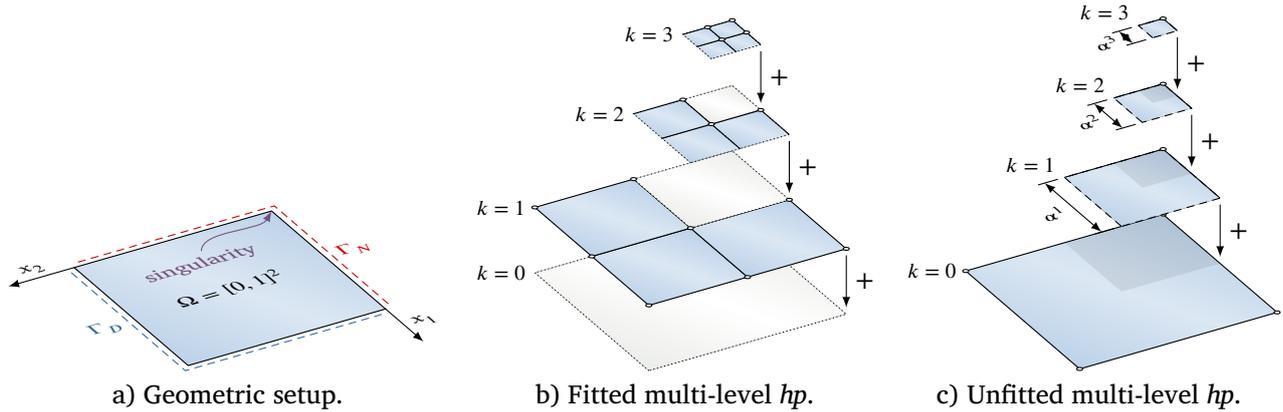

The benchmark problem, taken from \cite{Kopp:2022aa}, features a unit square $\Omega = [0, 1]^2$ governed by the Poisson equation.
The geometric setup is shown in \Cref{fig:benchmark_singular_square} (a).

The boundaries of the square aligned with the coordinate axes were subjected to Neumann (flux-free) conditions, whereas the remaining boundaries were set to the exact solution, resulting in the governing equations
\begin{subequations}
  \begin{align}
    \nabla^2 u &= s &&\forall \boldsymbol{x} \in \Omega, \\
    \boldsymbol{n} \cdot \nabla u &= 0 &&\forall \boldsymbol{x} \in \Gamma_N, \\
    u &= u_\mathrm{exact} &&\forall \boldsymbol{x} \in \Gamma_D.
  \end{align}
\end{subequations}

With the radial coordinate $\rho = \| \boldsymbol{x} \|_2$, the manufactured exact solution produces a singular gradient at $\boldsymbol{x} = [0, 0]$:
\begin{equation}
  u_\mathrm{exact} = \sqrt{\rho}.
\end{equation}
To obtain this solution, the following source term is applied in the interior:
\begin{equation}
  s = \frac{1}{4}\, \rho^{-\frac{3}{2}}.
\end{equation}
The exact energy for the problem can be derived analytically as:
\begin{equation}
  a\left( u_\mathrm{exact}, u_\mathrm{exact} \right)
  = \frac{1}{4} \int_\Omega \frac{1}{\rho}\, \mathrm{d}\Omega
  = \frac{1}{4} \log\left[ - \frac{\sqrt{2} + 2}{\sqrt{2} - 2} \right].
\end{equation}

The discretization strategies employed for this problem are shown in \Cref{fig:benchmark_singular_square} (b) and (c), both beginning with a single-element mesh.
In \Cref{fig:benchmark_singular_square} (b), the fitted multi-level \textit{hp} method is depicted, which recursively bisects elements toward the coordinate origin, where the gradient singularity occurs.
Each refinement level consequently forms a $2 \times 2$ overlay mesh.

The unfitted approach shown in \Cref{fig:benchmark_singular_square} (c) avoided the creation of three additional elements that the fitted method introduced away from the origin.
Instead, it utilized a single overlay element per refinement layer.
Since there is no topological coupling between the meshes, the overlay size could be controlled using the parameter $\alpha$, with each overlay mesh covering the domain $\Omega^{(k)} = [0, \alpha^k]^2$.

\begin{figure}[!ht]
  \centering
  \begin{tabular}{c @{} c @{} c @{} l}
    fitted ml-\textit{hp} & unfitted ml-\textit{hp} --- $\alpha = 0.5$ & unfitted ml-\textit{hp} --- $\alpha = 0.2$ & \\[2mm]
    \includegraphics[width = 0.3\textwidth]{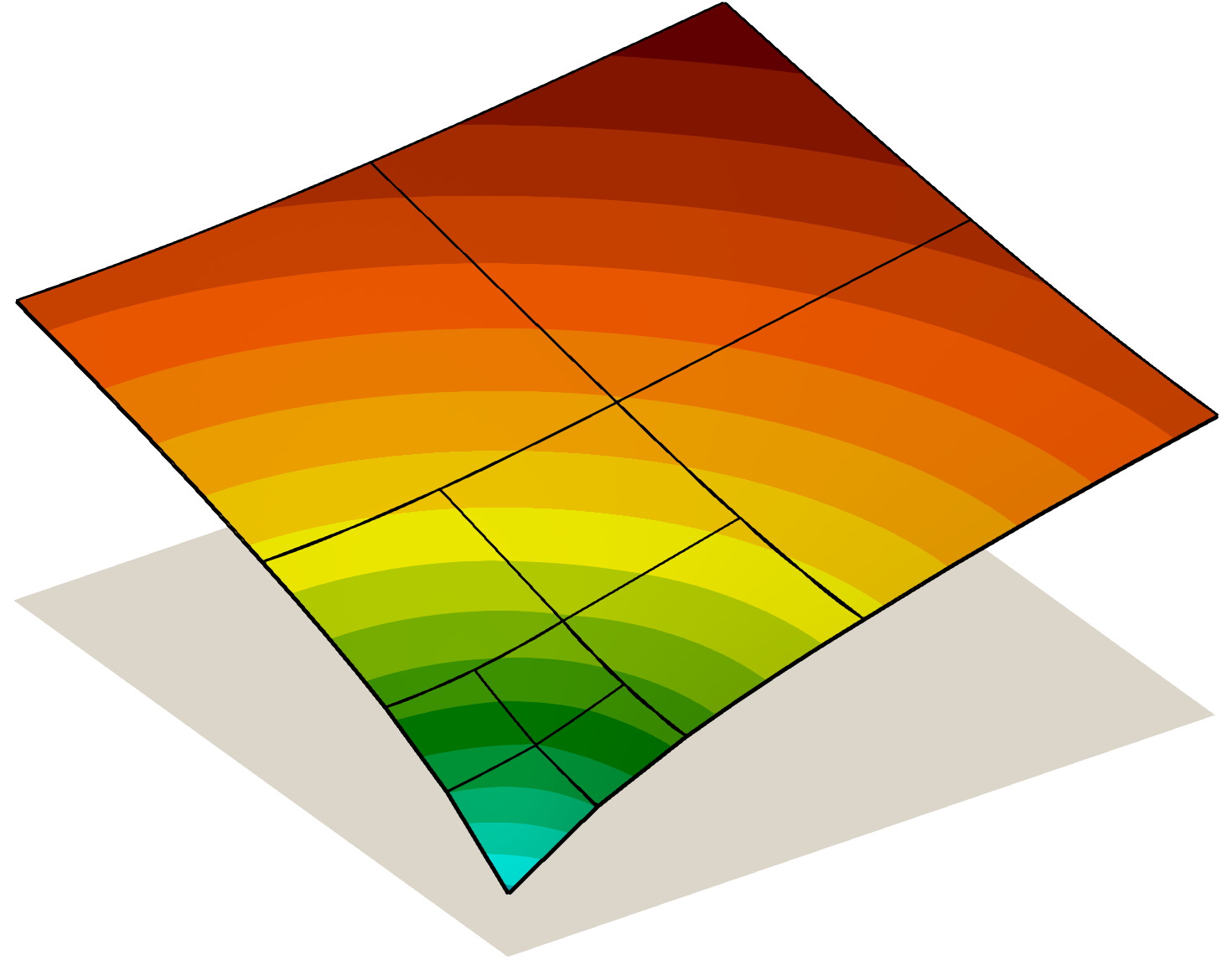} &
    \includegraphics[width = 0.3\textwidth]{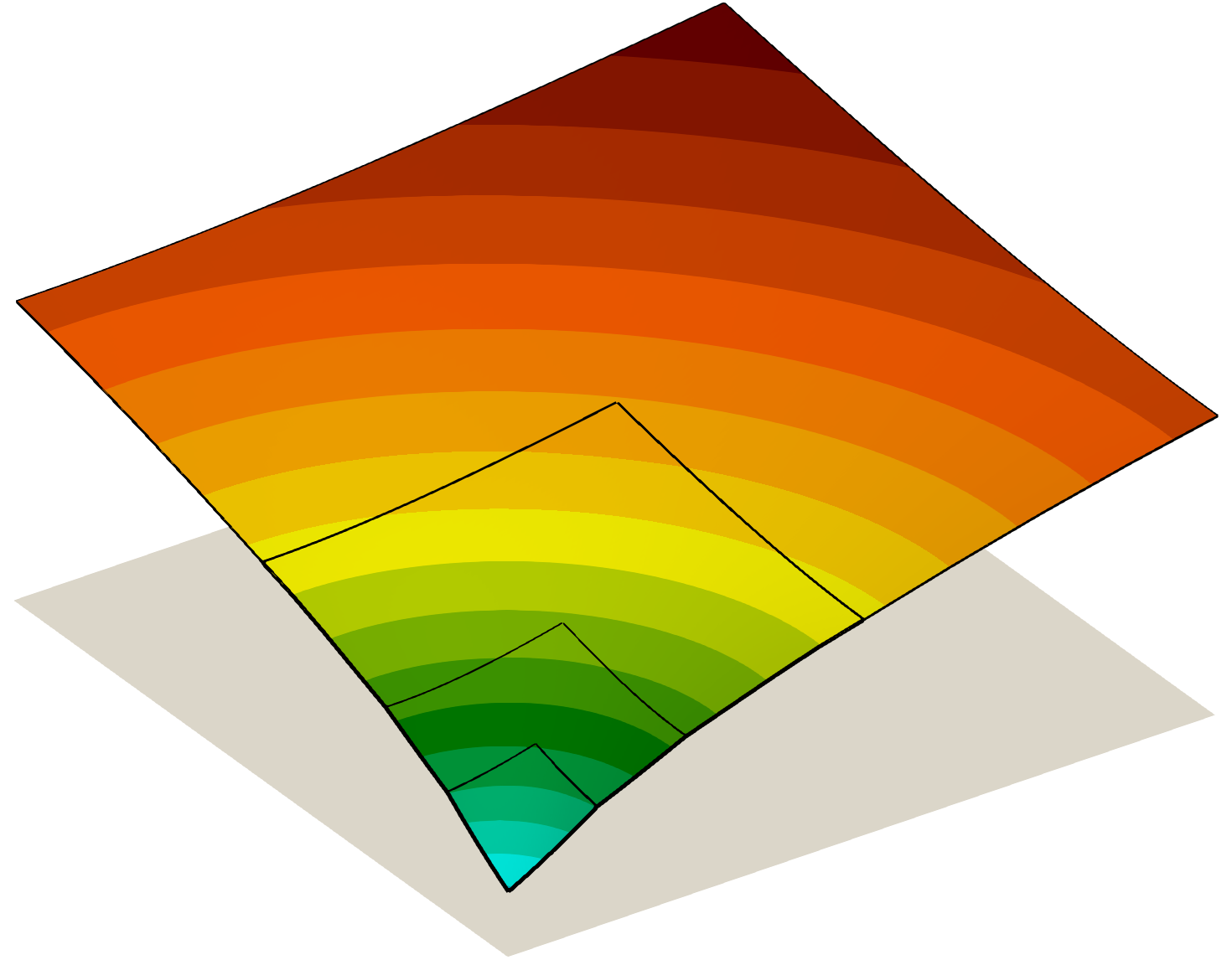} &
    \includegraphics[width = 0.3\textwidth]{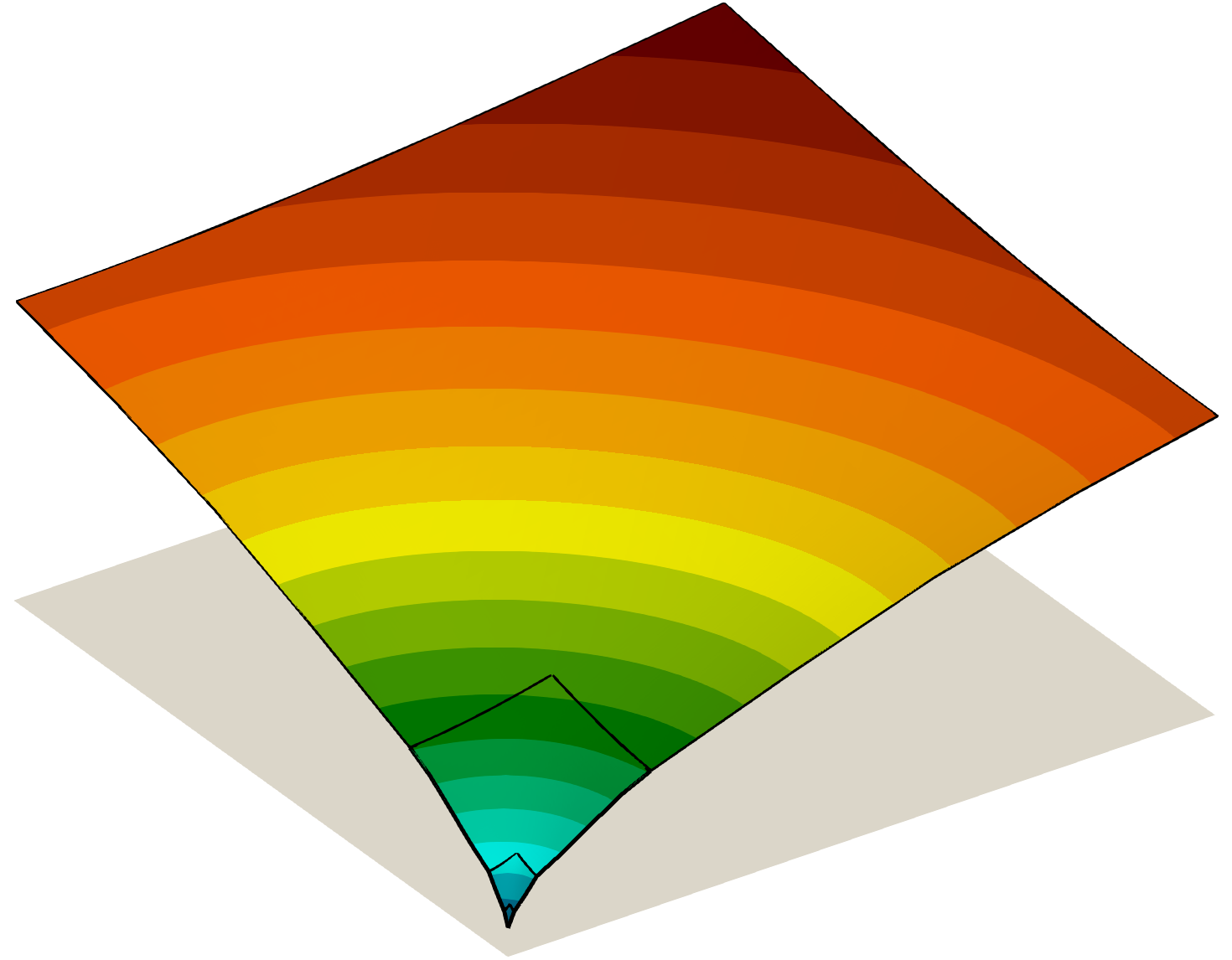} &
    \begin{tikzpicture}%
      \colorbar{0cm}{0cm}{0.0}{0.12}{2cm}{\phantom{$\| \nabla$}$u$\phantom{$\|_2$}}%
    \end{tikzpicture} \\
    \includegraphics[width = 0.3\textwidth]{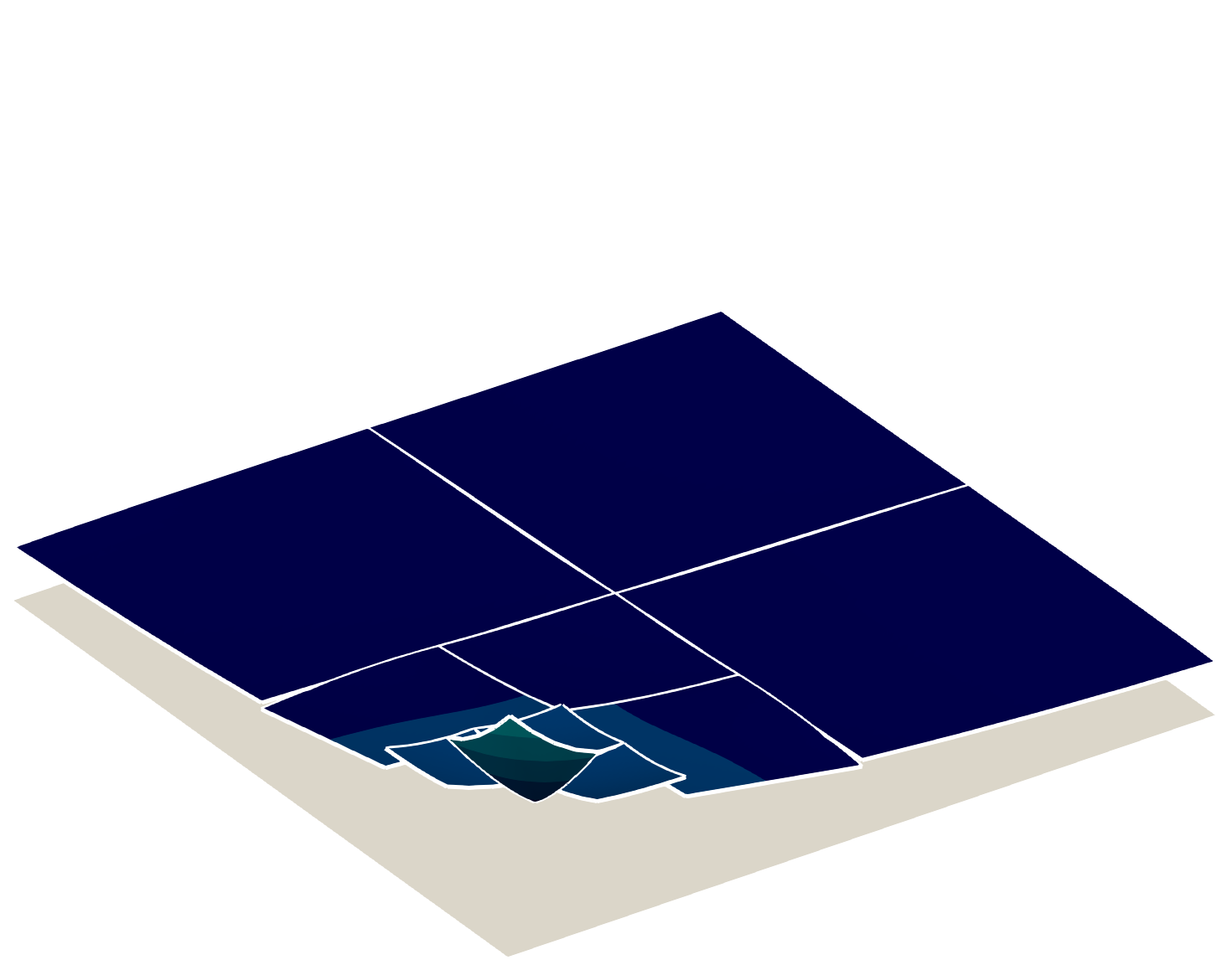} &
    \includegraphics[width = 0.3\textwidth]{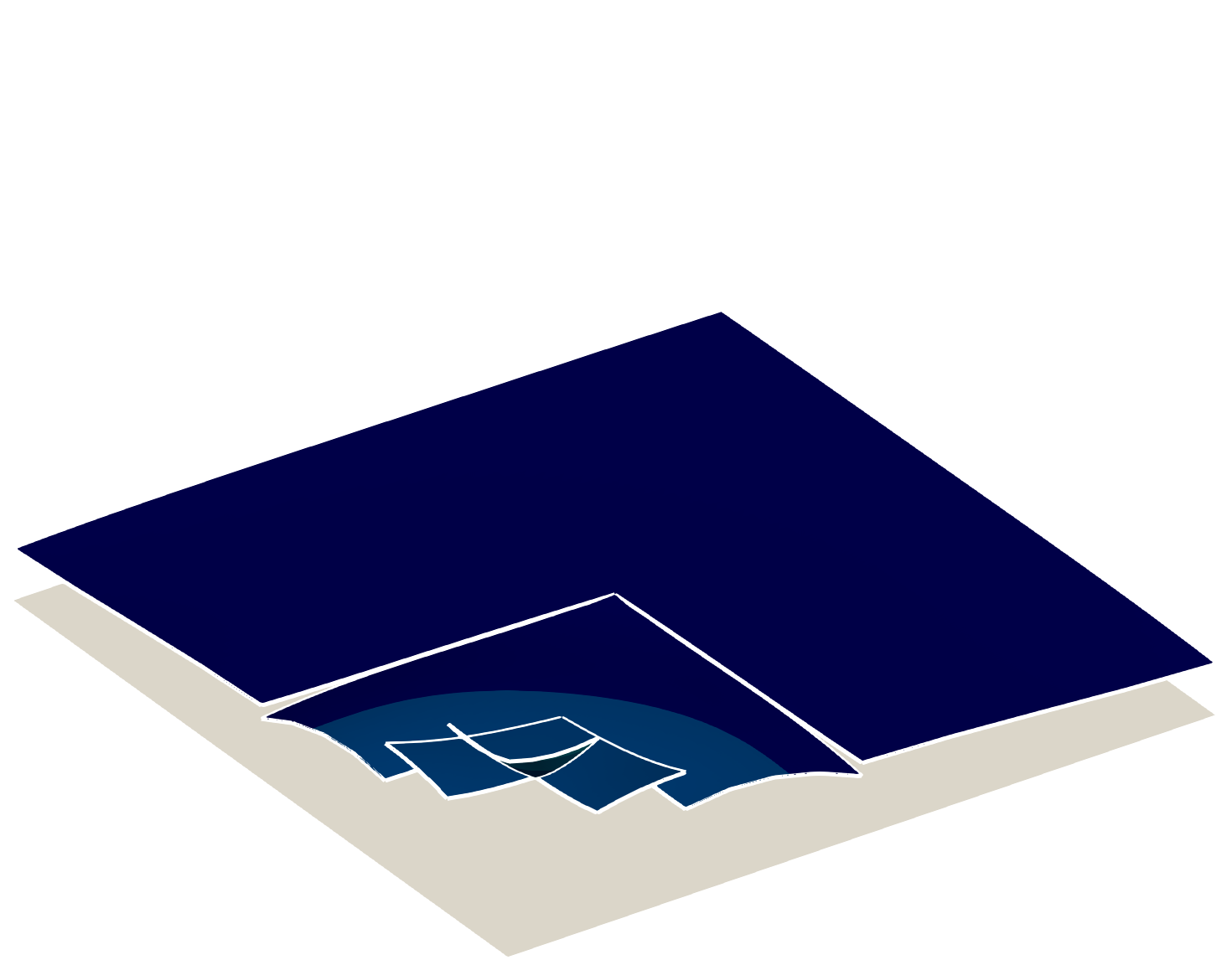} &
    \includegraphics[width = 0.3\textwidth]{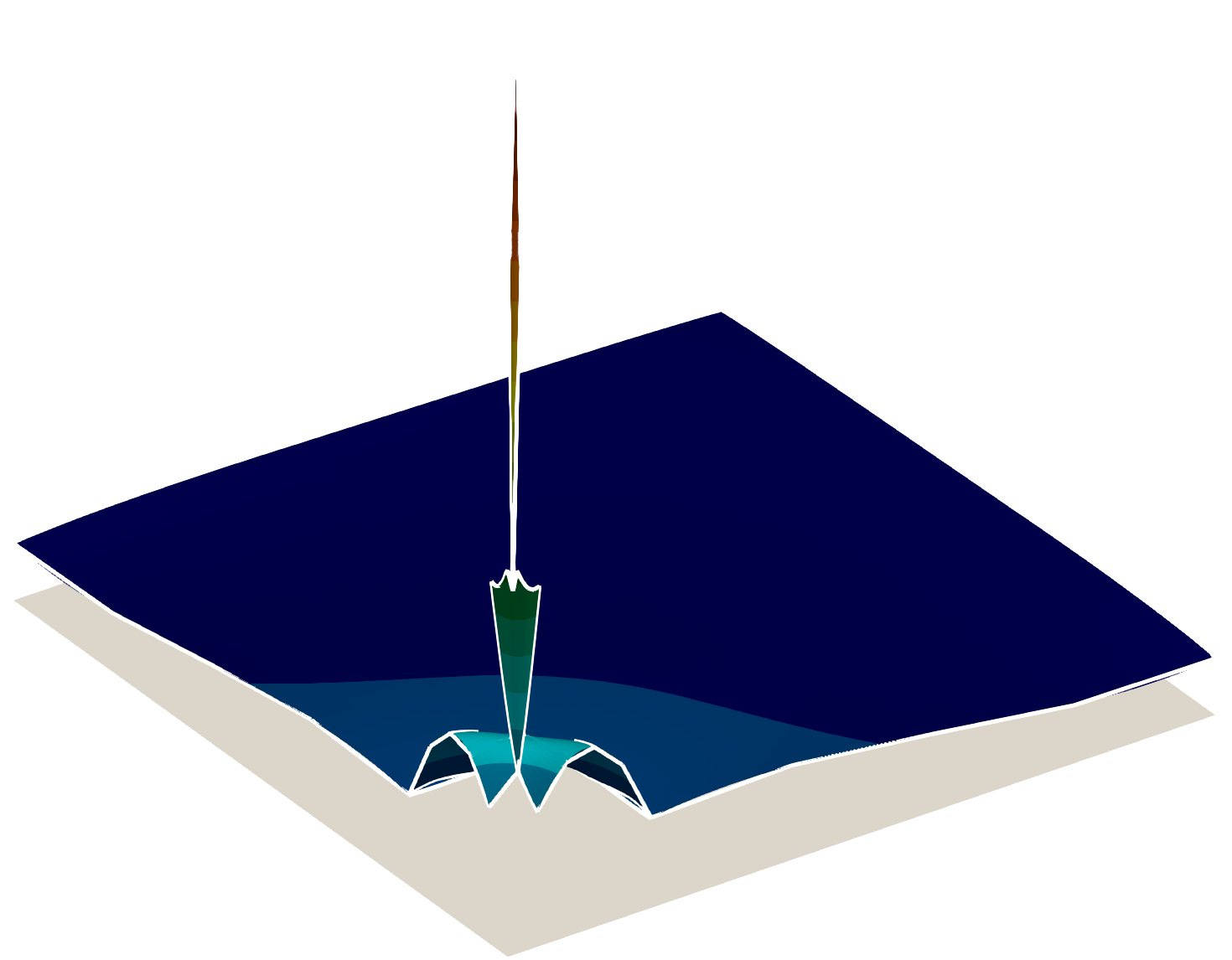} &
    \begin{tikzpicture}%
      \colorbar{0cm}{0cm}{0.0}{8}{2cm}{$\| \nabla u \|_2$}%
    \end{tikzpicture} \\
    28 DoFs & 17 DoFs & 17 DoFs & \\
  \end{tabular}
  \caption{Numerical solution of the singular corner problem for three different meshes. 
  The top row displays the solution field $u$, while the bottom row depicts the gradient magnitude $\|\nabla u\|_2$. 
  In all three solutions, a $1 \times 1$ base mesh and three subsequent refinement layers were used. 
  The first column used fitted multi-level \textit{hp} refinements, and the second and third used unfitted refinements with overlay size factors of $\alpha = 0.5$ and $\alpha = 0.2$, respectively.}
  \label{fig:benchmark_singular_square_visual_results}
\end{figure}

\paragraph{Visual assessment}
The visual results from three representative computations are shown in \Cref{fig:benchmark_singular_square_visual_results}.

The top row displays the solution $u$, while the bottom row shows the gradient magnitude $\|\nabla u\|_2$, both vertically scaled for visualization.
All results were obtained with three refinement layers.
The fitted approach (first column) results in a model with 28 degrees of freedom.
The unfitted approach in the second column corresponds to $\alpha = 0.5$ and uses the same refinement pattern.
While the visual quality is nearly identical, the model requires only 17 degrees of freedom because it omits the three unnecessary elements.

The rightmost model uses the same refinement structure as the second one but with a smaller overlay size factor of $\alpha = 0.2$.
This yields a substantially improved approximation of the singularity, at the expense of more pronounced oscillations in the neighboring elements.

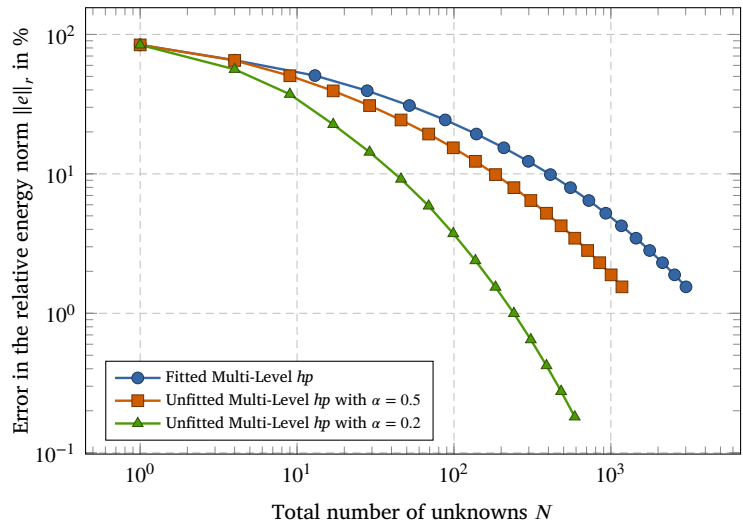
\begin{figure}[!ht]
  \centering
  \tikzsetnextfilename{singular_corner_convergence_plot}
\begin{tikzpicture}

	\begin{axis} [
		font={\footnotesize},
		axis lines = box,
		xlabel = Total number of unknowns $N$,
		ylabel = Error in the relative energy norm $\|e\|_r$ in \%,
		width = 0.6\textwidth,
		height = 7.5cm,
		max space between ticks = 40,
		grid = major,
		grid style = {densely dashed, line width = 0.1pt},
		minor x tick num = 9,
		minor y tick num = 9,
		legend pos = south west,
		legend style = {
			nodes = {scale = 0.75, transform shape},
			anchor = south west,
		},
		ylabel style = {
			yshift = -4mm
		},
		xmode = log,
		ymode = log,
		legend cell align={left},
	]

		\addplot table [x = N, y = E, col sep = comma] {04_Z_data_singular_corner_errors_mlhp.csv};
		\addplot table [x = N, y = E, col sep = comma] {04_Z_data_singular_corner_errors_ushp_r0.5.csv};
		\addplot table [x = N, y = E, col sep = comma, restrict expr to domain={\coordindex}{0:14}] {04_Z_data_singular_corner_errors_ushp_r0.2.csv};

		\legend{
      Fitted Multi-Level \textit{hp},
      Unfitted Multi-Level \textit{hp} with $\alpha = 0.5$,
      Unfitted Multi-Level \textit{hp} with $\alpha = 0.2$
		};

	\end{axis}

\end{tikzpicture}
  \caption{Error convergence of the singular corner problem for different refinement strategies.}
  \label{fig:benchmark_singular_square_errors}
\end{figure}

\paragraph{Quantitative assessment}
The convergence behavior of the refinement strategies provided a quantitative evaluation.
All refinement studies started with a single element with $p = 1$.
In each refinement step, the polynomial order was increased by one and an additional refinement layer was added.
The resulting errors, measured in the relative energy norm, are shown in \Cref{fig:benchmark_singular_square_errors}.

The three curves correspond to the strategies shown in \Cref{fig:benchmark_singular_square_visual_results}.
The red curve represents the fitted approach, which achieves an exponential rate of convergence.
The unfitted approach with $\alpha = 0.5$ yields nearly identical errors at each refinement level, but with fewer unknowns due to the absence of the three unnecessary elements.
The final fitted model yields an error of $1.54778\,\%$ with $3013$ unknowns, whereas the corresponding unfitted model achieves an essentially identical error of $1.54783\,\%$ using only $1177$ unknowns.

Further reduction of the overlay size to $\alpha = 0.2$ improves the results, resulting in a significantly faster convergence rate and a final error of $0.12734\,\%$ with $1160$ unknowns.

These computations were performed without over-integration of the singular source term.
While over-integration reduces the absolute errors, additional tests indicated that it does not alter the relative behavior or error ratios between the refinement strategies.

\subsection{Influence of small overlaps}
\begin{figure}[!ht]
  \centering
  \begin{tabular}{@{} c @{\enspace} c @{\enspace} c @{}}
    \tikzsetnextfilename{small_overlap_model}
\begin{tikzpicture}

  \foreach \x in {0, 1, 2}
  {
    \foreach \y in {0, 1, 2}
    {
      \draw[black, alternativesurface] (\x, \y) rectangle ++(1cm, 1cm);
    }
  }

  \foreach \x in {0, 1, 2, 3}
  {
    \foreach \y in {0, 1, 2, 3}
    {
      \draw[black, fill = white] (\x, \y) circle (0.4mm);
    }
  }

  \fill[black, opacity = 0.1] (0.35cm, 0.15cm) rectangle ++(2cm, 2cm);
  \foreach \x in {0.25cm, 1.25cm}
  {
    \foreach \y in {0.25cm, 1.25cm}
    {
      \draw[black, densely dotted, surface] (\x, \y) rectangle ++(1cm, 1cm);
    }
  }

  \fill[lightred, opacity = 0.1] (2cm, 0.25cm) -- ++(0cm, 1.75cm) -- ++(-1.75cm, 0cm) -- ++(0cm, 0.25cm) -- ++(2cm, 0cm) -- ++(0cm, -2cm) -- cycle;
  \fill[opacity = 0.5, pattern = north east lines, pattern color = lightred] (2cm, 0.25cm) -- ++(0cm, 1.75cm) -- ++(-1.75cm, 0cm) -- ++(0cm, 0.25cm) -- ++(2cm, 0cm) -- ++(0cm, -2cm) -- cycle;

  \draw[black, densely dotted] (0.25cm, 1cm) -- ++(2cm, 0cm);
  \draw[black, densely dotted] (0.25cm, 2cm) -- ++(2cm, 0cm);

  \draw[black, densely dotted] (1cm, 0.25cm) -- ++(0cm, 2cm);
  \draw[black, densely dotted] (2cm, 0.25cm) -- ++(0cm, 2cm);

  \draw[black] (0.25cm, 1.25cm) -- ++(2cm, 0cm);
  \draw[black] (1.25cm, 0.25cm) -- ++(0cm, 2cm);
  \draw[black, fill = white] (1.25cm, 1.25cm) circle (0.4mm);

  \draw[black] (3.1cm, 0cm) -- ++(0.3cm, 0cm);
  \draw[black] (3.1cm, 3cm) -- ++(0.3cm, 0cm);
  \draw[black, latex-latex] (3.3cm, 0cm) -- node[rotate = 90, below] {$3\, a$} ++(0cm, 3cm);

  \draw[black] (0cm, 3.1cm) -- ++(0cm, 0.3cm);
  \draw[black] (3cm, 3.1cm) -- ++(0cm, 0.3cm);
  \draw[black, latex-latex] (0cm, 3.3cm) -- node[above] {$3\, a$} ++(3cm, 0cm);

  \draw[black] (0cm, -0.1cm) -- ++(0cm, -0.3cm);
  \draw[black] (2.25cm, 0.15cm) -- ++(0cm, -0.55cm);
  \draw[black, latex-latex] (0cm, -0.3cm) -- node[below] {$2\, a + \eta$} ++(2.25cm, 0cm);

  \draw[black] (-0.1cm, 0cm) -- ++(-0.3cm, 0cm);
  \draw[black] (0.15cm, 2.25cm) -- ++(-0.55cm, 0cm);
  \draw[black, latex-latex] (-0.3cm, 0cm) -- node[above, rotate = 90] {$2\, a + \eta$} ++(0cm, 2.25cm);
\end{tikzpicture} &
    \tikzsetnextfilename{conditioning_condition_number_plot}
\begin{tikzpicture}

	\begin{axis} [
		font={\footnotesize},
		axis lines = box,
		xlabel = Overlap $\eta$,
		ylabel = Condition number $\kappa$,
		width = 6cm,
		height = 7.5cm,
		max space between ticks = 40,
		grid = major,
		grid style = {densely dashed, line width = 0.1pt},
		minor x tick num = 9,
		minor y tick num = 9,
		legend pos = north east,
		legend columns = 1,
		legend style = {
			nodes = {scale = 0.75, transform shape},
			anchor = north east,
		},
		y label style = {
			yshift = -0.3cm
		},
		xmode = log,
		ymode = log,
		legend cell align={left},
	]

		\addplot table [x = o, y = p1, col sep = comma] {04_Z_data_small_overlap_condition_number.csv};
		\addplot table [x = o, y = p2, col sep = comma] {04_Z_data_small_overlap_condition_number.csv};
		\addplot table [x = o, y = p3, col sep = comma] {04_Z_data_small_overlap_condition_number.csv};
		\addplot table [x = o, y = p4, col sep = comma] {04_Z_data_small_overlap_condition_number.csv};
		\addplot table [x = o, y = p5, col sep = comma] {04_Z_data_small_overlap_condition_number.csv};
		\addplot table [x = o, y = p6, col sep = comma] {04_Z_data_small_overlap_condition_number.csv};
		\addplot table [x = o, y = p7, col sep = comma] {04_Z_data_small_overlap_condition_number.csv};
		\addplot table [x = o, y = p8, col sep = comma] {04_Z_data_small_overlap_condition_number.csv};

    \legend{
      $p = 1$, 
      $p = 2$,
      $p = 3$,
      $p = 4$,
      $p = 5$,
      $p = 6$,
      $p = 7$,
      $p = 8$,
		};

	\end{axis}

\end{tikzpicture} &
    \tikzsetnextfilename{conditioning_pcg_iterations_plot}
\begin{tikzpicture}

	\begin{axis} [
		font={\footnotesize},
		axis lines = box,
		xlabel = Ovelap $\eta$,
		ylabel = PCG Iterations with Jacobi Preconditioning,
		width = 6cm,
		height = 7.5cm,
		max space between ticks = 40,
		grid = major,
		grid style = {densely dashed, line width = 0.1pt},
		minor x tick num = 9,
		minor y tick num = 9,
		legend pos = north east,
		legend columns = 4,
		legend style = {
			nodes = {scale = 0.6, transform shape},
			anchor = north east,
		},
		y label style = {
			yshift = -0.4cm
		},
		xmode = log,
		ymode = log,
		legend cell align={left},
	]

		\addplot table [x = o, y = p1, col sep = comma] {04_Z_data_small_overlap_pcg_iterations.csv};
		\addplot table [x = o, y = p2, col sep = comma] {04_Z_data_small_overlap_pcg_iterations.csv};
		\addplot table [x = o, y = p3, col sep = comma] {04_Z_data_small_overlap_pcg_iterations.csv};
		\addplot table [x = o, y = p4, col sep = comma] {04_Z_data_small_overlap_pcg_iterations.csv};
		\addplot table [x = o, y = p5, col sep = comma] {04_Z_data_small_overlap_pcg_iterations.csv};
		\addplot table [x = o, y = p6, col sep = comma] {04_Z_data_small_overlap_pcg_iterations.csv};
		\addplot table [x = o, y = p7, col sep = comma] {04_Z_data_small_overlap_pcg_iterations.csv};
		\addplot table [x = o, y = p8, col sep = comma] {04_Z_data_small_overlap_pcg_iterations.csv};

	\end{axis}

\end{tikzpicture} \\
    a) Geometric setup. & b) Condition number. & c) Solver convergence. \\
  \end{tabular}
  \caption{Setup and results of a test model used to investigate the influence of small overlaps between superimposed meshes. 
  (a) shows the geometric setup, consisting of a $3 \times 3$ base mesh and a $2 \times 2$ overlay mesh offset in $x_1$ and $x_2$ by the scalar $\eta$. 
  (b) displays the condition number~$\kappa$ of the resulting system matrix for different values of $p$ and~$\eta$, and (c) shows the corresponding number of PCG iterations required to solve the linear system.}
  \label{fig:benchmark_small_overlaps}
\end{figure}
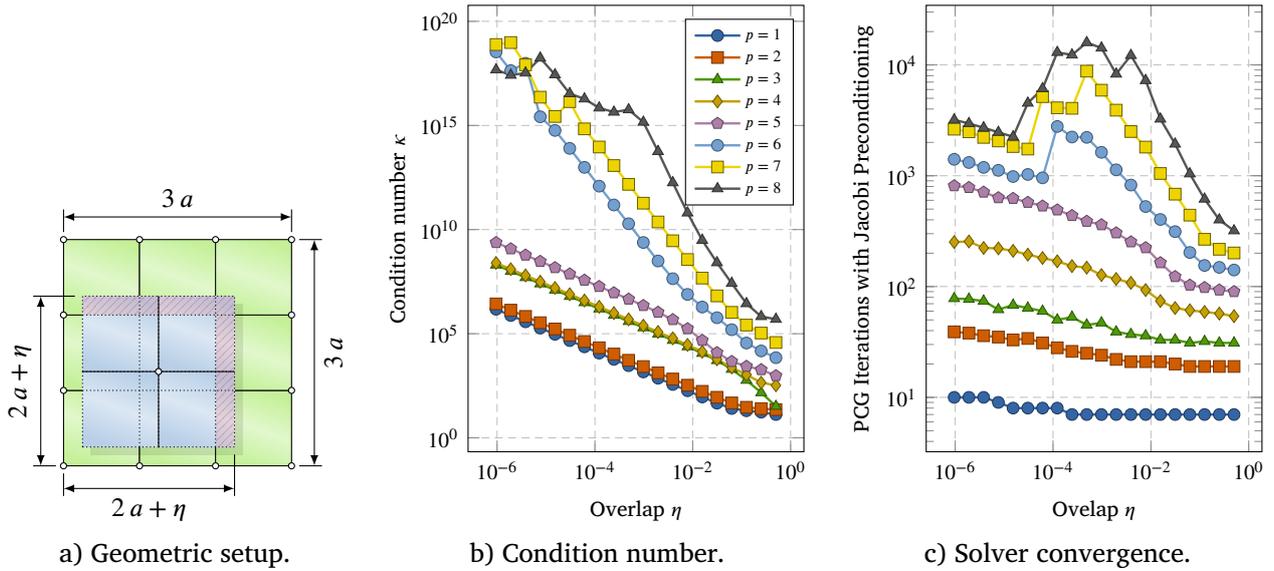

From research on fictitious-domain methods, it is well known that small cut elements, i.e., elements for which only a very small portion of their volume contributes to the analysis, can severely deteriorate the conditioning of the system matrices arising from finite element discretizations \cite{Prenter:2017aa, Prenter:2023aa, Stoter:2023aa}.
The unfitted multi-level \textit{hp} approach is subject to a similar issue in the context of partially overlapping elements.
To investigate this effect, a simple model problem, shown in \Cref{fig:benchmark_small_overlaps} (a), is used to study how small overlaps between superimposed elements influence numerical stability.

The geometric setup consists of a $3 \times 3$ base mesh covering the domain $\Omega = [0, 3a]^2$.
A $2 \times 2$ overlay mesh is superimposed, covering $\Omega^{(1)} = [\eta,\, \eta + 2a]^2$.
The parameter $\eta$ controls the amount of overlap between the base and overlay elements, creating the critical region highlighted in red in \Cref{fig:benchmark_small_overlaps} (a).

To assess the influence of the overlap, $\eta$ is varied over the range $\eta \in [10^{-6},\, 0.5]$.
For each configuration, the condition number of the system matrix associated with a Laplace problem is computed for several polynomial degrees, keeping the degree identical on both the base and overlay meshes.
The resulting condition numbers are shown in \Cref{fig:benchmark_small_overlaps} (b).

The results show that the condition number increases as the overlap decreases.
Moreover, the growth rate appears similar for $1 \leq p \leq 5$ and for $6 \leq p \leq 8$, although the latter group exhibits a substantially higher increase.
The reason for this change in behavior around $p \geq 6$ is not obvious and may require further investigation.

In addition to the condition number, the number of iterations required by the preconditioned conjugate gradient (PCG) solver is shown in \Cref{fig:benchmark_small_overlaps} (c).
A simple Jacobi preconditioner, using the inverse of the diagonal of the system matrix, is employed.
As expected, the number of iterations increases as $\eta$ decreases and as $p$ increases.
This effect becomes significantly more pronounced for higher polynomial degrees:
while the linear model exhibits only a mild increase (from 7 to 10 iterations), the case $p = 8$ ranges from 201 to 1592 iterations.

Although these results demonstrate that small overlaps can severely impact the conditioning of the system, the moving-overlay computations performed in this work did not encounter small-overlap configurations that dominated the overall computational cost.
The practical relevance of such configurations depends on the overlay motion, the chosen refinement sizes, and the solver strategy.
Potential mitigation strategies include scaling, preconditioning, or removing or aggregating overlap regions below a prescribed threshold, but these options require a systematic consistency study and are left for future work.

\subsection{Traveling heat source}
A primary motivation for the unfitted approach is the desire to avoid costly remeshing operations during transient analyses involving moving features, a setting also addressed by Chimera-FEM methods with moving local meshes \cite{Storti:2022aa}.
In this subsection, a simple test model motivated by additive manufacturing simulations is analyzed to assess the method's performance under such conditions.

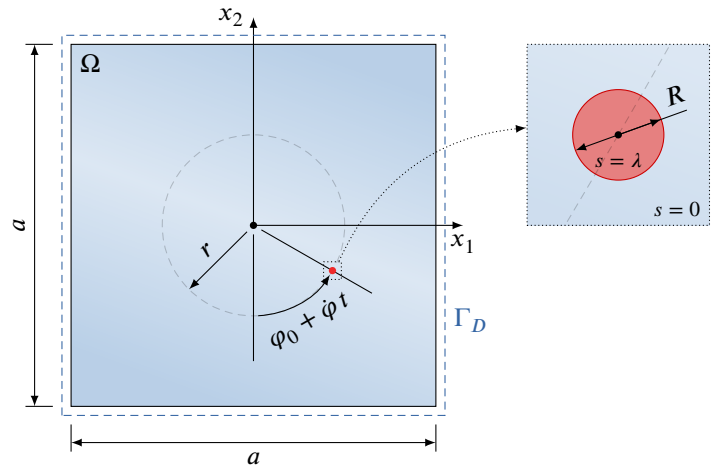
\begin{figure}[!ht]
  \centering
  \tikzsetnextfilename{traveling_heat_source_model}
\begin{tikzpicture}[scale = 1.2]

  \fill[surface] (-2cm, -2cm) rectangle (2cm, 2cm);
  \draw[foreground] (-2cm, -2cm) rectangle (2cm, 2cm);

  \fill[foreground] (0cm, 0cm) circle (0.04cm);
  \draw[foreground, -latex] (0cm, 0cm) -- node[pos = 1, below] {$x_1$} (2.3cm, 0cm);
  \draw[foreground, -latex] (0cm, 0cm) -- node[pos = 1, left] {$x_2$} (0cm, 2.3cm);

  \draw[foreground, densely dashed, opacity = 0.2]  (0cm, 0cm) circle (1cm);
  \draw[foreground, -latex] (-135:0.1cm) -- node[above, rotate = 45] {$r$} (-135:1cm);

  \draw[foreground, -latex] (-87:1cm) arc(-87:-33:1cm);
  \draw[foreground] (-90:0.1cm) -- (-90:1.5cm);
  \draw[foreground] (-30:0.1cm) -- (-30:1.5cm);
  \node[foreground, below, rotate = 30] at (-60:1cm) {$\varphi_0 + \dot{\varphi}\,t$};
  \fill[lightred]  (-30:1cm) circle (0.04cm);

  \draw[foreground] (-2.2cm, -2cm) -- ++(-0.3cm, 0cm);
  \draw[foreground] (-2.2cm,  2cm) -- ++(-0.3cm, 0cm);
  \draw[foreground, latex-latex] (-2.4cm, -2cm) -- node[above, rotate = 90] {$a$} ++(0cm, 4cm);

  \draw[foreground] (-2cm, -2.2cm) -- ++(0cm, -0.3cm);
  \draw[foreground] ( 2cm, -2.2cm) -- ++(0cm, -0.3cm);
  \draw[foreground, latex-latex] (-2cm, -2.4cm) -- node[below] {$a$} ++(4cm, 0cm);

  \draw[blue, densely dashed] (-2.1cm, -2.1cm) rectangle ++(4.2cm, 4.2cm);
  \node[right, blue] at (2.1cm, -1.05cm) {$\Gamma_D$};

  \node[below right] at (-2cm, 2cm) {$\Omega$};

  \begin{scope}[xshift = 4cm, yshift = 1cm]
    \fill[bottom color = lightblue!35!white, top color = lightblue!25!white, shading angle = -20] (-1cm, -1cm) rectangle ++(2cm, 2cm);
    \clip (-1cm, -1cm) rectangle ++(2cm, 2cm);

    \fill[red!50!background] (0cm, 0cm) circle (0.5cm);
    \draw[red] (0cm, 0cm) circle (0.5cm);

    \draw[foreground, densely dashed, opacity = 0.2] (0cm, 0cm) arc(-30:-60:125cm);
    \draw[foreground, densely dashed, opacity = 0.2] (0cm, 0cm) arc(-30:0:125cm);

    \fill[foreground] (0cm, 0cm) circle (0.04cm);
    \draw[foreground, latex-latex] (-160:0.5cm) -- (20:0.5cm);
    \draw[foreground] (0cm, 0cm) -- node[pos = 0.9, above, rotate = 20] {$R$} (20:0.8cm);
    \node[above left] at (1cm, -1cm) {\footnotesize$s = 0$};
    \node[below] at (0cm, -0.1cm) {\footnotesize$s = \lambda$};
  \end{scope}

  \node[minimum width = 0.2cm, minimum height = 0.2cm, draw = foreground, densely dotted] (probe) at (-30:1cm) {};
  \node[minimum width = 2.4cm, minimum height = 2.4cm, draw = foreground, densely dotted] (magnification) at (4cm, 1cm) {};

  \draw[foreground, densely dotted, -latex] (probe) to[bend left] (magnification);

\end{tikzpicture}
  \caption{Geometric setup of the traveling heat source problem. Left: global computational domain. Right: magnified view of the traveling circular heat source.}
  \label{fig:traveling_heat_source}
\end{figure}

The model was a time-dependent heat-transfer problem defined on a square domain, shown in \Cref{fig:traveling_heat_source}.

Inside this geometry, a small circular heat source moves along a circular path, thereby heating the domain.
The governing initial-boundary-value problem is defined by
\begin{subequations}
  \begin{align}
    \frac{\partial T}{\partial t} &= \nabla \cdot \left( \kappa \, \nabla T \right) + s 
      &&\forall \left( \boldsymbol{x} \in \Omega,\ t \in (0, t_\mathrm{max}] \right), \\
    T &= 0 
      &&\forall \left( \boldsymbol{x} \in \Gamma_D,\ t \in (0, t_\mathrm{max}] \right), \\
    T &= 0 
      &&\forall \left( \boldsymbol{x} \in \Omega,\ t = 0 \right).
  \end{align}
\end{subequations}

The geometric and physical properties were not intended to represent a specific material, but instead define an artificial test case:
\begin{equation}
  a = 10, \qquad \kappa = 1.
\end{equation}

The heat source is defined as a discontinuous function that takes the value $s = 0$ outside the circular region shown in the right panel of \Cref{fig:traveling_heat_source} and $s = \lambda$ inside.
Its size, location, and velocity parameters are
\begin{equation}
  r = 2.5, \qquad R = 0.1, \qquad \varphi_0 = -\sfrac{\pi}{2}, \qquad 
  \dot{\varphi} = \sfrac{\pi}{2}, \qquad \lambda = 10.
\end{equation}

Time integration was performed using the classical $\theta$-method with $\theta = \sfrac{1}{2}$, resulting in the semi-discrete form
\begin{equation}
  \left[ \frac{\boldsymbol{M}}{\Delta t} + \theta\,\boldsymbol{K} \right] \boldsymbol{T}_{n+1}
  = \left[ \frac{\boldsymbol{M}}{\Delta t} - (1 - \theta)\boldsymbol{K} \right] \boldsymbol{T}_n
  + \theta\,\boldsymbol{F}_{n+1}
  + (1 - \theta)\boldsymbol{F}_n.
\end{equation}

The analysis was performed using two static discretizations and one dynamic discretization.
The first static discretization used a $501 \times 501$ element mesh with $p = 1$ as a highly refined reference solution, and the second used an $11 \times 11$ mesh with $p = 4$ to represent a case without local \textit{h}-refinement.
The first discretization comprised $n = 250000$ unknowns, with the second comprising $n = 881$ unknowns.

The dynamic discretization also employed an $11 \times 11$, $p = 4$ base mesh, augmented by three overlay meshes of the same resolution ($11 \times 11$).
Each overlay mesh was reduced in size and polynomial degree by one, resulting in a total of $n = 1841$ unknowns.
The three overlay meshes were centered at the location of the heat source and moved with it in time steps of $\Delta t = 0.1$.
After each movement, the solution was transferred between successive mesh configurations using an $L^2$ projection, which corresponds to a weighted projection with $w = 1$ for the present constant-coefficient setup:
\begin{equation}
  \int \limits_\Omega w\, (T_t - T_s)\, v_t\, \mathrm{d}\Omega = 0 
  \qquad \forall\, v_t \in V_t.
\end{equation}
Here, $T_s$ denotes the temperature field in the source space before the transfer, $T_t$ denotes the projected field in the target space, and $V_t$ is the target test space.

\begin{figure}[!ht]
  \centering
  \begin{tabular}{@{} c @{\space} c @{\space} c @{} l @{}}
    $11 \times 11$, $p = 4$ 
      & unfitted ml-\textit{hp} 
      & $501 \times 501$, $p = 1$ & \\
    \includegraphics[width = 0.28\textwidth]{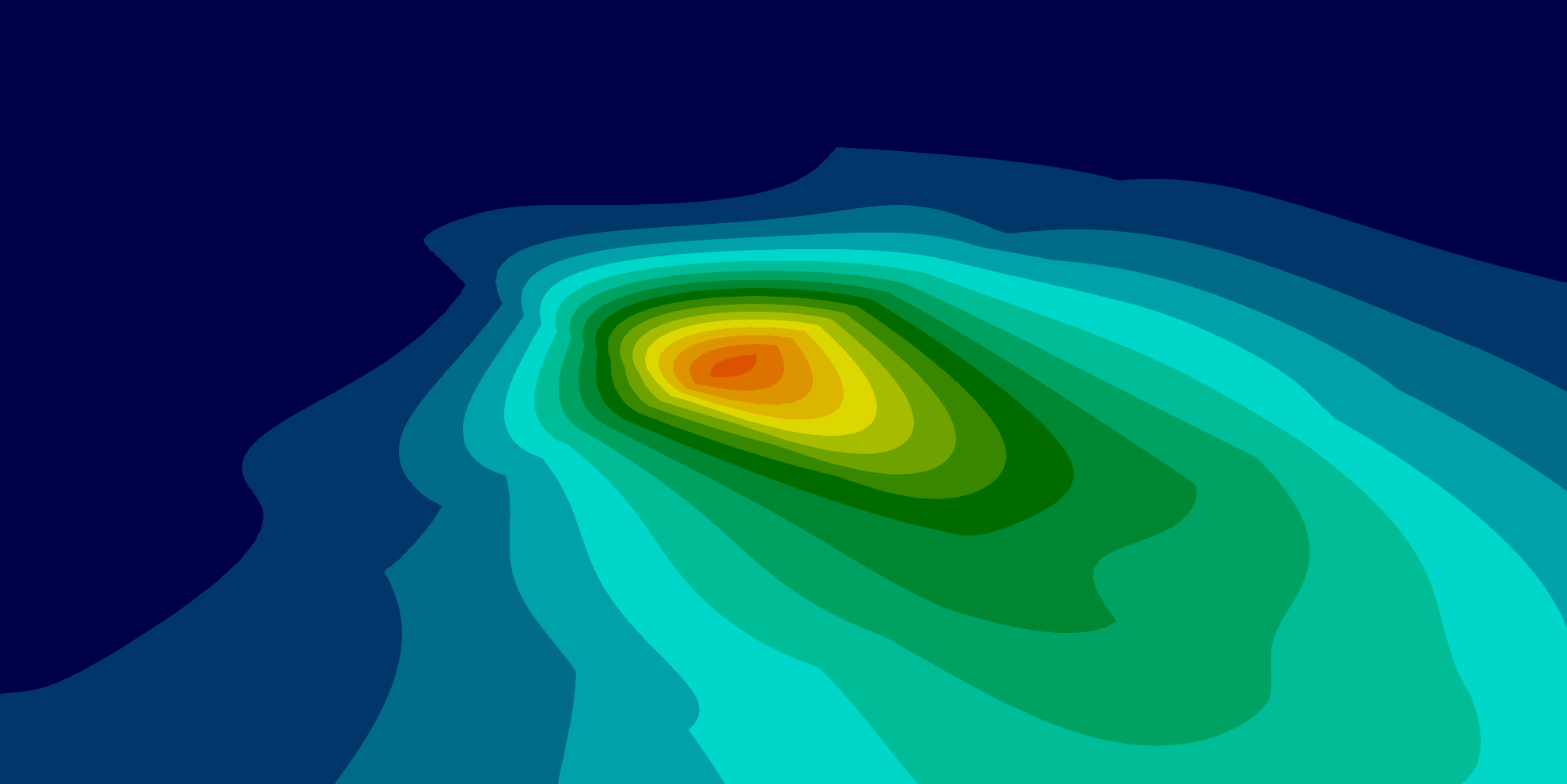} &
    \includegraphics[width = 0.28\textwidth]{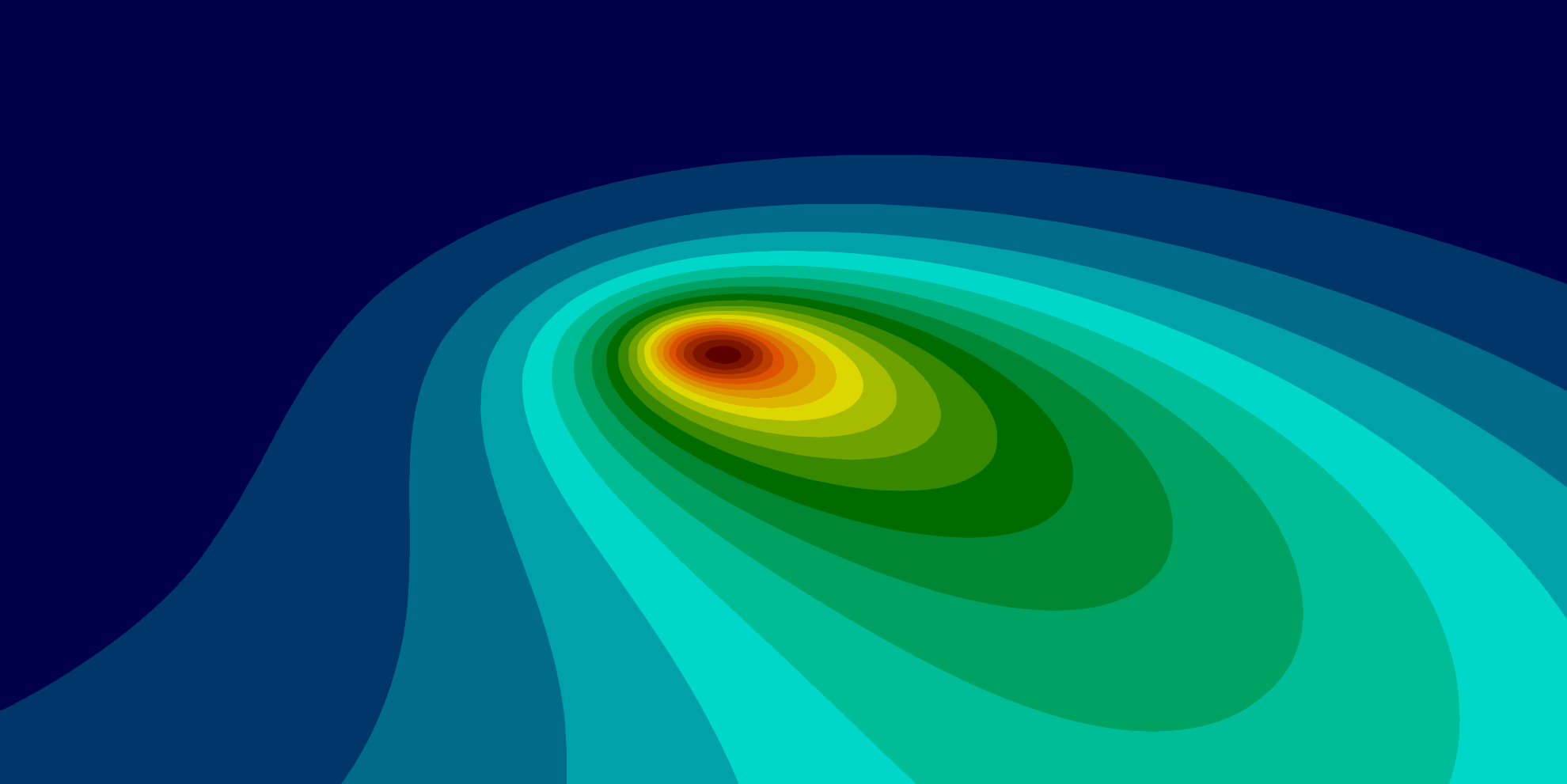} &
    \includegraphics[width = 0.28\textwidth]{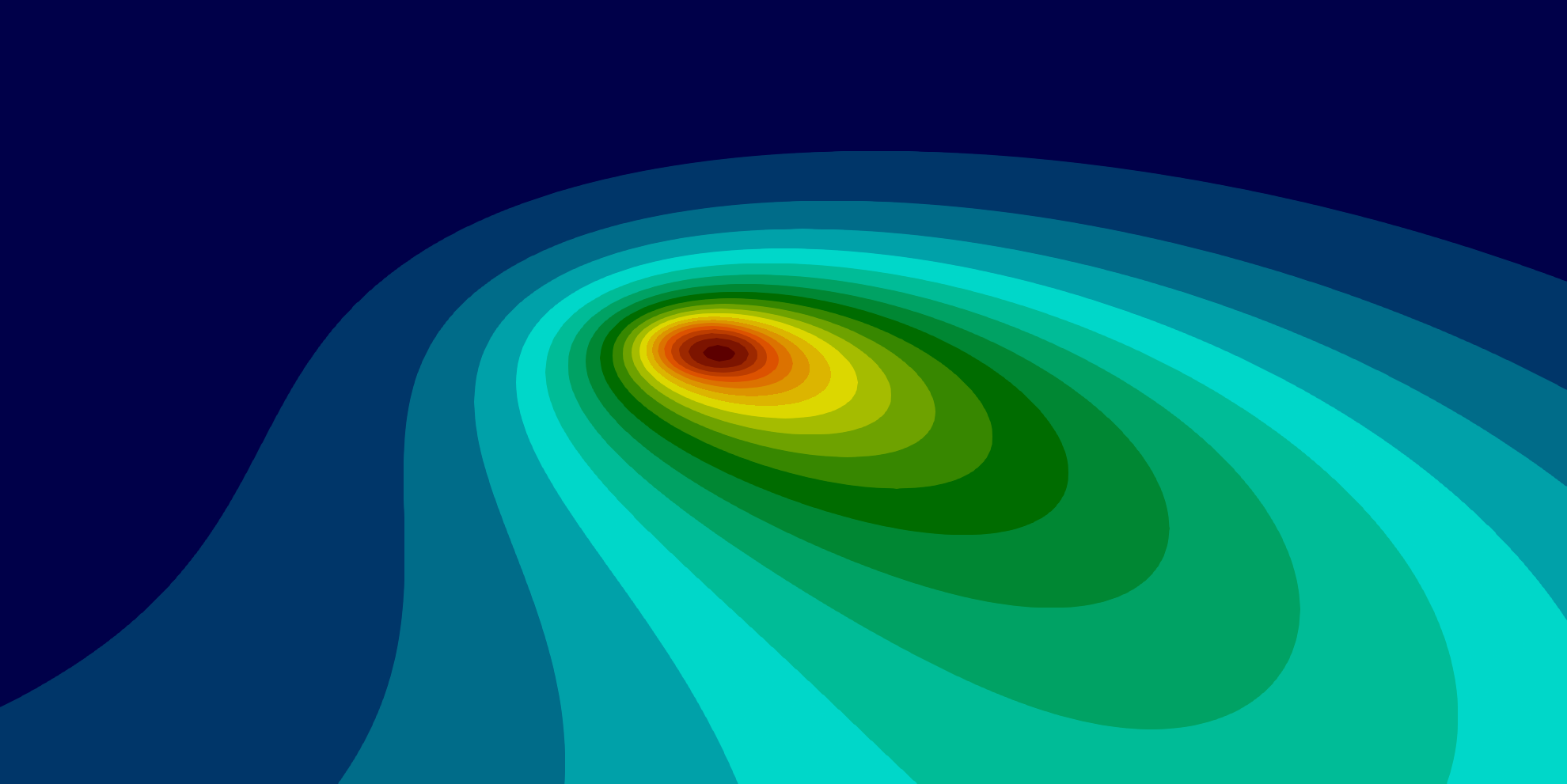} &
    \begin{tikzpicture}%
      \colorbar{0cm}{0cm}{0.0}{0.12}{1.8cm}{\phantom{$\|\nabla$}$T$\phantom{$\|_2$}}%
    \end{tikzpicture} \\[0.4mm]
    \includegraphics[width = 0.28\textwidth]{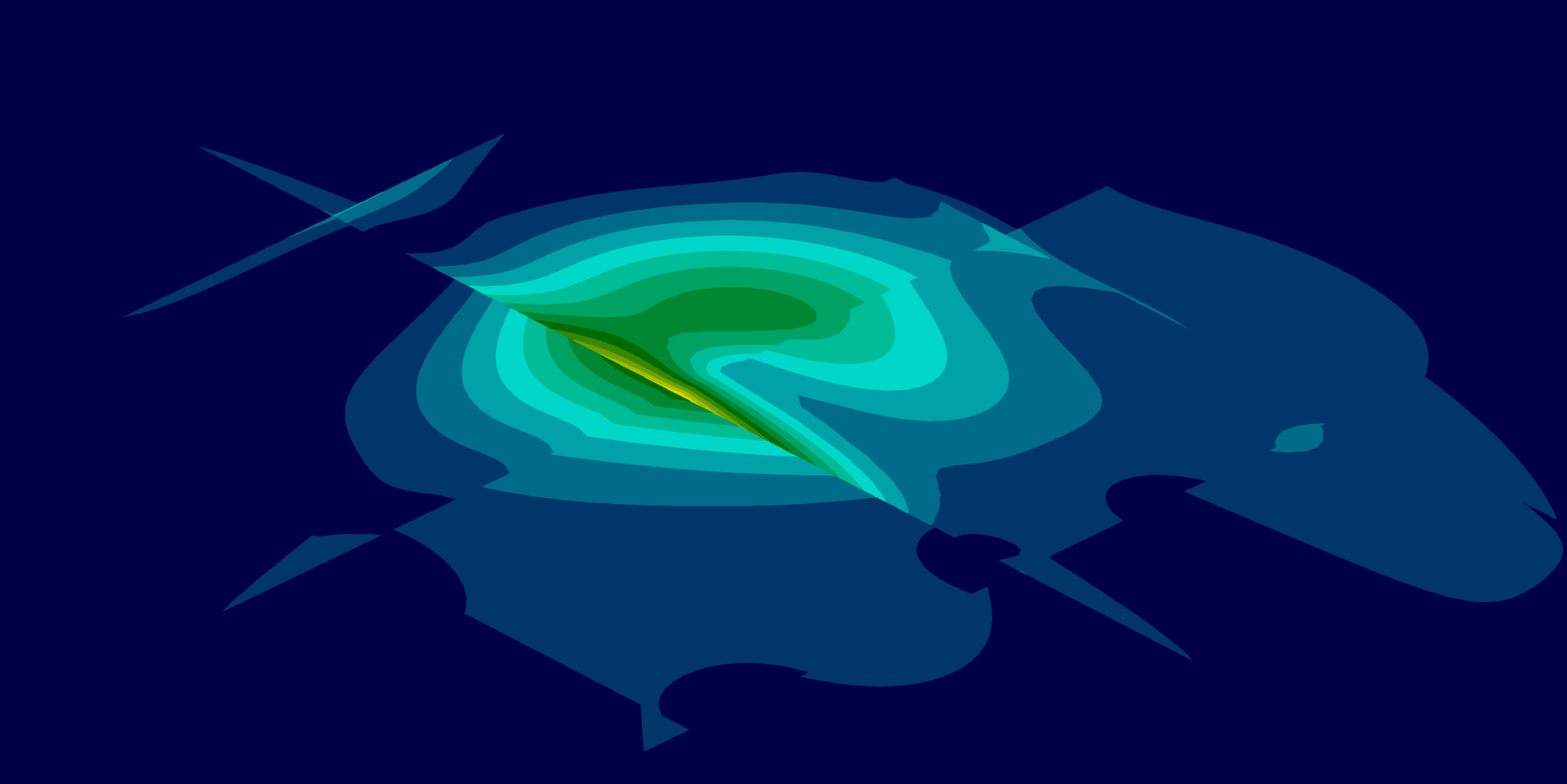} &
    \includegraphics[width = 0.28\textwidth]{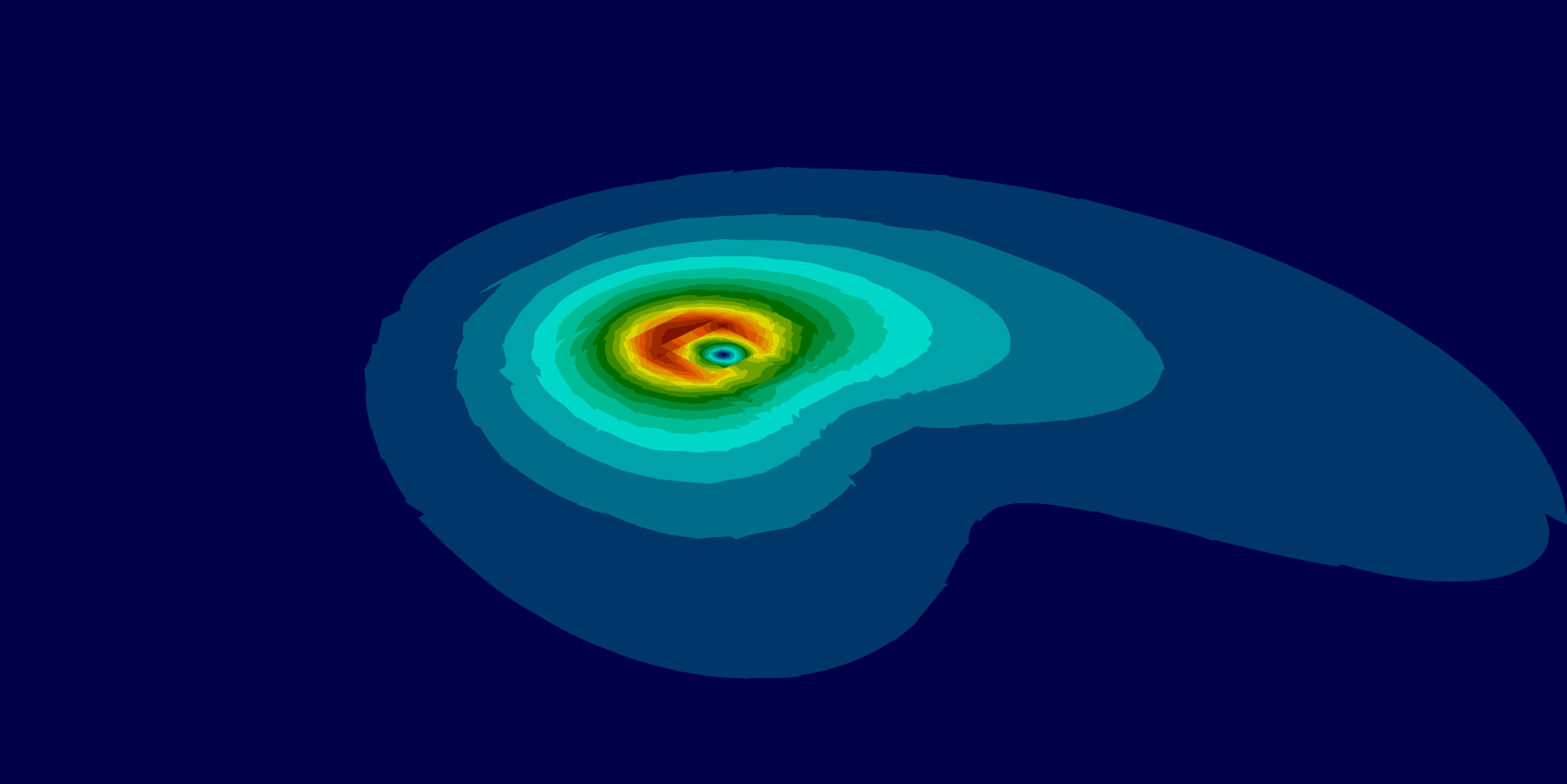} &
    \includegraphics[width = 0.28\textwidth]{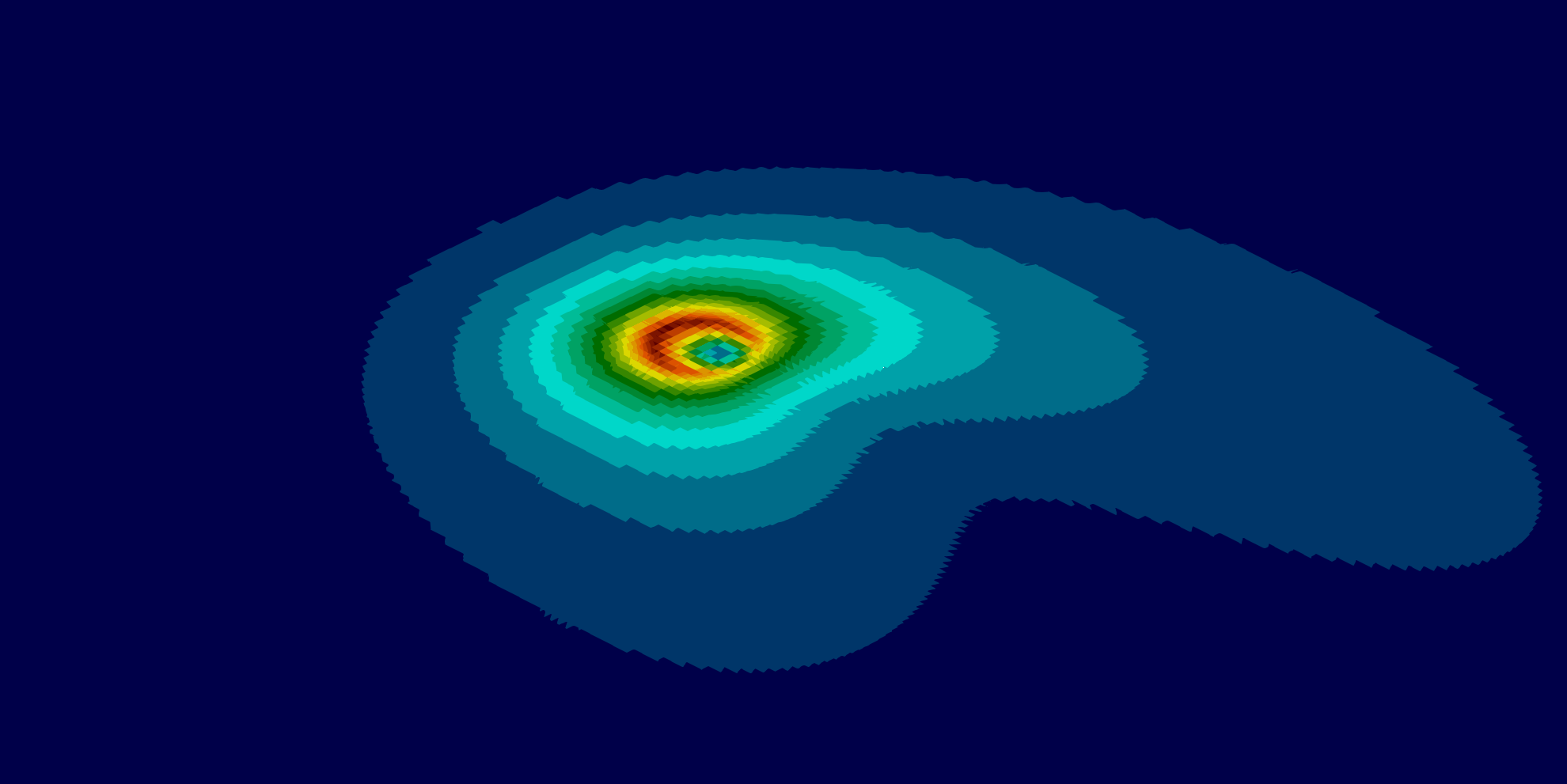} &
    \begin{tikzpicture}%
      \colorbar{0cm}{0cm}{0.0}{0.55}{1.8cm}{$\| \nabla T \|_2$}%
    \end{tikzpicture} \\[0.4mm]
    \includegraphics[width = 0.28\textwidth]{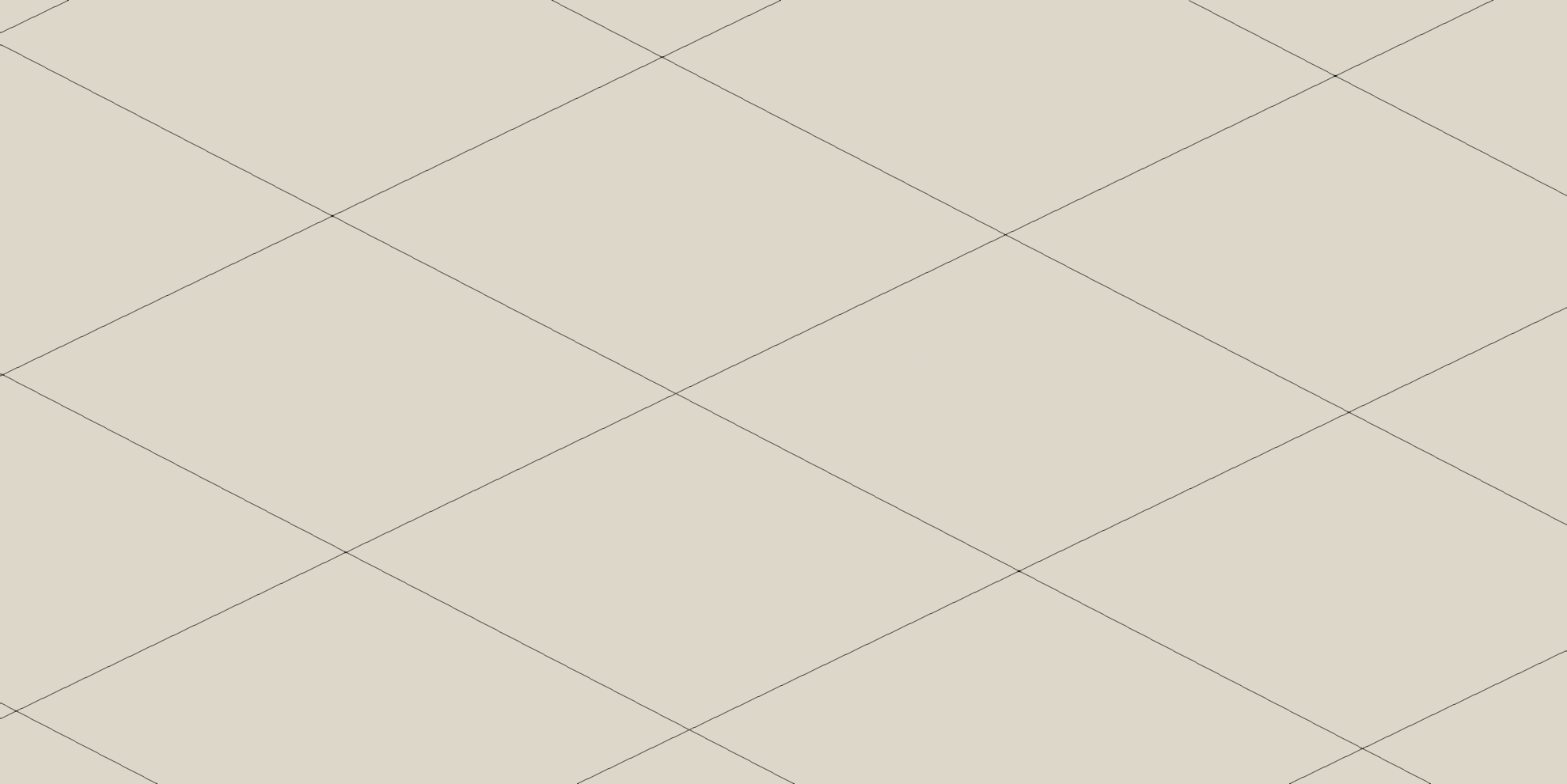} &
    \includegraphics[width = 0.28\textwidth]{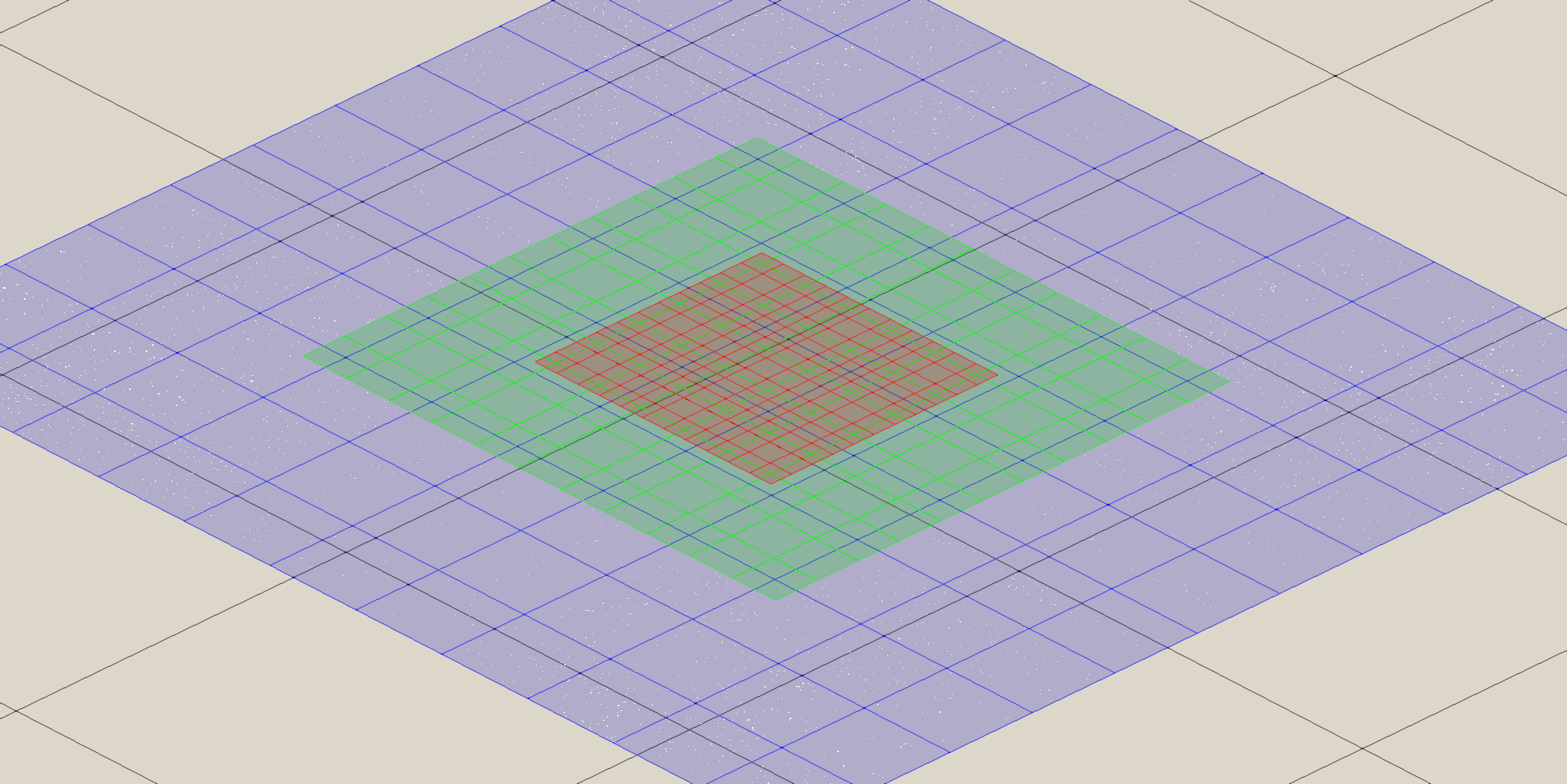} &
    \begin{tikzpicture}[spy using outlines={white, circle, size=1cm, magnification=6, connect spies}]
      \node[outer sep = 0, inner sep = 0] at (0, 0) {\includegraphics[width = 0.28\textwidth]{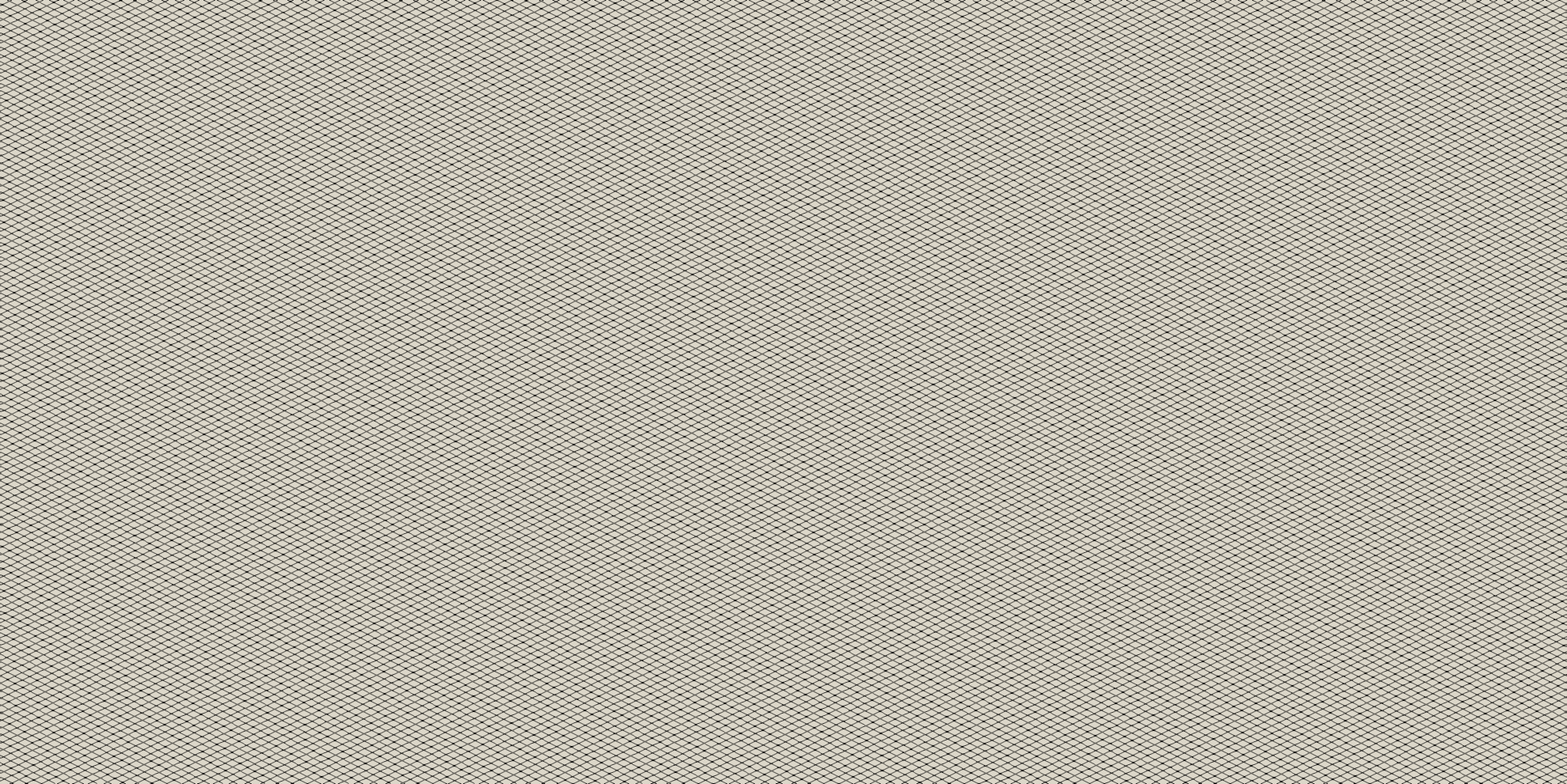}};
      \spy[height = 1cm, width = 1cm] on (-1.0cm, -0.5cm) in node[fill = white] at (1.0cm, 0.2cm);
    \end{tikzpicture} & \\
    $881$ unknowns & $1841$ unknowns & $250000$ unknowns & \\
  \end{tabular}
  \caption{Numerical results for different meshing strategies of the traveling heat source problem at $t = 2.134$. 
  The top row shows the temperature field $T$, the second row the temperature gradient magnitude $\|\nabla T\|_2$, and the bottom row the corresponding finite element meshes. 
  The first column shows the unrefined $11 \times 11$ mesh with degree $p = 4$, the second column the same base mesh with three unfitted overlay meshes shrinking toward the heat source, and the third column the low-order reference solution computed on a $501 \times 501$ mesh with $p = 1$.}
  \label{fig:traveling_heat_source_visual_results}
\end{figure}

\paragraph{Visual assessment}
\Cref{fig:traveling_heat_source_visual_results} shows the visual results obtained from the three meshing strategies at time $t = 2.134$ with $t_\mathrm{max} = 4$.

The upper row visualizes the temperature field $T$, the second row shows the corresponding temperature gradient magnitude $\| \nabla T \|_2$, and the third row displays the meshes used for the respective solutions.
The three columns correspond to the unrefined high-order mesh, the locally refined high-order mesh, and the low-order reference solution.

Comparison with the reference solution shows that the unrefined high-order mesh cannot accurately represent either the temperatures or their gradients.
Notably, the unrefined high-order mesh required substantial over-integration to correctly evaluate the heat source, an adjustment not required for the other two models.

In contrast, the locally refined high-order mesh provides a solution in excellent agreement with the reference solution in both $T$ and $\|\nabla T\|_2$, while using only $0.74\,\%$ of the reference model's unknowns.

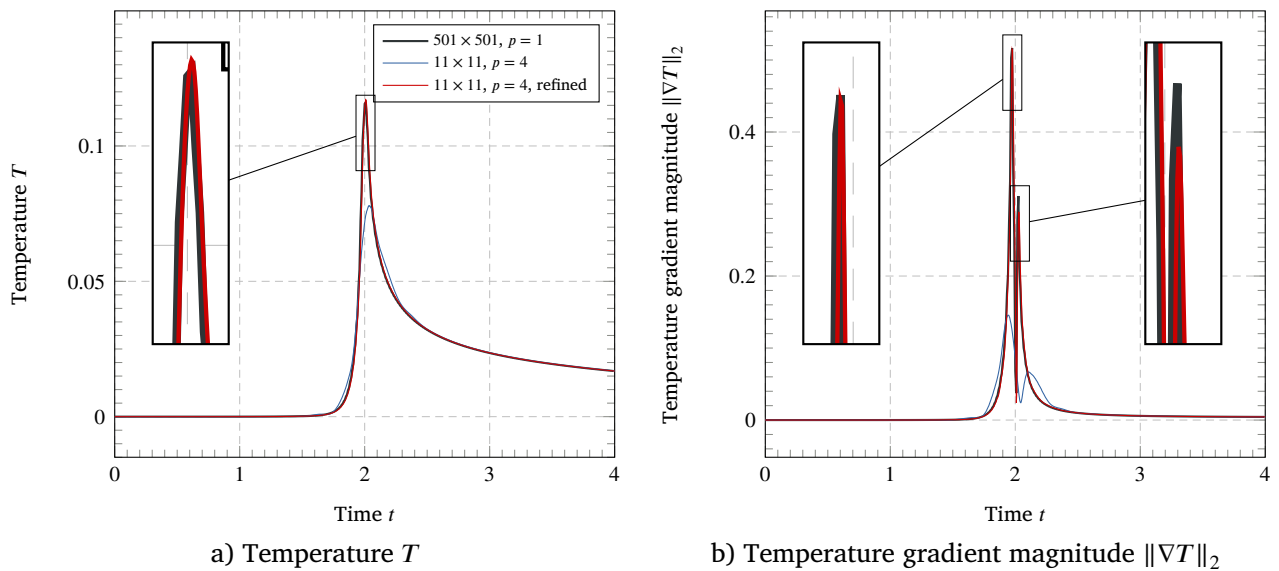
\begin{figure}[!ht]
  \centering
  \begin{tabular}{c @{\space} c}
    \tikzsetnextfilename{traveling_heat_source_t_plot}
\begin{tikzpicture}[spy using outlines={black, rectangle, size=2cm, magnification=4, connect spies}]

	\begin{axis} [
		font={\footnotesize},
		axis lines = box,
		xlabel = Time $t$,
		ylabel = Temperature $T$,
		width = 0.48\textwidth,
		height = 7.5cm,
		max space between ticks = 40,
		grid = major,
		grid style = {densely dashed, line width = 0.1pt},
		minor x tick num = 9,
		minor y tick num = 9,
    xmin = 0,
    xmax = 4,
    ymax = 0.1499,
    yticklabel style = {/pgf/number format/fixed},
		legend pos = north east,
		legend columns = 1,
		legend style = {
			nodes = {scale = 0.75, transform shape},
			anchor = north east,
		},
		legend cell align={left},
	]

		\addplot[darkgray, thick] table [x expr = \thisrow{Row ID} / 120, y = avg(T), col sep = comma] {04_Z_data_traveling_heat_source_probe_501x501_p1.csv};
		\addplot[blue] table [x expr = \thisrow{Row ID} / 360, y = avg(T), col sep = comma]            {04_Z_data_traveling_heat_source_probe_11x11_p4_r0.csv};
		\addplot[red] table [x expr = \thisrow{Row ID} / 360, y = avg(T), col sep = comma]             {04_Z_data_traveling_heat_source_probe_11x11_p4_r3.csv};

    \legend{
      $501 \times 501${,} $p = 1$,
      $11 \times 11${,} $p = 4$,
      $11 \times 11${,} $p = 4${,} refined
		};

	\end{axis}

  \spy[height = 4cm, width = 1cm] on (3.3cm, 4.3cm) in node[fill = white] at (1.0cm, 3.5cm);

\end{tikzpicture} &
    \tikzsetnextfilename{traveling_heat_source_nabla_t_plot}
\begin{tikzpicture}[spy using outlines={black, rectangle, size=2cm, magnification=4, connect spies}]
	\begin{axis} [
		font={\footnotesize},
		axis lines = box,
		xlabel = Time $t$,
		ylabel = Temperature gradient magnitude $\| \nabla T \|_2$,
		width = 0.48\textwidth,
		height = 7.5cm,
		max space between ticks = 40,
		grid = major,
		grid style = {densely dashed, line width = 0.1pt},
		minor x tick num = 9,
		minor y tick num = 9,
    xmin = 0,
    xmax = 4,
		legend pos = north east,
		legend columns = 4,
		legend style = {
			nodes = {scale = 0.6, transform shape},
			anchor = north east,
		},
		legend cell align={left},
	]

		\addplot[darkgray, thick] table [x expr = \thisrow{Row ID} / 120, y = avg(T Gradient (Magnitude)), col sep = comma] {04_Z_data_traveling_heat_source_probe_501x501_p1.csv};
		\addplot[blue] table [x expr = \thisrow{Row ID} / 360, y = avg(T Gradient (Magnitude)), col sep = comma]            {04_Z_data_traveling_heat_source_probe_11x11_p4_r0.csv};
		\addplot[red] table [x expr = \thisrow{Row ID} / 360, y = avg(T Gradient (Magnitude)), col sep = comma]             {04_Z_data_traveling_heat_source_probe_11x11_p4_r3.csv};

	\end{axis}

  \spy[height = 4cm, width = 1cm] on (3.25cm, 5.1cm) in node[fill = white] at (1.0cm, 3.5cm);
  \spy[height = 4cm, width = 1cm] on (3.35cm, 3.1cm) in node[fill = white] at (5.5cm, 3.5cm);

\end{tikzpicture} \\
    a) Temperature $T$ & b) Temperature gradient magnitude $\|\nabla T\|_2$ \\
  \end{tabular}
  \caption{Temperature $T$ (a) and temperature gradient magnitude $\|\nabla T\|_2$ (b) probed at $\boldsymbol{x} = (0, 2.5)$ for all computed time steps. 
  The black curve shows the low-order reference solution, blue the unrefined high-order solution, and red the locally refined high-order solution.}
  \label{fig:traveling_heat_source_time_plots}
\end{figure}

\paragraph{Quantitative assessment}
A quantitative assessment was performed by probing the temperature and temperature-gradient values from all three models at the spatial location $\boldsymbol{x} = [0,\ 2.5]$ over the full simulation time.
The results are shown in \Cref{fig:traveling_heat_source_time_plots}.

Panel (a) shows the temperature values over time: the reference solution in black, the unrefined high-order solution in blue, and the locally refined solution in red.
The unrefined high-order model severely underestimates the maximum temperatures near $t = 2$, whereas the refined model reproduces the reference values almost exactly.

Panel (b) displays the corresponding gradient magnitudes.
The unrefined high-order solution dramatically underestimates the peak gradients both before and after the heat source passes, leading to unreliable results.
In contrast, the locally refined solution closely matches the reference gradients, although the deviations are slightly larger than those observed in the temperature field.
At the probe point, the reference solution reaches a maximum temperature of $0.11605$ and a maximum gradient magnitude of $0.51653$.
The unrefined high-order model reaches only $0.07798$ and $0.14564$, corresponding to underestimations of $32.8\,\%$ and $71.8\,\%$, respectively.
The locally refined model reaches $0.11712$ and $0.51449$, corresponding to relative deviations of $0.92\,\%$ in the peak temperature and $0.39\,\%$ in the peak gradient magnitude.

  \section{Summary \& Conclusion}

This manuscript presents an unfitted multi-level \textit{hp}-refinement strategy for the numerical solution of problems with localized features.
The method extends classical refinement-by-superposition approaches by allowing non-matching overlay meshes, thereby reducing the topological constraints inherent to fitted multi-level \textit{hp} formulations.

The main contributions and findings are summarized as follows:
\begin{itemize}
  \item We introduced a refinement approach based on the superposition of independently positioned overlay meshes, which generalizes classical fitted multi-level \textit{hp} approaches.
  In contrast to fitted strategies, the proposed method permits partial element coverage and does not transfer the deactivation rules used in fitted methods for linear independence directly to the unfitted setting, thereby increasing the flexibility of the overlay mesh design.

  \item The numerical studies demonstrated that the proposed approach achieves exponential convergence behavior for the investigated problems with discontinuities and singularities.
  In particular, benchmark results show that the unfitted strategy attains substantially lower errors per degree of freedom compared to fitted multi-level \textit{hp}-refinement, for example by reducing the error by orders of magnitude while using less than half the unknowns in the one-dimensional benchmark reported in \Cref{fig:benchmark_elastic_bar_errors}. 

  \item By enabling strict localization of refinement regions, the method avoids the introduction of unnecessary degrees of freedom in unaffected parts of the domain.
  This is particularly evident in the singular corner problem, where equivalent accuracy was achieved with significantly fewer unknowns due to the omission of superfluous elements. 

  \item The introduction of a refinement parameter governing the size of the overlay mesh provides an additional degree of freedom to tune approximation properties.
  The results indicate that appropriately chosen overlay mesh sizes can substantially accelerate convergence beyond that of classical fitted strategies. 

  \item The method was applied to a time-dependent heat-transfer problem with a moving heat source.
  The unfitted formulation enables dynamic refinement without remeshing, yielding solutions in close agreement with a highly resolved reference while using less than 1\% of the degrees of freedom.
  This highlights the method's suitability for applications such as additive manufacturing. 

  \item The investigation of partially overlapping meshes characterizes the conditioning behavior of the proposed formulation across a wide range of configurations.
  Although very small overlap regions may affect the conditioning of the system matrix, particularly at higher polynomial degrees, this behavior is consistent with established observations in immersed methods.
  In the computations performed here, no significant increase in computational cost was observed for the moving-overlay examples.
\end{itemize}

The current study establishes the basic framework for extending fitted multi-level \textit{hp}-refinement to non-matching overlay meshes.
Future work should address automated refinement indicators for localized and moving features, a more systematic sensitivity analysis of conditioning in small-overlap configurations, and stabilization or preconditioning techniques for cases in which such configurations cannot be avoided.

  \section*{Acknowledgments}
  Jan Niklas Schmäke and Martin Ruess gratefully acknowledge the support provided by the Deutsche Forschungsgemeinschaft (DFG, German Research Foundation) under the funding code RU 885/4-1, project number 515687474.

  \clearpage
  \bibliography{99_references.bib}

\end{document}